\documentclass[a4paper,11pt]{article}
\pdfoutput=1 

\usepackage[no-natbib-sort]{jheppub}

\usepackage{youngtab}
\usepackage{hyperref}
\usepackage{fancybox}
\usepackage{float}
\usepackage{amsfonts}
\usepackage{amsmath}
\usepackage{amsbsy}
\usepackage{graphicx}
\usepackage{amssymb}
\usepackage{amsthm}
\usepackage{bm}
\usepackage{comment}
\usepackage{epsfig}
\usepackage{latexsym}
\usepackage{pdflscape}
\usepackage{cancel} 
\usepackage{color}
\usepackage{cleveref}

\makeatletter
\def\@fpheader{\relax}
\makeatother

\DeclareSymbolFont{AMSa}{U}{msa}{m}{n}
\DeclareSymbolFont{AMSb}{U}{msb}{m}{n}
\DeclareMathSymbol{\fieldR}{\mathalpha}{AMSb}{"52}

\newcommand{\beq}{\begin{eqnarray}}
\newcommand{\eeq}{\end{eqnarray}}
\newcommand{\bea}{\begin{eqnarray}}
\newcommand{\eea}{\end{eqnarray}}
\newcommand{\be}{\begin{equation}}
\newcommand{\ee}{\end{equation}}
\newcommand{\bq}{\begin{equation}}
\newcommand{\eq}{\end{equation}}

\newcommand{\nn}{\nonumber}

\def\o{\omega}

\def\m{\mu}

\def\6{\partial}
\def\de{\partial}

\def\6{\partial}

\title {Hunting for fermionic instabilities in charged AdS black holes}

\author[a]{Oscar J. C. Dias,}
\affiliation[a]{STAG research centre and Mathematical Sciences, University of Southampton, UK}

\author[a]{Ramon Masachs,}

\author[a,b]{Olga Papadoulaki}
\affiliation[b]{International Centre for Theoretical Physics, Strada Costiera 11, 34151 Trieste, Italy}

\author[a]{and Paul Rodgers}

\emailAdd{ojcd1r13@soton.ac.uk}
\emailAdd{rmg1e15@soton.ac.uk}
\emailAdd{papadoulaki@ictp.it}
\emailAdd{pwr1u17@soton.ac.uk}

\abstract{Fermions scattering on a black hole background cannot develop an instability sourced by superradiance. 
However, in a global (or planar) AdS$_4$-Reissner-Nordstr\"om background fermions can violate the AdS$_2$ fermionic mass stability bound as measured by a near horizon observer at zero temperature. This suggests that AdS-Reissner-Nordstr\"om black holes might still be unstable to Dirac perturbations. Motivated by this observation we search for linear mode instabilities of Dirac fields in these backgrounds but find none. This is in contrast with the scalar field case, where a violation of the near-horizon Breitenl\"ohner-Freedman stability bound in the  AdS-Reissner-Nordstr\"om background triggers the already known scalar condensation near-horizon linear instability (in the  planar limit this is Gubser's instability that initiated the holographic superconductor programme). We consider both the standard and alternative AdS/CFT quantizations (that preserve the conformal invariance of AdS). These are reflective boundary conditions that have vanishing energy flux at the asymptotic boundary.      
}

\begin{document} 
\maketitle
\flushbottom

\section{Introduction}\label{sec:Intro}

It is a well known fact that bosonic waves impinging charged or rotating black holes can be amplified via superradiant scattering (see {\it e.g.} the review \cite{Brito:2015oca} and references therein). It follows that black holes perturbed by bosonic fields in the presence of a gravitational potential well $-$ provided by, for example, an asymptotic anti-de Sitter (AdS) potential, a physical cavity at finite radius or by the mass of the field $-$ can develop a superradiant instability. However, this is not the only instability that can be present in such systems. Indeed, the family of superradiant black holes always has a configuration with zero temperature. Typically, such extreme black holes have a near-horizon geometry that is the direct product (or a fibration) of a base space ({\it e.g.} a sphere) and an AdS$_2$ space \cite{Bardeen:1999px}. These extreme (and near-extreme) black holes, when confined in a gravitational wall, are unstable if the `{\it effective} mass of the perturbation (as seen by an AdS$_2$ observer) violates the 2-dimensional Breitenl\"ohner-Freedman (BF) bound for stability \cite{Breitenlohner:1982jf,Breitenlohner:1982bm,Klebanov:1999tb,Ishibashi:2004wx} (even though the asymptotic AdS$_4$ BF bound is obeyed). Recall that perturbations with a mass below this bound are normalizable (i.e. they have finite conserved energy) but their energy is negative and, consequently, they can trigger an instability. The superradiant and near-horizon instabilities have a different physical nature. But, in a non-extremal black hole they are usually entangled. However, if $L$ is the typical dimension of the gravitational well ({\it e.g.} the radius of the cavity or the AdS radius), they disentangle for small dimensionless horizon radius $r_+/L$ since the near-horizon instability is  suppressed for $r_+/L\ll 1$, while the superradiant instability is still present \cite{Gubser:2008px,hartnoll2008building,Faulkner:2009wj,Dias:2010ma,dias2012hairyBHs,Dias:2016pma,Dias:2018zjg}.  

Not less well known is the fact that fermionic waves, unlike bosons, cannot suffer from superradiant scattering amplification  \cite{Klein:1929zz,Sauter:1931zz,Winter:1959,Unruh:1973bda,Holstein:1998,Brito:2015oca}. Perhaps less familiar is the fact that fermions, like bosons  \cite{Gubser:2008px,hartnoll2008building,Faulkner:2009wj,Dias:2010ma}, can also violate the 2-dimensional stability bound of the near-horizon geometry of near-extremal black holes, if the fermion charge is high enough (for fermions the stability bound is lower than the BF bound). In particular, this can happen in a Reissner-Nordstr\"om (RN) black hole in an asymptotically planar AdS background  \cite{Faulkner:2009wj,Iqbal:2009fd,Lee:2008xf,Liu:2009dm,Guarrera:2011my,Iqbal:2011ae,Hartnoll:2016apf,Cubrovic:2009ye} (see in particular section V of \cite{Faulkner:2009wj} and sections 4.4 and 4.5 of \cite{Hartnoll:2016apf}) or in an asymptotically global AdS geometry (section \ref{sec:DiracRN} below). This suggests that such a RN black hole {\it might} be unstable to condensation of a fermionic cloud around the horizon. Moreover, {\it if} the features from the bosonic field extend to the fermion case, then this might  be a  linear instability. 

Motivated by these considerations, in the present manuscript we will search for {\it linear} mode instabilities of Dirac fields in a global AdS$_4$ RN black hole. We will not find any such instabilities.  The absence of linear mode instabilities in the global AdS RN background 
is consistent with the fact that they are also not present in the planar AdS limit, $r_+/L\to \infty$, as found previously in \cite{Faulkner:2009wj,Iqbal:2009fd,Lee:2008xf,Liu:2009dm,Guarrera:2011my,Iqbal:2011ae,Hartnoll:2016apf,Cubrovic:2009ye}. In view of these  `no-go' findings, in the conclusion remarks of section \ref{sec:Conc}, we will argue that the absence of {\it linear} mode instabilities in the  AdS RN background still leaves room (but not necessarily) for the following possibility:  for a large number of fermions and in the semiclassical limit, the violation of the AdS$_2$ BF bound might signal a {\it non}-linear instability of the system. 

In this manuscript, we also take the opportunity to improve the understanding of the near-horizon  condensation instability of scalar fields. In particular, following a similar analysis for scalar fields in a Minkowsky cavity \cite{Dias:2018zjg}, we will explicitly show that the BF bound criterion for instability in AdS-RN is {\it quantitatively sharp} (section \ref{sec:NH}). This scalar field analysis will fit smoothly in our presentation since reviewing its details will also allow to make a direct comparison with the Dirac field case and pinpoint major differences between them. We further take the opportunity to show that the unstable modes belong to a family of {\it near-extremal modes} that are connected to the {\it normal modes} of AdS when the dimensionless horizon radius shrinks to zero. (This is an interesting observation because in {\it de Sitter} black holes the `normal mode de Sitter family' is {\it distinct} from the `near-extremal family of modes', where we are using the nomenclature of \cite{Cardoso:2017soq,Dias:2018ynt,Dias:2018etb}).

Necessarily, we will also clarify some misleading analyses and interpretations that were presented in previous literature. Namely, the authors of \cite{Wang:2017fie}\footnote{See also Ref. \cite{Wang:2019qja}, which appeared after the present manuscript was submitted to the ArXiv.} missed that the vanishing energy flux boundary conditions at the asymptotic boundary of AdS $-$ that they propose to be ``novel" $-$ are nothing else but the AdS/CFT correspondence no-source boundary conditions $-$ often denoted as `{\it standard}' or, if allowed, `{\it alternative}' quantizations (discussed, for $s=1/2$, originally in \cite{Mueck:1998iz,Henningson:1998cd,Henneaux:1998ch,Contino:2004vy,Cubrovic:2009ye} and specially in  \cite{Breitenlohner:1982jf,Breitenlohner:1982bm,Klebanov:1999tb,Ishibashi:2004wx,Amsel:2008iz}; this discussion applies both to bosonic and fermionic fields). The {\it vanishing energy flux} boundary condition is fundamental to guarantee that the energy is {\it conserved}. If and only if this is the case, the Schr\"odinger operator that describes the (bosonic or fermionic) wave equation in AdS is Hermitian\footnote{Without the zero flux or energy conservation condition, the Schr\"oedinger operator is only symmetric \cite{Ishibashi:2004wx}.}. It follows that {\it finiteness of energy} then boils down to simply require that the wavefunctions of the system have finite norm in the usual quantum mechanical sense, i.e. that the solutions are {\it normalizable} (square integrable). In these conditions, the Schr\"odinger operator of the system is self-adjoint (the associated matrix is Hermitian) and we have a well-posed initial value problem (after imposing regularity at the inner boundary). That is, the dynamical evolution of the system is deterministic. 

If the energy is positive, the evolution is {\it stable} (this happens if we are above the stability mass bound \cite{Ishibashi:2004wx,Amsel:2008iz}); on the other hand, if the energy is negative we should have a dynamical evolution that develops an {\it instability} (this is the case if the mass of the perturbation is below the  stability mass bound  \cite{Ishibashi:2004wx,Amsel:2008iz}).\footnote{The stability mass bound for scalars is the BF bound $m^2=m^2_{BF}$  \cite{Breitenlohner:1982jf,Breitenlohner:1982bm,Klebanov:1999tb,Ishibashi:2004wx}, while for Dirac fields it is $m^2=0$  \cite{Amsel:2008iz}; see discussion of \eqref{BFfermionic}.}   
The aforementioned homogeneous Dirichlet (standard) or Neumann (alternative) AdS/CFT boundary conditions are special in the sense that, by construction, they yield zero energy flux normalizable modes that preserve the conformal symmetry group of AdS (and thus do not deform the boundary conformal field theory). The zero-flux boundary conditions of \cite{Wang:2017fie,Wang:2019qja} are nothing but these single-trace homogeneous boundary conditions \cite{Cubrovic:2009ye,Mueck:1998iz,Henningson:1998cd,Henneaux:1998ch,Contino:2004vy,Amsel:2008iz,Amsel:2009rr,Andrade:2011dg}. Besides these, the AdS/CFT correspondence literature identifies, for a certain range of the boson/fermion masses, other {\it normalizable} modes (finite conserved energy modes and thus with vanishing energy flux). This is the case of the inhomogeneous Dirichlet and Neumann boundary conditions but also of the often denoted mixed, Robin or multi-trace boundary conditions (see \cite{Ishibashi:2004wx} for bosons and \cite{Amsel:2008iz,Amsel:2009rr,Andrade:2011dg} for fermions). These zero-flux boundary conditions break the AdS conformal symmetry while still preserving its Poincar\'e symmetry subgroup.

The plan of this manuscript is the following. In sections \ref{sec:AdS-RN}-\ref{sec:DiracAdS-RN}, we will review the Dirac equation in a AdS-RN background and we will do the necessary field redefinitions in the (physical) Dirac field that allow it to separate and even decouple. Then, in section \ref{sec:BCsDiracAdS-RN} we will analyse in detail the AdS/CFT standard and alternative quantizations of a Dirac field. In particular, we will check that the requirement that the source vanishes implies that the energy flux at the conformal boundary also vanishes. In this sense, these boundary conditions can also be denoted as {\it reflective} boundary conditions.  For a massive Dirac field these no-source boundary conditions translate into homogeneous Dirichlet or Neumann boundary conditions in the auxiliary decoupled Dirac radial fields. However, for a massless fermion (a Weyl field) these no-source boundary conditions $-$ which are homogeneous Dirichlet or Neumann conditions on appropriate projections of the original {\it physical} Dirac field $-$ translate into mixed (Robin) boundary conditions for the {\it auxiliary} decoupled  Dirac radial fields. The misleading focus on the boundary condition for the {\it auxiliary} fields (and associated consequences) occurs recurrently. It is the case in \cite{Wang:2017fie} and it is similar to the one taken on the boundary conditions of the Regge-Wheeler$-$Zerilli master fields (aka Kodama-Ishibashi fields) in the case of {\it gravitational} perturbations of AdS black holes (as discussed in \cite{Michalogiorgakis:2006jc,Dias:2013sdc}).\footnote{For a discussion that AdS/CFT no-source boundary conditions for bosonic fields yield (`reflective') solutions with vanishing energy (and momentum) flux at the conformal AdS boundary see Appendix A of \cite{Cardoso:2013pza}.} In this context it is also important to clarify that in \cite{Giammatteo:2004wp,Jing:2005ux} massless Dirac quasinormal modes in Schwarschild-AdS were computed imposing Dirichlet boundary conditions in the auxiliary decoupled fields. These boundary conditions do not have vanishing energy flux at the asymptotic boundary, the energy of the system is thus not conserved,  and it is not known what deformation they produce. 
Finally, in section \ref{sec:DiracRN3} we will describe our strategy to search (unsuccessfully) for linear mode instabilities (eventually sourced by the 2-dimensional stability bound violation) of Dirac fields in global AdS$_4$-RN black holes. We consider both the standard and alternative quantizations and we will highlight the differences between the scalar and fermion systems. For a reader interested in a future detailed analysis of the frequency spectra, we also compute (analytically) the normal modes of massive and massless Dirac field in global AdS.

\section{Global AdS Reissner-Nordstr\"om black hole and the Dirac equation}\label{sec:WarmUp}

\subsection{AdS-RN black holes and an orthogonal {\it vierbein}}\label{sec:AdS-RN}

The gravitational $g_{\mu\nu}$ and Maxwell $A_\mu$ fields of the AdS-RN BH are described by\footnote{\label{foot:action}This is a solution of the Lagrangian ${\cal L}=\sqrt{-g}\left(R - 2\Lambda  -\frac{1}{2}F^2\right)$ with $F=dA$. Note that if we rescale $A\to \kappa A$ then the charge  $q$ of the perturbation field (to be discussed in later sections) rescales as $q\to \kappa^{-1} q$ so that $qA$, and thus the gauge covariant derivative, remain invariant.}
\begin{align}
 \label{RNAdS}
&{\rm d}s^2= -f {\rm d}t^2+ f^{-1}dr^2+ r^2 \, {\rm d} \Omega_{2}^2, \qquad f = \frac{r^2}{L^2}+1-\frac{2 M}{r} +\frac{Q^2}{2 r^2};  \nonumber \\
&A_{\mu} \, dx^{\mu} = A(r) \hspace{0.3em} {\rm d}t, \qquad A(r)= -\frac{Q}{r}+C;
\end{align}
where ${\rm d} \Omega_{2}^2$ is the line element of a unit radius 2-sphere, $M$ and $Q$ are the mass and charge parameters. 
 We will find convenient to replace $M$ and $Q$ by the event horizon radius $r_+$ (where $f(r_+)=0$) and chemical potential $\mu$. The relation between these two pairs of parameters is
\be
M=r_+\left(1+\frac{r_+^2}{L^2}\right)+\frac{1}{2}\,r_+ \mu^2\,, \qquad  Q=\mu \,r_+\,.
\ee
In \eqref{RNAdS}, $C$ is an arbitrary integration constant and fixing it amounts to choosing a particular gauge. One common gauge choice is $C=0$ where one has $A |_{\infty}=0$ and the chemical potential is $\mu=-A |_{r_{+}}=Q/r_+$ (we will typically use this one when presenting our results). Another gauge that is also commonly used is  $C=\mu$ whereby  $A |_{r_{+}}=0$ and $A |_{\infty}=\mu$.

The temperature of this black hole is 
\be
T_H=\frac{1}{8 \pi }\frac{1}{r_+}\left(2+\frac{6 r_+^2}{L^2}-\mu ^2\right).
\ee
Thus, AdS-RN black holes exist for $\mu \leq \mu_{\rm ext}$ where $\mu=\mu_{\rm ext}$ with
\be \label{RNextreme}
\mu_{\rm ext}=\sqrt{2} \sqrt{1+3\,\frac{r_+^2}{L^2}}
\ee
describes the extremal AdS-RN black hole with zero temperature.

Later, we will  consider the Dirac equation coupled to the curved spacetime \eqref{RNAdS} \cite{Birrell:1982ix,Tong:2007,Pollock:2010zz,Yepez:2011bw}. For that, it will be useful to introduce the  tetrad vector basis (\emph{vierbein}) $e^{(a)}=e^{(a)}_{\:\:\:\:\mu} {\rm d}x^\mu$, with non-coordinate curved bracket Latin indices $(a)=(0),\cdots,(3)$: 
 \be
 e^{(0)}=f^{1/2}{\rm d}t \,,\quad  e^{(1)}=f^{-1/2}{\rm d}r\,,\quad  e^{(i)}=r \hat{e}^{(i)}\,,
 \ee
where $\hat{e}^{(i)}$ is the tetrad on the $S^2$ manifold. This is an orthonormal basis since $g^{\mu \nu} e^{(a)}_{\:\:\:\:\mu}e^{(b)}_{\:\:\:\:\nu}=\eta^{(a)(b)}$ where $\eta^{(a)(b)}={\rm diag}(-1,1,1,1)$ is the Minkowski metric. The tetrad dual basis $e_{(a)}=e_{(a)}^{\:\:\:\:\mu} \partial_{\mu}$ becomes $e_{(a)}=\eta_{(a)(b)}e^{(b)}$. Latin (Greek) letters will always be used for tetrad (coordinate basis) indices.

The components of any tensor in the coordinate basis $\{{\rm d}x^{\mu}\}$ can be obtained from the components on the tetrad basis using the projectors $e^{(a)}$ and $e_{(a)}$. For example, $T_{\mu}^{\:\:\nu}=g^{\nu \sigma}e^{(b)}_{\:\:\:\: \sigma}T_{\mu(b)}=g^{\nu \sigma}e^{(a)}_{\:\:\:\:\mu}e^{(b)}_{\:\:\:\: \sigma}T_{(a)(b)}$. This example also illustrates that we can have mixed-index tensors with mixed components in the coordinate and tetrad bases.

The spin connection of non-coordinate based differential geometry can be introduced in terms of the affine connection $\Gamma^\beta_{\:\:\mu \nu} $ of coordinate based differential geometry as $\gamma _{(c) (a) (b)}= e_{(c)}^{\:\:\:\:\mu } \left(\partial_{\nu } e_{(a) \mu }  - \Gamma^\beta_{\:\:\mu \nu} e_{(a) \beta }\right) e_{(b)}^{\:\:\:\:\nu }$. Equivalently, one can define the spin connection as
\be\label{def:spincoef}
 \omega_{\mu(a)(b)}= e^{(c)}_{\:\:\:\:\mu } \gamma _{(a) (b) (c)}= \frac{1}{2}  e^{(c)}_{\:\:\:\:\mu }  \left(  \lambda_{(a)(b)(c)} + \lambda_{(c)(a)(b)} - \lambda_{(b)(c)(a)} \right) 
\ee
with  $\lambda_{(a)(b)(c)} = e_{(a)}^{\:\:\:\:\alpha } \left(\partial_{\beta} e_{(b) \alpha} - \partial_\alpha e_{(b)\beta} \right) e_{(c)}^{\:\:\:\:\beta }$, which allows to compute the spin connections without the use of the affine connection. 

For a multi-index tensor with tetrad and coordinate indices the mixed-index covariant derivative is defined as
\be \label{def:DerMixed}
\nabla_\alpha T^{(a)\nu}_{\phantom{(a)\nu}(b)\mu} = \partial_\alpha T^{(a)\nu}_{\phantom{(a)\nu}(b)\mu} 
- \Gamma^\beta_{\:\:\mu \alpha} T^{(a)\nu}_{\phantom{(a)\nu}(b)\beta}
+\Gamma^\nu_{\:\: \beta \alpha} T^{(a)\beta}_{\phantom{(a)\nu}(b)\mu}
 +\omega_{\alpha\phantom{a}(c)}^{\phantom{i}(a)} T^{(c)\nu}_{\phantom{(a)\nu}(b)\mu}
 -\omega_{\alpha\phantom{c}(b)}^{\phantom{i}(c)} T^{(a)\nu}_{\phantom{(a)\nu}(c)\mu}.
\ee
Onwards, we take the affine connection $ \Gamma^\nu_{\:\:\mu \alpha}$ to be given by the Christoffel symbols that covariantly conserve the metric, $\nabla_\alpha g_{\mu\nu}=0$. It follows that the spin coefficients $\omega_{\mu(a)(b)}$ defined in \eqref{def:spincoef} are such that the \emph{vierbein} is also covariantly conserved, $\nabla_\alpha e^{(a)}_{\:\:\:\:\mu}=0$. That the latter implies the first conservation follows from $g_{\mu\nu}=\eta_{(a)(b)}e^{(a)}_{\:\:\:\:\mu }e^{(b)}_{\:\:\:\:\nu }$. Further note that the spin coefficient is anti-symmetric in the tetrad pair of indices, $\omega_{\mu(a)(b)}=-\omega_{\mu(b)(a)}$.

To discuss the Dirac equation one necessarily needs to introduce the (coordinate independent) Dirac gamma matrices $\gamma^{(a)}$.
Let $\sigma^{i}$ be the Pauli matrices and $I_{n} $ the $n\times n$ identity matrix. 
We choose to work with the Weyl (chiral) spinor representation of the 4-dimensional Clifford algebra:
\be \label{def:gammaTetrad}
\gamma^{(0)}= 
\begin{pmatrix}
0 & i I_{2}\\
i I_{2} & 0
\end{pmatrix},\quad
\gamma^{(1)} =
\begin{pmatrix} 
0 & i \sigma^{3}\\
-i \sigma^{3}&  0 
\end{pmatrix},\quad
\gamma^{(2)}=
\begin{pmatrix}
0 & i \sigma^{1}\\
-i \sigma^{1}&  0 
\end{pmatrix},\quad
\gamma^{(3)}=
\begin{pmatrix}
0 & i \sigma^{2}\\
-i \sigma^{2}&  0 
\end{pmatrix}
\ee
which indeed obeys the anti-commutation relations of the  Clifford algebra: 
\be\label{Clifford}
\left\{\gamma^{(a)},\gamma^{(b)}\right\}= 2 \eta^{(a)(b)} I_{4}\,,
\ee
where $\{A,B\}=A B+B A$ is the usual anti-commutator, as well as the relations $(\gamma^{(0)})^2=-I_4$ and  $(\gamma^{(i)})^2=I_4$ for $i=1,2,3$.

Let us also introduce the pseudoscalar
\be \label{def:gamma5}   
\gamma^{(5)}=i \gamma^{(0)} \gamma^{(1)} \gamma^{(2)} \gamma^{(3)} = \begin{pmatrix}  -I_{2} & 0\\  0 & I_{2} \end{pmatrix}\,, \ee
which obeys the relations $(\gamma^{(5)})^2=1$ and $\{\gamma^{(5)},\gamma^{(a)}\}=0$.\footnote{$\gamma^{(5)}$ satisfies the 5-dimensional Clifford algebra $\left\{\tilde{\gamma}^{(A)},\tilde{\gamma}^{(B)}\right\}= 2\eta^{(A)(B)} I_{5}$ if we define $\tilde{\gamma}^{(A)}=(\gamma^{(a)},\gamma^{(5)})$, which justifies its label (from an Euclidean perspective).}

The components of the (coordinate dependent) Dirac gamma matrices in the coordinate basis can be obtained from the tetrad basis components \eqref{def:gammaTetrad} using 
\be\label{def:gammaCov}
\gamma^{\mu}=\gamma^{(a)} e_{(a)}^{\:\:\:\:\mu}.
\ee
and they obey the covariant Clifford algebra $\left\{\gamma^{\mu},\gamma^{\nu}\right\}= 2 g^{\mu\nu} I_{4}$.

\subsection{Dirac equation in the AdS-RN background}\label{sec:DiracAdS-RN}

One of our main goals will be to compare scalar (spin-$0$) and fermionic (spin 1/2) perturbations on the AdS-RN background. Therefore, below we briefly review the equations that these fields have to obey in a curved background, {\it e.g.} in the AdS-RN spacetime \eqref{RNAdS}. For more detailed discussions see \cite{Birrell:1982ix,Tong:2007,Pollock:2010zz,Yepez:2011bw,Cotaescu:2003be}.

To start, consider spin-$s$ fields in Minkowski spacetime. The spin of a field can be identified looking into how the field transforms under a Lorentz transformation, $x^\mu \to \tilde{x}^\mu=\Lambda^\mu_{\:\:\:\nu} x^\nu$. Let ${\cal M}^{\alpha\beta}=-{\cal M}^{\beta\alpha}$ be the generators of Lorentz transformations ({\it i.e.} a basis of six $4\times 4$ antisymmetric matrices obeying the Lorentz Lie algebra). A finite Lorentz transformation is described by $\Lambda=\exp{\left(\frac{1}{2}\Omega_{\alpha\beta}{\cal M}^{\alpha\beta}\right)}$ where $\Omega_{\alpha\beta}=-\Omega_{\beta\alpha}$ are six parameters describing the particular  transformation $\Lambda$ (boost, rotations) of the Lorentz group $SO(3,1)$.

Under a Lorentz transformation a spin-$s$ field $\Psi(x)$ transforms as $\Psi(x)\to \tilde{\Psi}(x) =S[\Lambda]\Psi \left(\Lambda^{-1}x \right)$. Here, the matrices $S[\Lambda]$ form a representation of the Lorentz  group (i.e. $S[\Lambda_1]S[\Lambda_2]=S[\Lambda_1 \Lambda_2]$, $S[\Lambda^{-1}]=S[\Lambda]^{-1}$ and $S[I]=I$) with group generators $S^{\alpha\beta}=-S^{\beta\alpha}$ such that a finite Lorentz transformation is described by $S[\Lambda]=\exp{\left(\frac{1}{2}\Omega_{\alpha\beta}S^{\alpha\beta}\right)}$.\footnote{Note that we are applying the same Lorentz transformation to $x$ and $\Psi$; thus the coefficients $\Omega_{\alpha\beta}$ of the transformations $\Lambda$ and $S[\Lambda]$ are the same although the bases of generators ${\cal M}^{\alpha\beta}$ and $S^{\alpha\beta}$, respectively, are different.}  

In Minkowski spacetime, under Lorentz transformations the derivative of a spin-$s$ field  transforms as  $\partial_\mu \Psi\to  \partial_\mu\tilde{\Psi}(x) =\Lambda_\mu^{\:\:\:\nu} S[\Lambda]\partial_\nu \Psi(\Lambda^{-1}x)$. If we want to couple the spin-$s$ field to a curved background, while preserving general covariance, one needs to promote the partial derivative $\partial_\mu$ to a covariant derivative ${\cal D}_\mu$. This promotion is chosen such that any function of $\Psi$ and  ${\cal D}_\mu\Psi$ that is a scalar under Lorentz transformations in Minkowski spacetime remains a scalar $-$ under general coordinate transformations and local changes in the \emph{vierbein} $-$ in the curved background. 
This is the case if, under an arbitrary Lorentz transformation, the covariant derivative  ${\cal D}_\mu$  still transforms as a derivative of a spin-$s$ field:
\be\label{CovDeriv}
{\cal D}_\mu \Psi(x)\to  {\cal D}_\mu\tilde{\Psi}(x) =\Lambda_\mu^{\:\:\:\nu} S[\Lambda]{\cal D}_\nu \Psi \left(\Lambda^{-1}x \right).
\ee
It follows that the covariant derivative of a spin-s field that preserves Lorentz invariance in a curved background is 
\be\label{Dspin}
{\cal D}_\mu=\partial_\mu-\Gamma_\mu-i q A_\mu\,,
\ee
where we took the opportunity to allow the spin-s field to have a charge $q$ (that couples to the Maxwell background field $A_\mu$), and $\Gamma_\mu$ is a covariant spin connection
\be \label{def:GammaU}
\Gamma_\mu = -\frac{1}{2} \omega_{\mu (a)(b)} S^{(a)(b)},
\ee
which {\it depends} on the spin of the field it acts on since the generators $S^{(a)(b)}$ depend on whether we are looking into, for example, the scalar or spinor representation of the Lorentz group.

In more detail, for a scalar (spin-$0$) field $\Psi\equiv \Phi$ the Lorentz group generator is simply $S^{(a)(b)}=0$. Therefore $S[\Lambda]=1$ and the scalar covariant derivative \eqref{Dspin} is simply ${\cal D}_\mu=\partial_\mu-i q A_\mu$. The action for a massive charged complex scalar field is given by ${\cal S}_{\Phi}=\int_{\cal M} d^4x \,\sqrt{-g}\left( {\cal D}_\mu\Phi {\cal D}^\mu\Phi^* +m^2 \Phi \Phi^*\right)$ where $^*$ stands for complex conjugation. The factor of $\sqrt{-g}$ is introduced to ensure that the Lagrangian ${\cal L}_{\Phi}$ is a scalar density and thus the  action ${\cal S}_{\Phi}=\int_{\cal M} d^4x {\cal L}_{\Phi}$ is a scalar. Varying this action w.r.t. $\Phi^*$ one gets the Klein-Gordon equation for the scalar field 
\be
{\cal D}_\mu{\cal D}^\mu\Phi -m^2\Phi=0\,  
\ee
and similarly for $\Phi^*$.
 
On the other hand, for a (spin-$\frac{1}{2}$) Dirac 4-spinor field\footnote{The Dirac spinor is a 4-component field $\psi^{A}$ with complex components $A=1,2,3,4$. In our study we will typically omit the spinorial indices and simply write $\psi\equiv \psi^{A}$, $\gamma^{(a)}\equiv (\gamma^{(a)})^A_{\:\:B}$ and $S[\Lambda]\equiv S[\Lambda]^A_{\:\:B}$.} $\Psi\equiv \psi$, out of the gamma matrices $\gamma^{(a)}$ \eqref{def:gammaTetrad} that satisfy the covariant Clifford algebra \eqref{Clifford} one can build the commutator  ($[A,B]=AB-BA$)
\be\label{Sspinorial} 
S^{(a)(b)}=\frac{1}{4}\left[\gamma^{(a)},\gamma^{(b)} \right],
\ee
that satisfies the Lorentz Lie algebra.\footnote{More precisely, Dirac 4-spinor fields are invariant under internal local Lorentz transformations of the spinor representation of the $SU(4)$ group. There are 15 Dirac matrices that provide a $4\times 4$ fundamental representation of the SU(4) group \cite{Yepez:2011bw}. These are the four vectors $\gamma^{(a)}$ introduced in \eqref{def:gammaTetrad}, the six tensors $S^{(a)(b)}$ defined \eqref{Sspinorial}, the pseudoscalar $\gamma^{(5)}$ introduced in \eqref{def:gamma5}, and four axial vectors $\gamma^{(5)}\gamma^{(a)}$.} 
 $S^{(a)(b)}$ is the generator of the Lorentz group in the spinor representation and replacing this \eqref{Sspinorial} into \eqref{def:GammaU} and then the latter $\Gamma_\mu$ into \eqref{CovDeriv} one gets the spinor covariant derivative ${\cal D}_\mu$, namely \eqref{Dspin}, that acts on the Dirac spinor $\psi$.

The action that is Lorentz invariant and describes the coupling of a spin-$\frac{1}{2}$ fermion field $\psi$ to a curved background ${\cal M}$ is
\be \label{DiracAction}
{\cal S}_{D}=\int_{\cal M} d^4x \,\sqrt{-g} \left(\frac{i}{2}\Big[\bar{\psi}\gamma^\mu {\cal D}_\mu \psi -  \left({\cal D}_\mu \bar{\psi}\right)\gamma^\mu \psi \Big] -m \bar{\psi} \psi \right)
\ee
where we have introduced the Dirac adjoint $\bar{\psi}=\psi^\dagger \gamma^{(0)}$ with $\psi^\dagger=(\psi^*)^T$ being the Hermitian adjoint of the multi-component field $\psi$. One needs to work with the Dirac adjoint because  the Fermi bilinears $\bar{\psi} \psi$ and $\bar{\psi} \gamma^{\mu} \psi$ transform covariantly (as a scalar and as a vector, respectively) under the Lorentz group (while the Hermitian partner objects do not). 

Varying the action ${\cal S}_{D}$ w.r.t. $\bar{\psi}$ and $\psi$, respectively, one gets the Dirac equations 
\beq \label{DiracEOM}
&&  \left( i\,\overrightarrow{\cancel{\cal D}}_\mu -m\right)\psi=0 \quad \Leftrightarrow \quad i \gamma^{\mu}{\cal D}_\mu \psi -m\psi=0\,, \nonumber \\
&&  \bar{\psi} \left( i \,\overleftarrow{\cancel{\cal D}}_\mu +m\right)=0 \quad \Leftrightarrow \quad
 i \left({\cal D}_\mu \bar{\psi}\right)\gamma^{\mu} +m\bar{\psi}=0\,.
\eeq

To find solutions of \eqref{DiracEOM} it is advantageous to write the Dirac 4-spinor $\psi$ in terms of the left-handed and right-handed 2-spinors $\Psi_-$ and $\Psi_+$, respectively, as
\be
\psi=\begin{pmatrix}  \Psi_- \\  \Psi_+ \end{pmatrix}\,.
\ee
The chiral 2-spinors $\Psi_{\pm}$ emerge naturally when we note that the pseudoscalar $\gamma^{(5)}$ defined in \eqref{def:gamma5} obeys $(\gamma^{(5)})^2=1$. Therefore we can introduce the Lorentz invariant projection operators $P_{\pm}$ that project the Dirac 4-spinor $\psi$ into the chiral spinors:
\be
P_{\pm}\psi =\Psi_{\pm}\,, \qquad \hbox{with} \qquad P_{\pm}=\frac{1}{2}\left(I_4 \pm \gamma^{(5)} \right)
\ee
and such that $P_{\pm}^2=P_{\pm}$ and $P_+ P_-=0$.\footnote{In 4 spacetime dimensions and in the chiral representation \eqref{def:gammaTetrad} in which we work, $\Psi_{\pm}$ are nothing but the Weyl 2-spinors which transform in the same way under Lorentz rotations but oppositely under Lorentz boosts, and obey the Weyl equations if the fermion mass vanishes.}
Moreover, the task of finding solutions of the Dirac equation in the AdS-RN black hole gets considerably simplified by the fact that under the separation {\it anstaz} \cite{McKellar:1993ej,Dolan:2015eua}: 
\begin{eqnarray}\label{sepAnsatz}
&& \Psi_+(t,r,\theta,\phi)= e^{-i\o t}e^{im_{\phi}\phi}(-g f)^{-\frac{1}{4}} 
\begin{pmatrix}
R_{1}(r)S_{1}(\theta)\\
R_{2}(r)S_{2}(\theta) 
\end{pmatrix}, \nonumber\\
&& \Psi_-(t,r,\theta,\phi)= e^{-i\o t}e^{im_{\phi}\phi}(-g f)^{-\frac{1}{4}} 
\begin{pmatrix}
-R_{2}(r)S_{1}(\theta)\\
-R_{1}(r)S_{2}(\theta)
\end{pmatrix},
\end{eqnarray}
the Dirac equations \eqref{DiracEOM} reduce to a set of equations where the radial and angular functions of the fermion field are decoupled. This separation {\it ansatz} exploits the fact that $\partial_t$ and $\partial_{\phi}$ are Killing vectors of the background AdS-RN solution. This allows to do a Fourier decomposition in these directions which introduces the frequency $\omega$ and azimuthal angular momentum $m_\phi$ of the fermionic wavefunction.\footnote{\label{foot:gauge2}This frequency $\omega$ is measured in the gauge $A |_{\infty}=0$ $-$ see \eqref{RNAdS} $-$ and we will work preferentially on this gauge unless otherwise stated. In particular, all our numerical results will be using it. Note that in the alternative gauge $A |_{r_+}=0$ (also often used) the associated frequency is $\tilde{\omega}=\omega-q\mu$.}

Concretely, the radial functions $R_1(r), R_2(r)$ obey the coupled system of first order ODEs
\bea \label{radial:1storder}
\begin{split}
&& r \sqrt{f(r)} \Big( \frac{d}{dr} - i \,\frac{\omega + q A(r)}{f(r)} \Big) R_{1}(r) = ( \lambda + i \, m\, r ) R_{2}(r),  \\
&& r \sqrt{f(r)} \Big( \frac{d}{dr} + i \,\frac{\omega + q A(r)}{f(r)}\Big) R_{2}(r) = ( \lambda - i \,m\, r ) R_{1}(r), 
\end{split}
\eea
where $\lambda$  is a separation constant, while the angular functions $S_1(\theta), S_2(\theta)$ satisfy the coupled system of first order ODEs
\bea \label{ang1}
\begin{split}
& &\left(\frac{d}{d\theta}+\frac{m_{\phi}}{\sin\theta} +\frac{\cot\theta}{2}\right) S_{2}(\theta)=-\lambda S_{1}(\theta), \\
&& \left(\frac{d}{d\theta}-\frac{m_{\phi}}{\sin\theta} +\frac{\cot\theta}{2}\right) S_{1}(\theta)=\lambda S_{2}(\theta).
\end{split}
\eea

Furthermore, the coupled pair of first order radial equations \eqref{radial:1storder} can be decoupled in a pair of second order ODEs, one for $R_1(r)$ and the other for $R_2(r)$. For that we solve the  first (second) equation in \eqref{radial:1storder} w.r.t. $R_2$ ($R_1$) and replace it in the second (first) equation. We end up with two decoupled second order ODEs for $R_1$ and $R_2$,  
\bea \label{radial:2ndorder}
\begin{split}
&r \sqrt{f(r)} \frac{d}{dr}\Big( r \sqrt{f(r)} \frac{d}{dr}R_{1}(r) \Big) + H_{1}(r) \frac{d}{dr}R_{1}(r) + H_{2}(r) R_{1}(r) = 0,\\
&r \sqrt{f(r)} \frac{d}{dr}\Big( r \sqrt{f(r)} \frac{d}{dr}R_{2}(r) \Big) + H_{1}^{*}(r) \frac{d}{dr}R_{2}(r) + H_{2}^{*}(r) R_{2}(r) = 0,
\end{split}
\eea
where $^{*}$ denotes complex conjugation and we have defined
\bea 
&&H_{1}(r) = -\frac{ m\, r^2 f(r)}{m\, r - i \lambda}, \nonumber\\
&&H_{2}(r) = - i r \sqrt{f(r)} \frac{d}{dr}K(r) + K(r) \left( K(r) - i \frac{H_{1}(r)}{r \sqrt{f(r)}} \right) - \lambda^2 - m^2 r^2, \\
&&K = \frac{r}{\sqrt{f(r)}}\Big(\o + q A(r)\Big). \nn
\eea
Of course, we are only interested in solutions of \eqref{radial:2ndorder} that also solve the original first order system \eqref{radial:1storder}. The requirement that \eqref{radial:1storder} is solved imposes extra constraints on solutions of \eqref{radial:2ndorder}. This is best illustrated if we consider the Taylor expansion about the boundaries of the integration domain: the ODE pair \eqref{radial:2ndorder} has four integration constants about each boundary but only two of them are independent when we further require that the solution solves the two first order ODEs \eqref{radial:1storder}; see discussion of \eqref{AsympDecaysR} below.
  
Similarly, the coupled pair of first order ODEs for $S_{1,2}(\theta)$ can be written as a decoupled set of two second order ODEs for $S_1(\theta)$ and $S_2(\theta)$. They are hypergeometric equations and $S_{1,2}(\theta)$ are the spin-$\frac{1}{2}$ weighted spherical harmonics. Regularity at $\theta =0$ and $\theta=\pi$ quantizes the angular separation constant as ($\ell$ is a harmonic number related to the number of zeros of the wavefunction) 
\be\label{lambda:quantization}
\lambda = \ell + \frac{1}{2}, \qquad \ell =  \frac{1}{2}, \frac{3}{2}, \frac{5}{2},\cdots 
\ee
with the azimuthal number being constrained as $m_\phi\leq \ell$.

Unfortunately, the radial ODEs cannot be solved analytically\footnote{For global AdS, {i.e.} $M=0=Q$ these ODEs are hypergeometric equations and can be solved analytically: see section \ref{sec:DiracNormalModes}.}. We can however do a Frobenius analysis about the asymptotic boundary $r\to \infty$ to find the asymptotic behaviours of $R_1(r)$ and $R_2(r)$. One finds that (for $m\neq 0,\frac{1}{2}$)\footnote{For $m=1/2$ one of the two independent solutions decays asymptotically as a power law in $r$ and the other as a power law multiplied by a $\log r$. For this reason (since a similar logarithmic solution appears in the scalar field case when $m^2=m_{BF}^2$), this case is often called the BF solution of the Dirac system. We do not discuss further this special case (see \cite{Amsel:2008iz,Iqbal:2009fd} for more details). It is however important to emphasize that for the scalar field, $m^2=m_{BF}^2$ corresponds to $\Delta_+=\Delta_-$ and is thus also the bound for stability while in the Dirac case, the mass stability bound is \eqref{BFfermionic} not the BF mass $m=1/2$.}
\bea\label{AsympDecaysR}
\begin{split}
& (-g f)^{-\frac{1}{4}}R_1\big|_{r\to \infty}\sim r^{-\Delta_-}\left(\alpha_{1}+\cdots \right)+  r^{-\Delta_+}\left(\beta_{1}+\cdots \right),  \\
& (-g f)^{-\frac{1}{4}}R_2 \big|_{r\to \infty}\sim r^{-\Delta_-}\left(-i \,\alpha_{1}+\cdots \right)+  r^{-\Delta_+}\left(i \,\beta_{1}+\cdots \right),
\end{split}
\eea
where we used $(-g f)^{-\frac{1}{4}}|_{r\to \infty}\sim L^{1/2}r^{-3/2}$ and, anticipating the AdS/CFT discussion below, we have introduced the conformal dimensions 
\be\label{ConfDim}
\Delta_{\pm}=\frac{3}{2}\pm \sqrt{m^2 L^2}\,.
\ee
As expected for a coupled system \eqref{radial:1storder} of two first order ODEs,  there are two independent arbitrary constants $(\alpha_1,\beta_1)$ in the asymptotic decay \eqref{AsympDecaysR}, that is to say, the decays of $R_2$ are fixed by the equations of motion as a function of $(\alpha_1,\beta_1)$. The dots in \eqref{AsympDecaysR} represent subleading terms that depend only on $\alpha_{1}$ (in the $\Delta_-$ contribution) or $\beta_{1}$  (in the $\Delta_+$ terms). 

Before proceeding, one unavoidably needs to discuss the range of Dirac fermion masses that allow for  normalizable solutions, i.e with conserved finite energy. We also have to distinguish the positive energy solutions (which are stable) from those  negative  energy states (which should trigger an instability). It was proven in section~II/Appendix ~B of \cite{Amsel:2008iz} (see also 
\cite{Amsel:2009rr,Andrade:2011dg,Ishibashi:2004wx}) that the fermionic bound for stability (in {\it any} dimension) is given by 
\begin{equation}\label{BFfermionic} 
m^2\geq 0  \qquad \hbox{(Dirac stability bound condition)}\,,
\end{equation} 
with the lower bound being the solution for which $\Delta_+=\Delta_-$ in \eqref{ConfDim}.\footnote{So, for $m^2\geq 0$, $\Delta_\pm$ are real; otherwise they are complex numbers. Note that for a scalar field the configuration $\Delta_+=\Delta_-$ corresponds to the BF bound where one of the independent solutions is logarithmic. However, for the Dirac field, the state $\Delta_+=\Delta_-$ is {\it not} the BF logarithmic solution (which occurs instead for $m=1/2$). If follows that for the scalar field case the BF bound coincides with the bound for stability, $m^2\geq m_{BF}^2$, but {\it not} in the Dirac case. Moreover, in the scalar case, there is a 1-parameter family of boundary conditions that yield stable normalizable solutions for $m_{BF}^2\leq m^2< m_{BF}^2+1/L^2$ and a unique boundary condition that generates stable normalizable solutions for  $m^2\geq m_{BF}^2+1/L^2$ \cite{Breitenlohner:1982jf,Breitenlohner:1982bm,Klebanov:1999tb,Ishibashi:2004wx}. However, in the Dirac case, normalizable stable states exist for: 1) a 1-parameter choice of boundary conditions for $0\leq m^2< m_{BF}^2$ (with $m_{BF}^2=1/4$), and 2) a unique boundary condition for $m^2\geq m_{BF}^2$ \cite{Amsel:2009rr}. Further note that, unlike in the scalar case, the Dirac stability mass bound is independent of the dimension of the spacetime.}
To understand this bound it is useful to rewrite the radial Dirac equation \eqref{radial:2ndorder} as a Schr\"odinger equation \cite{Ishibashi:2004wx,Amsel:2008iz}. Without further conditions, the associated Schr\"odinger operator is not self-adjoint (hermitian). It becomes self-adjoint if and only if we impose as boundary condition that the energy-momentum flux at the asymptotic AdS boundary vanishes. That is to say, it becomes Hermitian if and only if the energy is conserved. In these conditions looking for (conserved) finite energy solutions boils down to look for normalizable states in the standard quantum mechanical sense. That is to say, normalizable solutions are those that are square integrable. 

For $m^2< 0$ there are normalizable solutions but they have negative energy. In a mathematical language,  if $m^2<0$, the  Schr\"odinger  operator of the Dirac equation is unbounded below and thus it  does not allow for a positive self-adjoint extension  \cite{Ishibashi:2004wx,Amsel:2008iz}. Alike in any other negative energy Schr\"odinger states, this signals the existence of an instability. We will explore further this in section \ref{sec:DiracRN1}.
 
On the other hand, if the mass is real, i.e. if it satisfies the bound \eqref{BFfermionic}, there are stable normalizable Dirac fermion solutions that are selected by a choice of boundary conditions. We will discuss in detail this issue of the boundary conditions in the next section. The upshot is that if $m L\geq 1/2$ there is an unique complete set of normalizable modes (and the non-normalizable modes must be fixed by boundary conditions; {\it e.g.} no-source/homogeneous boundary conditions that eliminate them) \cite{Amsel:2008iz}. On the other hand, for $0\leq m L<1/2$ there is a non-unique set of normalizable modes and thus a wider band of  boundary conditions that yield normalizable solutions (e.g. the no-source/homogeneous Dirichlet or Neumann boundary conditions that we will use later but also more general multi-trace boundary conditions) \cite{Amsel:2008iz}.
Further note that if we take $m \to -m$, we simply trade the role of the $\Delta_\pm$ contributions while preserving condition \eqref{BFfermionic}. Therefore onwards we assume, without any loss of generality, $m\geq 0$ in our discussion.

For our purposes, but without loss of generality, we will be particularly interested in the lower bound case of \eqref{BFfermionic}. For this $m=0$ case and choosing the gauge $A |_{\infty}=0$, a Frobenius analysis of the first order equations of motion about the asymptotic boundary yields\footnote{\label{foot:gauge}For a Dirac field (or scalar field) with phase $\varphi$, $\psi=|\psi| e^{i\varphi}$,  $U(1)$ gauge transformations with gauge parameter $\chi$ leave the action and equations of motion invariant and transform the Dirac (scalar) and Maxwell fields as 
$\varphi \to \tilde{\varphi}=\varphi +q \,\chi\,,\:A_t \to \tilde{A}_t = A_t +\nabla_t \chi$. Thus, if in the gauge  $A |_{\infty}=0$ ({\it i.e.} $C=0$) we denote the frequency of the Dirac (scalar) field by $\omega$ then a transformation with gauge parameter $\chi= \mu \,t$ into the gauge $\tilde{A} |_{\infty}=\mu$ ($C=\mu$) changes the frequency into $\tilde{\omega}=\omega-q\mu$. Thus, if we had  chosen the gauge $\tilde{A} |_{\infty}=\mu$, then we would have to make the replacement $\omega\to \tilde{\omega}+q\mu$ in \eqref{AsympDecaysRm0} (and later in the boundary conditions \eqref{StandardQuant}-\eqref{AlternativeQuant} and \eqref{HorizonSolns}). Further note that in \eqref{radial:1storder}-\eqref{radial:2ndorder} we are leaving the gauge choice arbitrary because we do not fix $A(r)$ introduced in \eqref{RNAdS}.}
\bea\label{AsympDecaysRm0}
\begin{split}
& (-g f)^{-\frac{1}{4}}R_1\big|_{r\to \infty}\sim  r^{-\frac{3}{2}}\left(
\alpha_1+\beta_1\,\frac{L}{r}
 +{\cal O}(r^{-2}) \right),\\
& (-g f)^{-\frac{1}{4}}R_2 \big|_{r\to \infty}\sim r^{-\frac{3}{2}}\left(
-\frac{\beta_1+i \alpha_1 \omega L}{\lambda }
+\frac{\alpha_1 \left(\omega^2 L^2-\lambda^2\right)-i \beta_1 \omega L}{\lambda }\,\frac{L}{r}
+{\cal O}(r^{-2}) \right),  
\end{split}
\eea
{\it i.e.} we can take the two independent integration constants associated to the coupled pair of first order ODEs to be $\alpha_1$ and $\beta_1$ and the equations of motion then fix the decay of $R_2$ as a function of $\alpha_1$ and $\beta_1$.

\subsection{Boundary conditions for the Dirac spinor in AdS-RN}\label{sec:BCsDiracAdS-RN}

To find the solution of the Dirac spinor field $\psi$ and its Dirac adjoint $\bar{\psi}$ in the AdS-RN background we have to solve a system of two equations that are first order, namely \eqref{radial:1storder}, subject to boundary conditions imposed at the event horizon $r=r_+$ and at the asymptotic boundary $r\to \infty$. Before imposing boundary conditions,  such a system of two first order differential equations necessarily has two independent constants at the horizon boundary and another two independent constants at the asymptotic boundary (namely, $\alpha_1$ and $\beta_1$ in \eqref{AsympDecaysR}), which can be identified doing a Frobenius analysis at these two boundaries. To have a well posed formulation of the elliptic problem one should impose two boundary conditions that fix two of the independent constants and solve the equations of motion to find the other two. We certainly want the Dirac solutions to be regular at the event horizon: this boundary condition  fixes one of the constants \footnote{Alternatively, since we have a ODE system, we could use two boundary conditions to fix the two asymptotic independent constants and solving the equations of motion would  yield the behaviour of the Dirac fields at the event horizon. However, this is not a good strategy because in general these solutions would not be smooth at the event horizon.}. One should then fix one of the asymptotic constants $\alpha_1$ or $\beta_1$ (or a relation between them) with an appropriate boundary condition \cite{Amsel:2008iz,Amsel:2009rr,Andrade:2011dg}. But we certainly cannot fix both  asymptotic independent constants: once the first is fixed, the second one must be found by solving the equations of motion in the bulk subject to the two aforementioned boundary conditions.
This poses the question: how do we choose a boundary condition at the asymptotic boundary that is physically relevant? We should choose one that conserves the energy and thus yields a self-adoint Schr\"odinger operator for the system that ensures that we have a well-posed hyperbolic evolution if we let the perturbed system evolve in time. Next, we will review how two boundary conditions with these properties can be identified. They single out in the AdS-CFT context because they are single-trace (no-source) boundary conditions that preserve the conformal symmetry group of AdS (and thus do not deform the boundary conformal field theory) \cite{Mueck:1998iz,Henningson:1998cd,Henneaux:1998ch,Contino:2004vy,Amsel:2008iz,Amsel:2009rr,Cubrovic:2009ye,Andrade:2011dg} .


Dirac spinor fields $\psi$ are intrinsically quantum fields. The dynamics of such fields can be naturally described by a path integral formulation whereby one sums over all possible field configurations in configuration space to get the transition amplitude between two states. In particular, the partition function $Z$ ({\it i.e.} the generating functional of correlation functions between operators) can also be naturally computed using the path integral formulation. Schematically one has,
\be 
Z=\int[D\psi][D\bar{\psi}] e^{\frac{i}{\hslash}{\cal S}[\psi,\bar{\psi}]},
\ee
where $[D\psi] [D\bar{\psi}]$ represents the integration measure, $\hslash$ is Planck's constant and ${\cal S}_D[\psi,\bar{\psi}]$ is the action \eqref{DiracAction} of the Dirac field. 

In the classical limit,  $\hslash\to 0$, the path integral reduces simply to $Z\sim e^{\frac{i}{\hslash}{\cal S}_{\rm cl}[\psi,\bar{\psi}]}$, where ${\cal S}_{\rm cl}[\psi,\bar{\psi}]$ is the action evaluated on a solution of the classical equations of motion, that follow from the variation $\delta {\cal S}=0$ subject to the boundary conditions. As emphasised in \cite{Mueck:1998iz,Henningson:1998cd,Henneaux:1998ch,Contino:2004vy,Amsel:2008iz,Amsel:2009rr,Cubrovic:2009ye,Andrade:2011dg} this statement that the action must be {\it stationary} when evaluated on a classical solution severely constrains the type of boundary conditions we can impose on the field $\psi$. Indeed, if $\delta {\cal S}=0$ then it is not necessarily true that  $\delta ({\cal S} + {\cal B}) =0$ where ${\cal B}$ is the boundary term describing the desired boundary conditions ({\it i.e.} a total derivative term that does not change the equations of motion). That is to say, the physical choice we make for the boundary conditions must be such that $\delta  {\cal B} =0$. In particular, in the context of the  AdS/CFT correspondence, this condition fixes the form of the boundary term that must be added to the standard Dirac action \eqref{DiracAction} to have stationary solutions. Vice-versa, this boundary term fixes the boundary field theory. 

To determine the boundary term ${\cal B}$, one first notes that the ``radial" Dirac gamma matrix $\gamma^{(1)}$ defined in  \eqref{def:gammaTetrad} satisfies $(\gamma^{(1)})^2=I_4$ and  $\gamma^{(1)}=\gamma^{(1) \,\dagger}$. It follows that we can decompose the Dirac spinor as 
\be
\psi=\psi_++\psi_-\,, \qquad \bar{\psi}=\bar{\psi}_++\bar{\psi}_-\,,
\ee
where $\psi_\pm$ ($\bar{\psi}_\pm$) are 4-eigenspinors of $\gamma^{(1)}$ with eigenvalue $\pm 1$ ($\mp 1$).\footnote{\label{foot}In more detail, $\gamma^{(1)} \psi_\pm = \pm \psi_\pm$ and $\gamma^{(1)} \bar{\psi}_\pm = \mp \bar{\psi}_\pm$ and thus 
$\psi_\pm=\frac{1}{2}\left(I_4\pm \gamma^{(1)} \right)\psi$ and $\bar{\psi}_\pm=\frac{1}{2}\bar{\psi}\left(I_4\mp \gamma^{(1)}\right)$. A few properties follow that are useful. For example, $\bar{\psi}\gamma^{(1)}=-\bar{\psi}_{+}+\bar{\psi}_{-}$, $\bar{\psi}_{\pm}\psi_{\pm}=\frac{1}{4}\bar{\psi}\left(I_4-(\gamma^{(1)})^2\right)\psi=0$, $\bar{\psi}_{\pm}\gamma^{(1)}\psi_{\pm}=\mp \bar{\psi}_{\pm}\psi_{\pm}=0$ and $\bar{\psi}_{\pm}\gamma^{\mu}{\cal D}_{\mu}\psi_{\pm}=\frac{1}{4}\bar{\psi}\left(I_4-(\gamma^{(1)})^2\right)\gamma^{\mu}{\cal D}_{\mu}\psi=0$.}
Using this property, including the associated properties listed in footnote \ref{foot}, one finds that the terms in the Dirac action \eqref{DiracAction} that contain radial derivatives of the spinor are 
\be \label{DiracAction:dr}
{\cal S}_{D}  \supset  {\cal S}_{D}\big|_{\partial_r}=i\int_{\cal M} d^4x \sqrt{-g} f^{1/2}\left(\bar{\psi}_{+}\partial_{r}\psi_{-}- \bar{\psi}_{-}\partial_{r}\psi_{+}  \right),
\ee
where we used the fact that ${\cal D}_r=f^{1/2}\partial_{r}$.
It follows that if we vary the Dirac action \eqref{DiracAction}  w.r.t. $\psi_{+}$ and $\psi_{-}$ one gets, after integration by parts,
\be 
\delta {\cal S}_{D} = \hbox{bulk terms}+\delta {\cal S}_{bdry}\,,
\ee
where the bulk terms describe a contribution that vanishes when the equations of motion $-$ which are equivalent to \eqref{DiracEOM} $-$ are satisfied and $\delta {\cal S}_{bdry}$ is a boundary term resulting from integrating by parts the radial derivative terms \eqref{DiracAction:dr} given by
\be \label{DiracAction:bdry}
\delta {\cal S}_{bdry}=i\int_{\partial {\cal M}} d^3x \sqrt{-g}  f^{1/2} \left(\bar{\psi}_{+}\delta\psi_{-}- \bar{\psi}_{-}\delta\psi_{+}  \right)\,.
\ee

As discussed above, to have a well-posed boundary value problem, after requiring that the solution is regular at the event horizon we no longer have the freedom to fix both $\psi_{+}$ and $\psi_{-}$ at the asymptotic boundary (these are the two independent asymptotic constants of our pair of first order ODEs). 
Instead, we can either fix $\psi_{+}$ at the asymptotic boundary (in which case $\delta\psi_{+}=0$) or fix the asymptotic value of $\psi_{-}$ (in which case $\delta\psi_{-}=0$ at the boundary).

Suppose we want to fix $\psi_{+}$ at the asymptotic boundary (a similar analysis would apply if we wanted to fix $\psi_{-}$). In order to have a well-defined variational problem one should add a boundary term that cancels the contribution $\bar{\psi}_{+}\delta\psi_{-}$ in \eqref{DiracAction:bdry}. Adding the  boundary term \cite{Mueck:1998iz,Henningson:1998cd,Henneaux:1998ch,Contino:2004vy,Cubrovic:2009ye}
\be
 {\cal S}_{\partial}=-i\int_{\partial {\cal M}} d^{3}x \sqrt{-g} f^{1/2}  \bar{\psi}_{+}\psi_{-},
\ee
produces the desired effect since the total on-shell action becomes
\be \label{bdryDiracAction}
\delta  {\cal S}_{tot}=\delta\left( {\cal S}_D +  {\cal S}_{\partial}\right)= -i\int_{\partial {\cal M}} d^3x \sqrt{-g} f^{1/2}  \left(\bar{\psi}_{-}\delta\psi_{+}+\delta\bar{\psi}_{+}\psi_{-}  \right),
\ee
which indeed vanishes when $\delta \psi_{+}=0$ (and thus $\delta \bar{\psi}_{+}=0$). We can also compute the momentum conjugate to $\psi_{+}$ and $\bar{\psi}_{+}$ by varying $ {\cal S}_{tot}$ w.r.t. $\psi_{+}$ and $\bar{\psi}_{+}$, respectively, yielding
\be \label{conjMom}
\Pi_+=\frac{\delta {\cal S}_{tot}}{\delta \psi_{+}}=-i \sqrt{-g} f^{1/2}\bar{\psi}_{-}\,, \qquad \hbox{and} \qquad \bar{\Pi}_+=\frac{\delta {\cal S}_{tot}}{\delta \bar{\psi}_{+}}=-i \sqrt{-g} f^{1/2}\psi_{-}\,.
\ee

In terms of the functions $R_1(r),R_2(r)$ and $S_1(\theta),S_2(\theta)$ introduced in the separation {\it ansatz} \eqref{sepAnsatz}, the 4-spinors $\psi_\pm$ are given by
\be\label{Radial-psi}
 \psi_+= \frac{e^{-i\o t}e^{im_{\phi}\phi}}{2(-g f)^{\frac{1}{4}}} 
\begin{pmatrix}
i(R_1+i R_2)S_{1}\\
-(R_1+i R_2)S_{2} \\
(R_1+i R_2)S_{1} \\
-i(R_1+i R_2)S_{2}
\end{pmatrix}, \qquad
 \psi_-= \frac{e^{-i\o t}e^{im_{\phi}\phi}}{2(-g f)^{\frac{1}{4}}} 
\begin{pmatrix}
-i(R_1-i R_2)S_{1}\\
-(R_1-i R_2)S_{2} \\
(R_1-i R_2)S_{1} \\
i(R_1-i R_2)S_{2}
\end{pmatrix}.
\ee
From the asymptotic decays of $R_{1,2}$ in \eqref{AsympDecaysR} (valid for $m\neq 0,\frac{1}{2}$) or in \eqref{AsympDecaysRm0} (valid for $m= 0$)  one  finds that $\psi_\pm$ decay as
\begin{eqnarray}\label{AsympDecays:psi+-}
&& \hspace{-1cm}
 \left\{
\begin{array}{ll}
 \psi_+\big|_{r\to \infty}\sim 2\alpha_{1} r^{-\Delta_-}+a(\alpha_1)\, r^{-\Delta_+-1} +{\cal O}\left( r^{-\Delta_--2} \right), & \\
 \psi_- \big|_{r\to \infty}\sim 2\beta_{1} r^{-\Delta_+}+b(\beta_1)\, r^{-\Delta_--1} +{\cal O}\left( r^{-\Delta_--2} \right), & \qquad \hbox{if $\, 0< m L<\frac{1}{2}$;} \label{AsympDecays:psi+-1}
\end{array}
\right. \\
&& \phantom{X}\nonumber \\
&& \hspace{-1cm} 
\left\{
\begin{array}{ll}
 \psi_+\big|_{r\to \infty}\sim  r^{-\frac{3}{2}}\left(
\frac{\alpha_1 (\lambda +\omega L)-i \beta_1}{\lambda }
-i (\lambda +\omega L) \frac{\alpha_1 (\lambda -\omega L)+i \beta_1}{\lambda }
 \frac{L}{r}
 +{\cal O}(r^{-2}) \right), & \\
 \psi_- \big|_{r\to \infty}\sim r^{-\frac{3}{2}}\left(
\frac{\alpha_1 (\lambda -\omega L)+i \beta_1}{\lambda }
+i (\lambda -\omega L)\frac{\alpha_1 (\lambda +\omega L)-i \beta_1}{\lambda }
\frac{L}{r}
+{\cal O}(r^{-2}) \right), & \:\: \hbox{if $\, m=0$;} \label{AsympDecays:psi+-m0}
\end{array}
\right. \\
&& \phantom{X}\nonumber \\
&& \hspace{-1cm} 
\left\{
\begin{array}{ll}
 \psi_+\big|_{r\to \infty}\sim 2\alpha_{1} r^{-\Delta_-}+\tilde{a}(\alpha_1)\, r^{-\Delta_--2} +{\cal O}\left( r^{-\Delta_+-1} \right), & \\
 \psi_- \big|_{r\to \infty}\sim \tilde{b}(\alpha_1)\, r^{-\Delta_--1} + 2\beta_{1} r^{-\Delta_+}+{\cal O}\left( r^{-\Delta_--2} \right), & \qquad \hbox{if $\, m L >\frac{1}{2}$;}\label{AsympDecays:psi+-2}
\end{array}
\right. 
\end{eqnarray}
where $\alpha_1,\beta_1$ are the free constants introduced in \eqref{AsympDecaysR} or \eqref{AsympDecaysRm0} and the constants $a(\alpha_1)$, $b(\beta_1)$, $\tilde{a}(\alpha_1)$ and $\tilde{b}(\alpha_1)$ are fixed as functions of $\alpha_1$ or $\beta_1$ (as described by their argument) by the equations of motion (details are irrelevant for our aim). The asymptotic decays of the Dirac adjoints $\bar{\psi}_\pm$ follow straightforwardly from \eqref{AsympDecays:psi+-}
with the exchange $\alpha_1\to \bar{\alpha}_1$, $\beta_1\to \bar{\beta}_1$, etc.

For $m L \geq \frac{1}{2}$ the only normalizable mode ({\it i.e.} with finite energy) is $\psi_+$ \cite{Klebanov:1999tb,Amsel:2008iz,Amsel:2009rr,Faulkner:2009wj,Iqbal:2009fd,Guarrera:2011my,Andrade:2011dg}. In the context of the AdS/CFT correspondence, the leading term of the asymptotic expansion $\lim_{r \to \infty} r^{\Delta_-}\psi_+=2\alpha_{1}$ is then identified with the source of the dual operator $\bar{\cal O}$ which has mass dimension $\Delta_+$.
We have a well-posed boundary value problem if we impose smoothness of $\psi_+$ at the event horizon and a Dirichlet boundary condition for $\alpha_{1}$ at the asymptotic boundary. In particular, if we do not want to deform the boundary field theory we impose the no-source/homogeneous Dirichlet boundary condition: $\alpha_{1}=0$. We have no freedom left to fix asymptotically $\psi_-$ {\it i.e.} $\beta_-$. Instead, $\beta_-$ and thus $\psi_-|_{\infty}$ is determined by solving the Dirac equations subject to the above boundary conditions.
The expectation value $\langle \bar{\cal O} \rangle$ of the dual operator is given by the conjugate momentum $\Pi_+$ defined in \eqref{conjMom}: $\langle \bar{\cal O} \rangle \propto \lim_{r \to \infty} r^{-\Delta_-}\Pi_+ \propto \bar{\beta}_1$. 

On the other hand for $0\leq m L<\frac{1}{2}$ both modes $\psi_\pm$ are normalizable \cite{Klebanov:1999tb,Amsel:2008iz,Amsel:2009rr,Faulkner:2009wj,Iqbal:2009fd,Guarrera:2011my,Andrade:2011dg}. Thus we can still impose the {\it standard} quantization where we identify the $\lim_{r \to \infty} r^{\Delta_-}\psi_+\equiv \psi_+^{(0)}$  as the source of the dual operator $\bar{\cal O}$. In particular, the no-source/homogeneous standard boundary condition for all possible masses:
\begin{eqnarray}\label{StandardQuant}
\psi_+^{(0)}=0 \quad \Leftrightarrow \quad
\left\{
\begin{array}{ll}
\alpha_1=0, & \qquad \hbox{if $\, 0< m L<\frac{1}{2}$ \,(or $m L\geq\frac{1}{2}$);} \\
\alpha_1 (\lambda +\omega L)-i \beta_1=0, & \qquad \hbox{if $\, m=0$;}
\end{array}
\right. 
\end{eqnarray}
But, since  for this range of masses both modes are normalisable,\footnote{Besides the single-trace standard/alternative boundary conditions, we can also impose multi-trace deformations which are mixed boundary conditions; see, e.g. \cite{Breitenlohner:1982jf,Breitenlohner:1982bm,Klebanov:1999tb,Ishibashi:2004wx,Amsel:2008iz,Amsel:2009rr,Andrade:2011dg}.} we can also impose the so-called {\it alternative} quantization; where we identify the $\lim_{r \to \infty} r^{\Delta_+}\psi_-\equiv \psi_-^{(0)}$ as the source of the dual operator ${\cal O}$ with mass dimension $\Delta_-$. In particular, if we do not want to deform the boundary field theory we impose the no-source alternative boundary condition:
 \begin{eqnarray}\label{AlternativeQuant}
\psi_-^{(0)}=0 \quad \Leftrightarrow \quad
\left\{
\begin{array}{ll}
\beta_1=0, & \qquad \hbox{if $\, 0< m L<\frac{1}{2}$;} \\
\alpha_1 (\lambda -\omega L)+i \beta_1=0, & \qquad \hbox{if $\, m=0$;}
\end{array}
\right. 
\end{eqnarray}
The two quantizations \eqref{StandardQuant} and \eqref{AlternativeQuant} yield two distinct boundary conformal field theories \cite{Klebanov:1999tb,Amsel:2008iz,Andrade:2011dg,Faulkner:2009wj,Iqbal:2009fd}. For $m=0$, note that the Dirichlet boundary condition on $\psi_+$, $\psi_+^{(0)}=0$, implies the Neumann condition in $\psi_-$ ({\it i.e.} the next-to-leading order term in the expansion for $\psi_-$ vanishes) and vice-versa. This follows straightforwardly from an inspection of \eqref{AsympDecays:psi+-m0}. 

We emphasize that the no-source standard and alternative boundary conditions  \eqref{StandardQuant}-\eqref{AlternativeQuant} that do not deform the boundary theory imply that the energy flux and fermion particle flux vanish at the asymptotic boundary (this is also the case for more elaborated normalizable AdS/CFT boundary conditions \cite{Amsel:2008iz,Andrade:2011dg}). In this sense we can regard these as `reflective' boundary conditions. 
The Dirac action \eqref{DiracAction} (and \eqref{bdryDiracAction}) is left invariant if we rotate the phase of the Dirac spinor, $\psi \to e^{-i \alpha} \psi $. The Dirac current associated to this symmetry is $j^\mu=\bar{\psi}\gamma^\mu \psi$
and one can check that it is conserved, $\nabla_\mu j^\mu=0$ after using the first order equations of motion \eqref{DiracEOM}. This is an internal vector symmetry since $\psi_\pm$ transform in the same way under this symmetry. This current gives the charge flux or particle number flux of fermions. The associated conserved charge is ${\cal Q}=\int_V dx^3 \sqrt{\gamma} j^\mu\xi_\mu=\int_V dx^3 \sqrt{\gamma} \psi^\dagger\psi$ where $V$ is the volume of a constant $t$ hypersurface, $\gamma_{ab}$ is the associated induced metric, and $\xi=\partial_t$ is the Killing vector describing time translations. In particular, $j^r|_{r\to\infty}$ gives the radial flux of particles at the asymptotic spacelike boundary $\Sigma$. One can also show that the energy flux across a spacelike boundary is proportional to the Dirac current. The energy flux across the asymptotic boundary $\Phi_{\partial_t}\rvert_{\infty}$ is proportional to the particle flux $j^r\rvert_{\infty}$ and is given by\footnote{Let again $\xi=\partial_t$ be the Killing vector field conjugate to the energy. The energy-momentum tensor for the Dirac field is $T_{\m\nu}=\frac{i}{2}\left[ \bar{\psi}\gamma_{(\mu}{\cal D}_{\nu)} \psi - \left({\cal D}_{(\mu} \bar{\psi}\right)\gamma_{\nu)}\psi \right]$ and it is conserved $\nabla_\mu T^{\mu\nu}=0$. This conservation law together with the Killing equation, $\nabla_{(\mu}\xi_{\nu)}=0$, imply that the 1-form $\mathcal{J}_\mu=-T_{\mu\nu} \xi^\nu$  is conserved, $d\star \mathcal{J}=0$, where $\star$ is the Hodge dual. 
We can then define the energy flux across the asymptotic hypersurface $\Sigma$ (like the asymptotic boundary) as   
$ \Phi_\xi \equiv -\int_\Sigma \star \mathcal{J}
 = -\int_{\Sigma} \!\!dV_{\Sigma} \;T_{\mu\nu}\xi^\mu n^\nu $
where $n^\nu$ is the unit normal vector to $\Sigma$ and $dV_\Sigma$ is the induced
volume on $\Sigma$.} 
\be\label{flux}
\Phi_{\partial_{t}} \rvert_{\infty} \propto  \rvert R_{1} \rvert^{2} -\rvert R_{2} \rvert^{2} .
\ee
Inserting the asymptotic decays \eqref{AsympDecaysR} for $R_{1,2}$ this yields
\be
\Phi_{\partial_{t}} \rvert_{\infty} \propto \alpha_1^*\beta_1+\alpha _1 \beta_1^*\,,\qquad \hbox{if $m\neq 0,\frac{1}{2}$}.
\ee
That is, the energy flux at the asymptotic boundary vanishes if we impose the above discussed no-source Dirichlet boundary conditions $\alpha_1=0$ or, for the alternative quantization,  $\beta_1=0$ which do not deform the boundary conformal field theory.

On the other hand, for $m=0$, inserting the asymptotic decays \eqref{AsympDecaysRm0} for $R_{1,2}$ into \eqref{flux} yields
\be\label{fluxm0}
\Phi_{\partial_{t}} \rvert_{\infty} \propto \lambda^2\alpha _1\alpha_1^*-\left(\beta_1+i \,\alpha_1 \omega L\right) \left(\beta_1^*-i \,\alpha_1^* \omega L \right)
\,,\qquad \hbox{if $m= 0$}.
\ee
Again, this flux vanishes if we impose the standard \eqref{StandardQuant} or alternative \eqref{AlternativeQuant} quantizations, $\beta_1=-i\left(\pm \lambda +\omega L\right)\alpha_1$ (and thus $\beta_1^*=i\left(\pm \lambda +\omega L\right)\alpha_1^*$). 

Here it is important to recall the clarification about AdS/CFT boundary conditions and vanishing flux conditions presented in the Introduction. The standard and alternative boundary conditions that we use have, by construction, zero energy flux at the asymptotic boundary, as reviewed above and originally discussed in \cite{Mueck:1998iz,Henningson:1998cd,Henneaux:1998ch,Contino:2004vy,Amsel:2008iz,Amsel:2009rr,Andrade:2011dg}. Without noticing, these standard/alternative boundary conditions are also the boundary conditions used in \cite{Wang:2017fie,Wang:2019qja}  where the ``generic physical principle of zero energy flux" was used to motivate the  boundary conditions originally established in  \cite{Mueck:1998iz,Henningson:1998cd,Henneaux:1998ch,Contino:2004vy,Amsel:2008iz,Amsel:2009rr,Andrade:2011dg} (using precisely the same rationale). But there is a broader family of zero-flux boundary conditions. The AdS/CFT standard and alternative quantizations are  a special class of zero-flux boundary conditions that, additionally, preserve the conformal symmetry group of AdS \cite{Mueck:1998iz,Henningson:1998cd,Henneaux:1998ch,Contino:2004vy,Amsel:2008iz,Amsel:2009rr,Andrade:2011dg}. It is this property that singles them out among other zero-flux boundary conditions that break this conformal symmetry \cite{Amsel:2008iz,Amsel:2009rr,Andrade:2011dg}. 
Further note that zero-flux boundary conditions are sometimes denoted as `reflective' boundary conditions in some literature and both set of words encode the familiar idea that `AdS behaves as a confining box' (under these boundary conditions).\footnote{Note however that in AdS/CFT there are other sets of boundary conditions that yield a well-defined boundary value problem but do not correspond to zero-flux boundary conditions ({\it e.g.} mass deformations describe sourced solutions with important physical interpretations where gauge field(s) have a non-vanishing asymptotic flux).}  

It is also important to emphasize that in the AdS/CFT language the standard classification of Dirichlet/Neumann/Robin boundary conditions applies to the physical fields that obey the original differential equation (in the present $s=1/2$ case, the Dirac equation). Often this classification does not then translate into the same type of boundary conditions on auxiliary (or even gauge invariant) fields that one might introduce. The classification should focus on the physical fields and not on auxiliary fields (we can fabricate many of these), unlike what is done for $s=1/2$ in \cite{Wang:2017fie,Wang:2019qja}. For example, for a massless Dirac fermion, no-source Dirichlet/Neumann boundary conditions $\psi_{\pm}^{(0)}=0$ translate into $\beta_1=-i\left(\pm \lambda +\omega L\right)\alpha_1$ not $\alpha_1=0$ or $\beta_1=0$. Facts like this are often missed:\footnote{This is {\it e.g.} the case in \cite{Giammatteo:2004wp,Jing:2005ux} where massless Dirac quasinormal modes of Schwarschild-AdS  are computed with the  Dirichlet boundary condition $\alpha_1=0$. This choice of boundary condition is not one of the AdS/CFT zero flux boundary conditions for a massless Dirac field.} zero-flux boundary conditions that preserve conformal symmetry require $\psi_{\pm}^{(0)}$ to vanish not $R_{1,2}|_\infty$.\footnote{Further note that there are other boundary conditions ({\it e.g.} $\beta_1=-i\left(\pm i\,\lambda +\omega L\right)\alpha_1$) that make the flux \eqref{fluxm0} vanish. These should correspond to multi-traced ({\it i.e.} mixed or Robin) AdS/CFT boundary conditions \cite{Amsel:2008iz,Andrade:2011dg} which deform the boundary theory in a way that might be interesting for other studies.}

\subsection{Near-horizon geometry of  the extreme AdS-RN black hole}\label{sec:NH}

The near-horizon geometry of the extremal AdS-RN black hole will play an important role in our discussions in sections \ref{sec:Scalar} and \ref{sec:DiracRN}. Therefore, we review it here. The limiting procedure described below was first presented in \cite{Bardeen:1999px}. 

The extremal AdS-RN black hole is given by \eqref{RNAdS} with $\mu=\mu_{\rm ext}$ given by \eqref{RNextreme}. To obtain the near-horizon geometry, it is convenient to work in the gauge $A|_{r_+}=0$ ($C=\mu$; otherwise we can do a gauge transformation in the end). One first zooms in around the horizon region by making the coordinate transformations:
\be\label{NHtransf}
r = r_{+} + \varepsilon \rho, \qquad
t = L_{AdS_{2}}^{2} \frac{\tau}{\varepsilon}\,,
\ee
where $L_{AdS_{2}}$ is the $AdS_{2}$ radius (to be defined below). Now the near-horizon geometry is obtained by taking $\varepsilon \to 0$ which yields
\beq \label{NHgeometry}
&& ds_{NH}^{2} = L_{AdS_{2}}^{2}\left(-\rho^{2} d\tau^{2} + \frac{d\rho^{2}}{\rho^{2}}\right) + r_{+}^2 d \Omega_{2}^{2}, \qquad L_{AdS_{2}} =\frac{ L r_{+}}{\sqrt{L^{2} + 6 r_{+}^2}}; \nonumber \\
&& A_{\mu}^{NH} dx^{\mu} = \alpha \,\rho \, d\tau ,  \hspace{1.5em} \alpha = L_{AdS_{2}} \sqrt{1 + \frac{L_{AdS_{2}}^{2}}{r_{+}^{2}}}\,.
\eeq
This geometry is the direct product of $AdS_2\times$S$^2$ and has a Maxwell potential that is linear in the radial direction. Remarkably, in spite of the limiting procedure, it is still a solution of the 4-dimensional Einstein-Maxwell-AdS theory. On the other hand, the AdS$_2$ metric solves the 2-dimensional Einstein-AdS equations, $R_{\mu\nu}=-L_{AdS_{2}}^{-2} g_{\mu\nu}$, if $L_{AdS_{2}}$ is identified as a function of the AdS$_4$ radius $L$ and the horizon radius $r_+$ as indicated in the first line of \eqref{NHgeometry}.

\section{Scalar fields in a AdS-RN background and their instabilities}\label{sec:Scalar}

Scalar fields confined inside the gravitational potential (like the AdS potential or a box in an asymptotically Minkowski background) of a black hole can condense creating near-horizon linear instabilities \cite{Gubser:2008px,hartnoll2008building,Faulkner:2009wj,Dias:2010ma,Dias:2011tj,dias2012hairyBHs,Dias:2016pma} (for planar AdS, this instability triggered the holographic superconductor programme  \cite{Gubser:2008px,hartnoll2008building,Faulkner:2009wj}). Essentially this happens because we can have scalar fields that obey the asymptotically AdS$_4$ UV Breitenl\"ohner-Freedman (BF) bound but violate the 2-dimensional BF stability bound associated to the AdS$_2\times S^2$ near-horizon geometry of the extremal black hole of the system. As we shall discuss in section \ref{sec:DiracRN1}, a similar violation of the 2-dimensional stability bound can occur for Dirac fields. In spite of this, as we will find in section \ref{sec:DiracRN3}, it turns out that Dirac fields are not linearly unstable to the near-horizon condensation mechanism. Therefore, before we discuss the fermionic case, it is important to revisit the scalar field case. This will allow to: 1) motivate the search of linear instabilities due to Dirac fields done in this manuscript, 2) eventually identify differences between the two spins that could help in understanding the opposite outcomes. We also take the opportunity to demonstrate: i) how remarkably sharp the near-horizon instability bound \eqref{scalar:qminNH} is by comparing it with the numerical solutions of the Klein-Gordon equation, and ii) that the unstable modes are both peaked near the horizon but also connected to the AdS normal modes (that is to say, in the language of \cite{Cardoso:2017soq,Dias:2018ynt,Dias:2018etb} the AdS and near-extremal families of modes coincide and describe the unstable modes).  

Using the fact that the AdS-RN background \eqref{RNAdS} is static and spherically symmetric we can consider a separation ansatz for the scalar field (with mass $m$ and charge $q$) with the Fourier decomposition
\be 
\Phi(t,r,\theta,\phi)=e^{-i\o t}e^{i m_\phi \phi}Y_{\ell}(\theta) \phi(r),
\ee
where $Y_{\ell}(\theta)$ are the familiar (spin-0) spherical harmonics $-$ which are regular when the separation constant of the system is quantized as $\lambda=\ell(\ell+1)$, $\ell=0,1,2,\cdots$ $-$ and $|m_\phi|\leq \ell$ is the azimuthal quantum number. The Klein-Gordon equation  yields the following equation for the radial function $\phi(r)$:
\be \label{ScalarEqn}
\frac{d}{dr}\left(r^2 f \frac{d\phi}{dr}\right)+\left(\frac{r^2}{f}\left(\o +q A\right)^{2}-r^2 m^2-\ell(\ell+1)\right)\phi=0.
\ee
A Taylor expansion around the asymptotic boundary yields the two independent solutions
\begin{equation}
\label{BFdecays}
 \phi(R)\simeq r^{-\Delta_-^{(\rm s)}}( a+\cdots)+ r^{-\Delta_+^{(\rm s)}}(b+\cdots)\,, \quad
 \hbox{with} \quad \Delta_\pm^{(\rm s)}=\frac{3}{2}\pm \sqrt{\frac{9}{4}+m^2L^2}
\end{equation}
being the conformal dimensions of the field.
Such a scalar field in AdS$_4$ is normalizable as long as its mass obeys the AdS$_4$  Breitenl\"ohner and Freedman (BF) bound, 
$m^2 \ge m^2_{\mathrm{BF}}\equiv -\frac{9}{4}\,\frac{1}{L^2}$ \cite{Breitenlohner:1982jf,Breitenlohner:1982bm}.
 
Such a scalar field that is stable in the UV region can however be unstable in the IR region. This is best understood if we take the near-horizon limit of \eqref{ScalarEqn}.
Concretely, applying the near-horizon coordinate transformation \eqref{NHtransf} together with the near-horizon frequency transformation $\tilde{\omega}\to \widehat{\omega} \,\varepsilon/L_{AdS_2}^2$ (so that $e^{-i\tilde{\omega} t}\to e^{-i\widehat{\omega} \tau}$) followed by the near-horizon limit $\varepsilon \to 0$ yields the radial Klein-Gordon equation in the near-horizon geometry \eqref{NHgeometry}:
\begin{equation}\label{NHlimKG}
\partial_{\rho}\left(\rho^2\partial_{\rho} R\right)+\left(\frac{\left(\widehat{\omega}+q \,\alpha \,\rho \right)^2}{\rho^2}-m^2 L_{AdS_2}^2\right)R=0,
\end{equation}
This is nothing else but the Klein-Gordon equation for a scalar field around AdS$_2$ with an electromagnetic potential $A_\tau=\alpha\, \rho$. A Frobenius analysis of \eqref{NHlimKG} yields
\begin{equation}
\label{BFdecaysAdS2}
R{\bigl |}_{\rho\to\infty}\simeq \rho^{\,-\widehat{\Delta}_-^{(\rm s)}}
(\widehat{a}+\cdots)+ + \,\rho^{\,-\widehat{\Delta}_+^{(\rm s)}}(\widehat{b}+\cdots)\,, \quad
 \hbox{with} \quad \widehat{\Delta}_\pm^{(\rm s)}=\frac{1}{2}\pm \frac{1}{2}\sqrt{1+m_{eff}^2L_{AdS_2}^2}\;,
\end{equation}
which determines the effective mass of the scalar field from the perspective of a near-horizon observer,
\begin{equation}
\label{nearmass}
m_{eff\,(\rm s)}^2 L_{AdS_2}^2 \equiv m^2 L_{AdS_2}^2 -q^2 \,\alpha^2
\end{equation}
Now, a scalar field with mass \eqref{nearmass} in AdS$_2$ has unstable modes if it violates the  AdS$_2$ BF bound $m_{eff\,(\rm s)}^2  \ge m^2_{\mathrm{AdS_2\, BF}}\equiv -\frac{1}{4}\,\frac{1}{L_{AdS_2}^2}$. It follows that extremal AdS-RN$_4$ black holes should be unstable whenever the charge of the scalar field obeys
\begin{equation}\label{scalar:qminNH}
q^2\geq \frac{1}{L^2} \left( 1+4 m^2 L^2 \frac{r_+^2}{L^2+6r_+^2} \right) \frac{\left(L^2+6 r_+^2\right)^2}{8 r_+^2 \left(L^2+3 r_+^2\right)}, \qquad \hbox{(near-horizon instability bound)}\,. 
\end{equation}
Note that scalar fields can also induce instabilities due to another mechanism that is known as  
superradiance. Unlike the near-horizon instability $-$ which is suppressed in the limit $r_+/L \to 0$; indeed \eqref{scalar:qminNH} goes as $q^2L^2\geq \frac{L^2}{8r_+^2}+\mathcal{O}(1)$  $-$ the superradiant instability is present for small $r_+/L\ll 1$ black holes. For example, for $m=0$, from the perturbative results of 
\cite{Dias:2016pma} one finds that the superradiant instability in extremal AdS-RN$_4$ is present for scalar charges\footnote{This bound can be obtained from the expression for the frequency obtained in section III.D of \cite{Dias:2016pma}. Namely, the onset charge \eqref{scalar:qminSR} of the superradiant instability is obtained by setting $\tilde{\omega}=0$ and $\mu=\mu_{\rm ext}$ in equation (55) of \cite{Dias:2016pma} and solving for the charge $q$. For further discussions between the entanglement of the superradiant and near-horizon instabilities and their different nature we ask the reader to see \cite{Dias:2016pma} and  \cite{Dias:2018zjg}.} 
\begin{align}\label{scalar:qminSR}
q L \geq \frac{3}{\sqrt 2}-\frac{9}{2\sqrt 2}\frac{r_+^2}{L^2}+{\cal O}\left( \frac{r_+^4}{L^4} \right), \qquad \hbox{(superradiant instability bound)}
\end{align}

Next, we solve the Klein-Gordon equation numerically to confirm that the near-horizon and superradiant instabilities are indeed present and to find how sharp the instability bounds  \eqref{scalar:qminNH} and \eqref{scalar:qminSR} are. We present results for scalar masses above the unitarity bound $m_{\mathrm{BF}}^2+1=-5/4$ so asymptotically we impose the Dirichlet boundary condition $a=0$; see \eqref{BFdecays}.\footnote{For $m_{\mathrm{BF}}^2<m^2<m_{\mathrm{BF}}^2+1$ both modes are normalizable and thus we could also impose the Neumann boundary condition $b=0$ (the so-called alternative quantization in the context of AdS/CFT) \cite{Breitenlohner:1982jf,Breitenlohner:1982bm}.} On the other hand at the horizon we require that the solution is regular in the future horizon which discards outgoing modes.
To present the results, note that our system has a scaling symmetry \cite{Dias:2016pma} which means that the physical  dimensionless quantities that are relevant for the problem are (this effectively sets $L\equiv 1$)
\be\label{dimensionless}
\left\{\frac{r_+}{L},\mu; m L, q L, \omega L,\ell \right\}.
\ee

\begin{figure}[th]
\centerline{
\includegraphics[width=.48\textwidth]{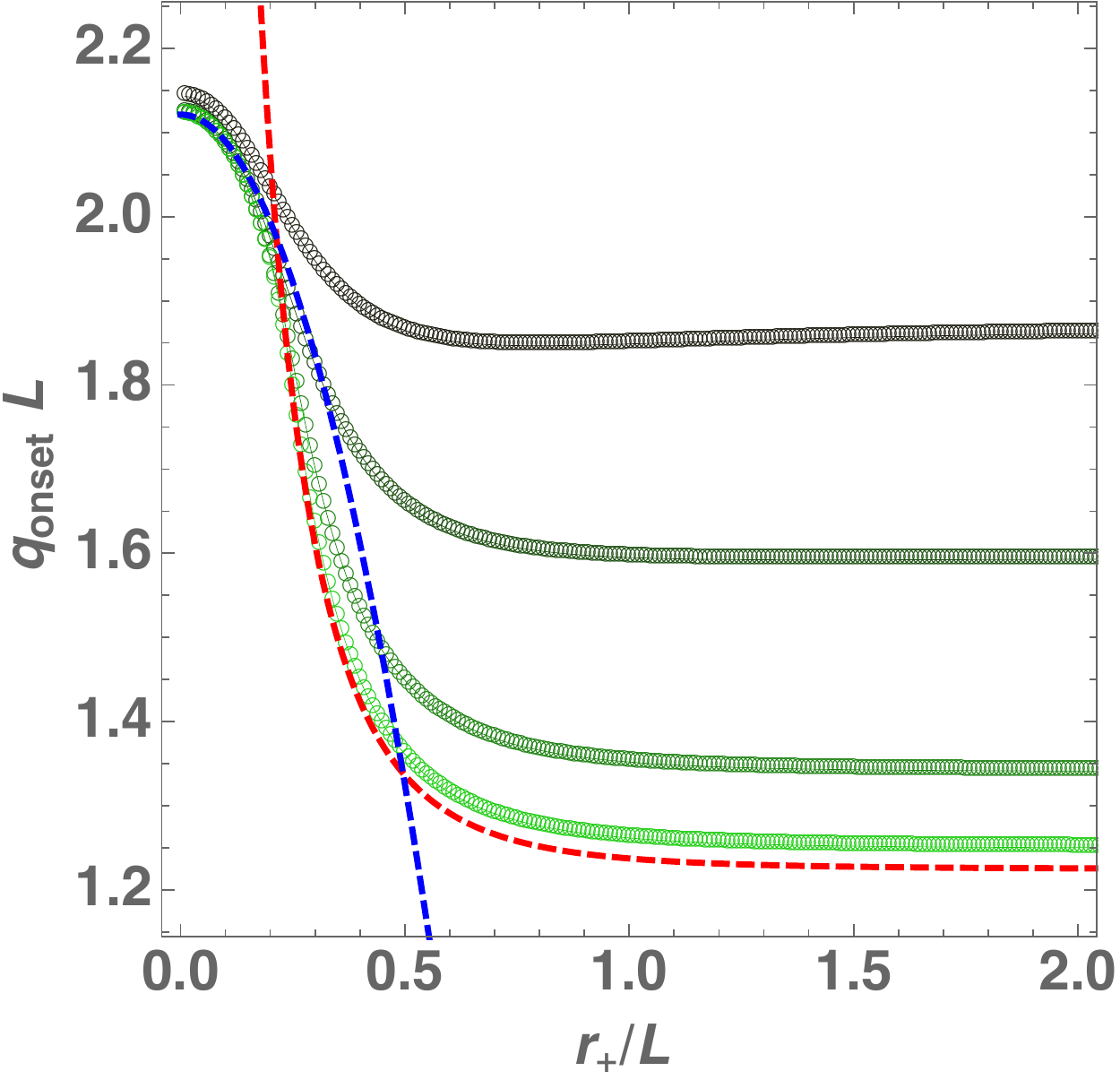}
\hspace{0.3cm}
\includegraphics[width=.48\textwidth]{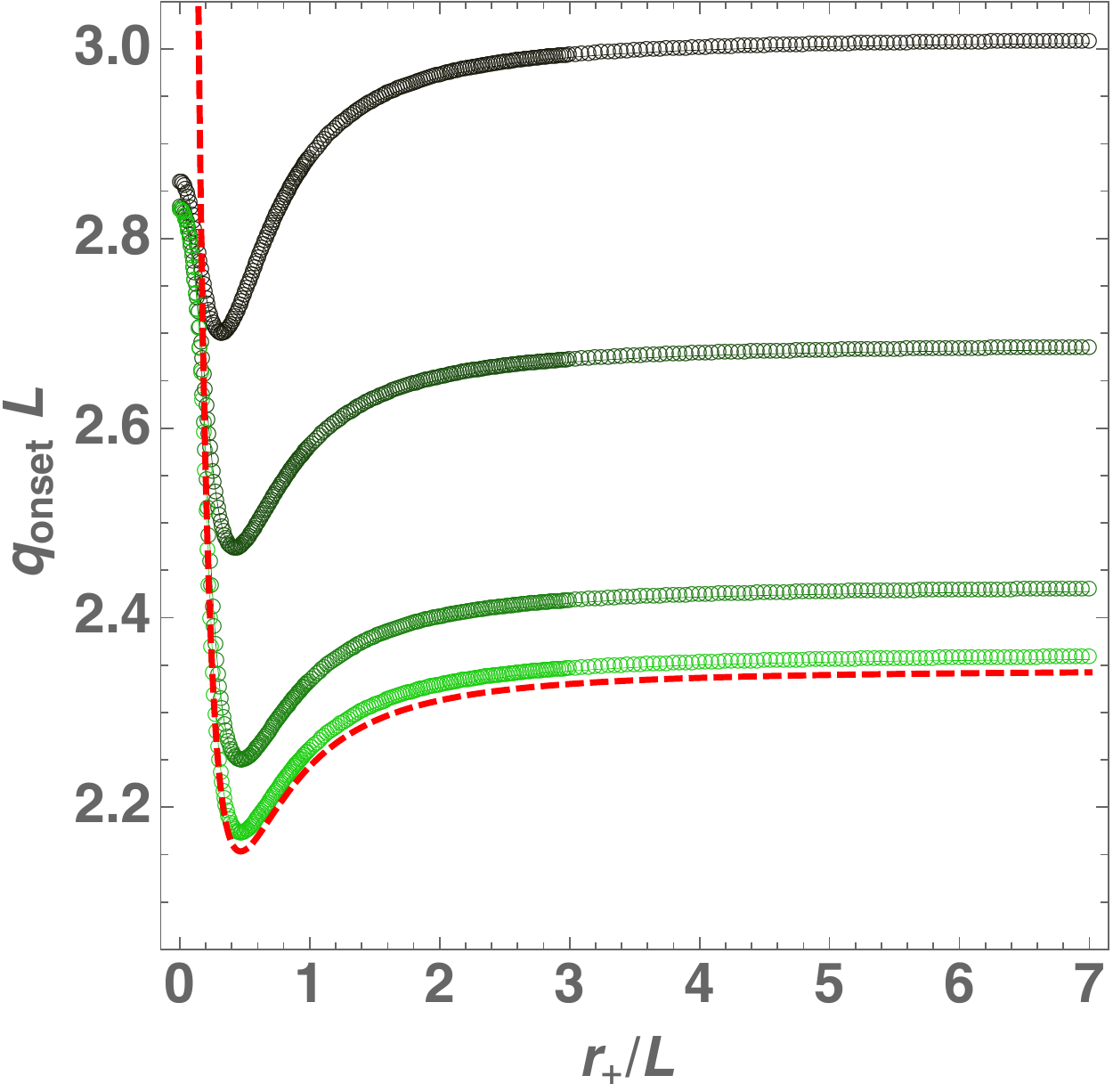}
}
\caption{Onset scalar field charge as a function of the horizon radius for chemical potential $\mu=\mu_{\rm ext}(1-10^{-x})$ with $x=2,3,6,15$ (from top to bottom on the right of each panel). The left panel corresponds to massless scalar fields; the right panel to massive scalar fields with $m L=2$. The red dashed line is the near-horizon condensation analytic bound \eqref{scalar:qminNH}. In addition in the left panel we have the dashed blue line (for small horizon radius) which is the superradiant bound \eqref{scalar:qminSR}.}
\label{scalar-fig:onset}
\end{figure} 

\begin{figure}[t]
\centerline{
\includegraphics[width=.48\textwidth]{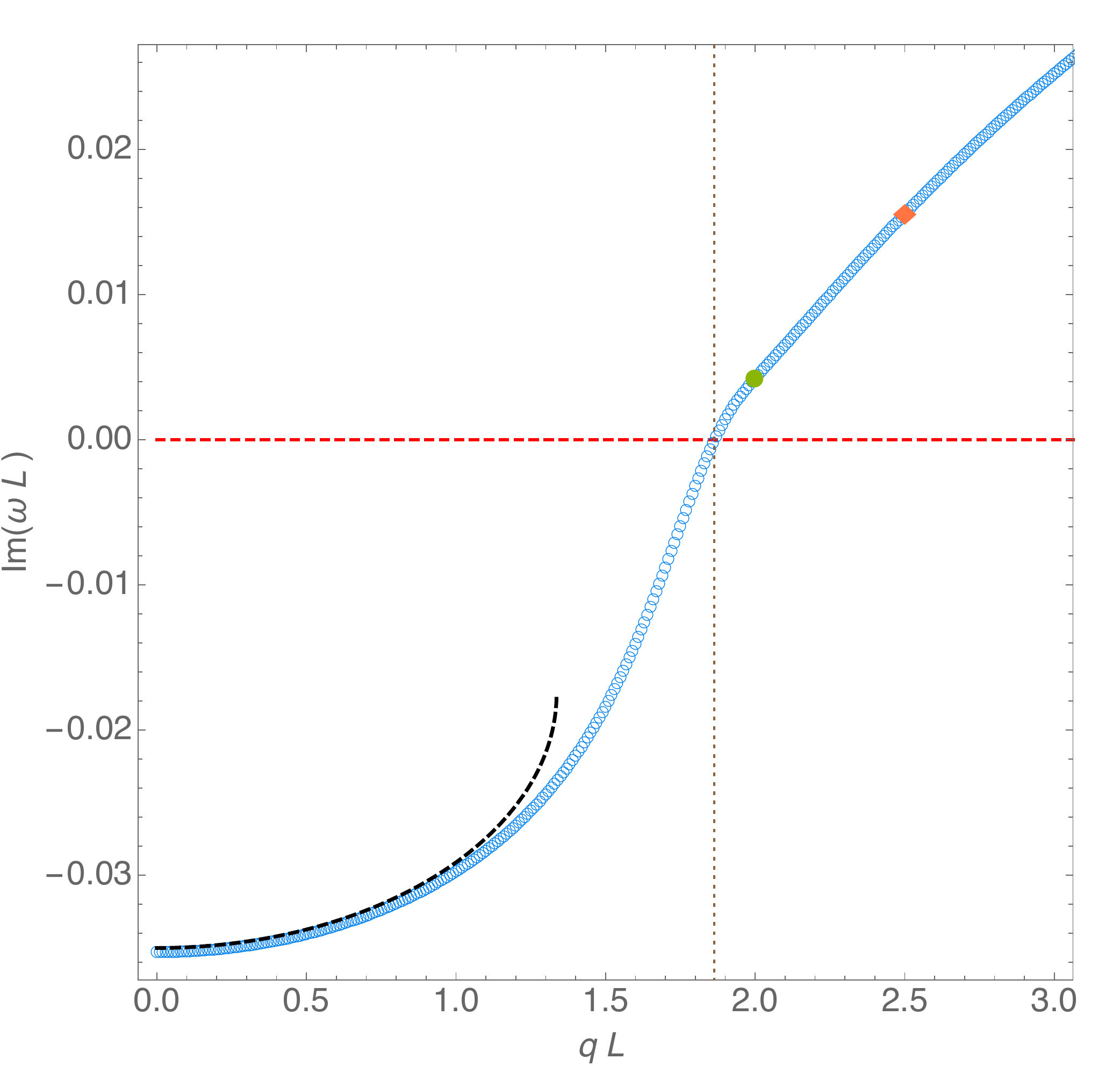}
\hspace{0.3cm}
\includegraphics[width=.46\textwidth]{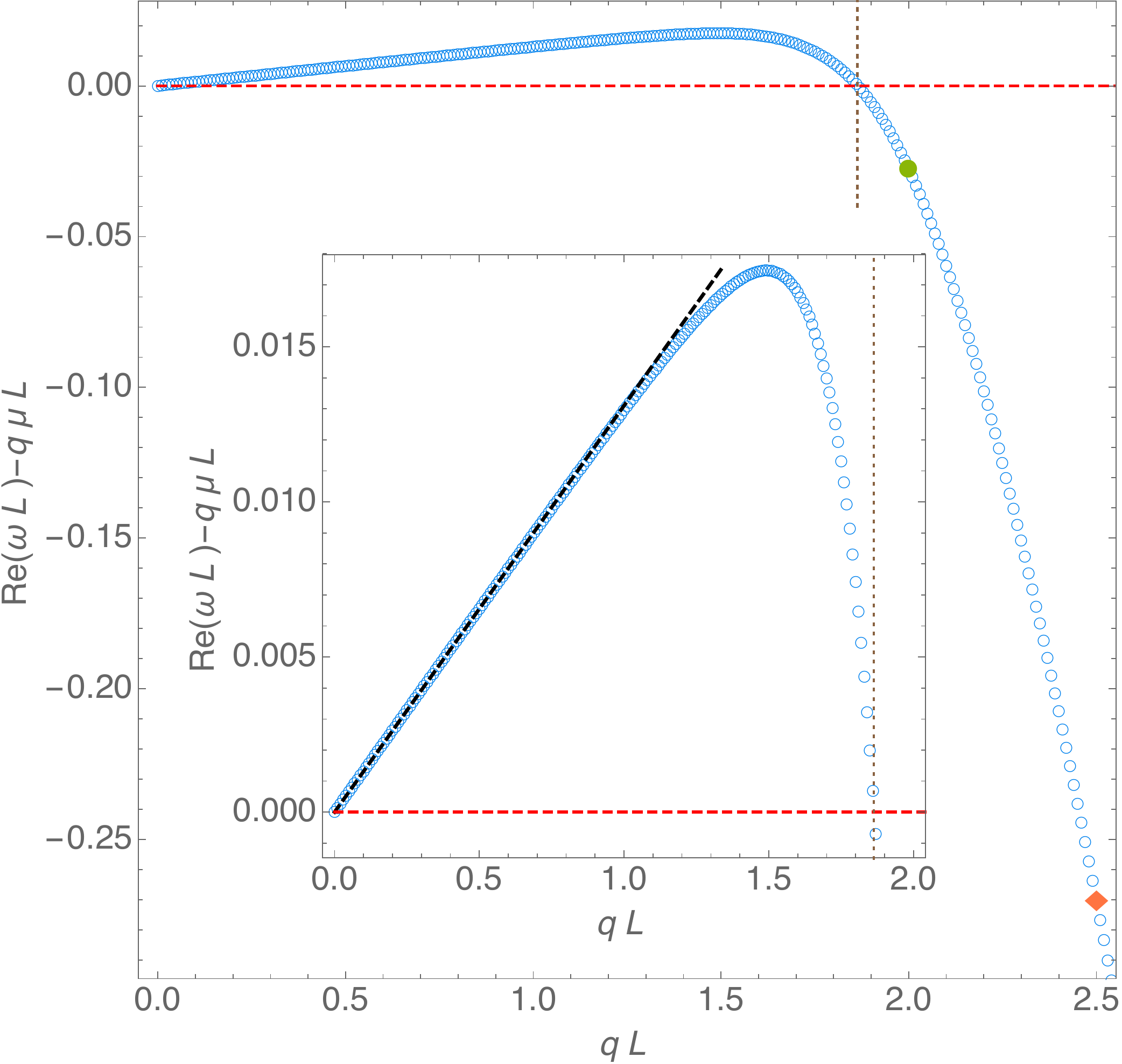}
}
\caption{Scalar field frequency as a function of the dimensionless scalar charge $q L$ for a AdS-RN black hole with $\mu = 0.99 \mu_{\rm ext}$ and $r_{+}/L = 0.5$ ($m L=0$ and $\ell=1$). {\it Left panel}: Imaginary part of the dimensionless frequency, ${\rm Im}(\omega L)$. The system becomes unstable for $q>q_\star$ where $q_\star L\sim 1.863$. {\it Right panel}: Real part of the dimensionless frequency, ${\rm Re}(\omega L)$, measured with respect to $q \mu L$. This quantity changes sign at $q=q_\star$, {\it i.e.} when ${\rm Im}(\omega L)$ changes sign. 
In both plots, the black dashed curves describe the analytic prediction of the asymptotic matching expansion \eqref{NExt_modes}. We find very good agreement with the numerical results (blue curves) for  $q L<1.1$ (say). This is a further justification of the crude assumption that we should match with 0 in the overlapping region. \textit{However} we find that these modes connect with the AdS normal modes as $r_{+} \rightarrow 0$.}
\label{scalar-fig:freqVSq}
\end{figure} 

First, we are interested in finding the onset of the instabilities namely, the scalar field charge $q_{onset}(\mu,r_+/L;mL,\ell)$ above which the system is unstable. This onset occurs when the frequency satisfies $\tilde{\omega}=\omega -q\mu =0$. The Klein-Gordon equation \eqref{ScalarEqn} is then solved as an eigenvalue problem for $q=q_{onset}$.
For concreteness, we fix $\ell=1$ (we need $m_\phi \geq 1$ to have an instability). In the left plot of Fig.~\ref{scalar-fig:onset} we set $m=0$ and we plot the dimensionless onset charge $q_{onset} L$ as a function of the dimensionless horizon radius $r_+/L$  for different values of the chemical potential $\mu=\mu_{\rm ext}(1-10^{-x})$ that increasingly approaches the extremal value. From top to bottom, the green numerical curves describe chemical potentials with $x=2,3,6,15$. We see that as we get closer to extremality these onset curves increasingly approach (for values of $r_+/L$ larger than $\sim 0.25$) the red dashed curve which describes the near-horizon bound \eqref{scalar:qminNH}. This strongly suggests that the instability, for large values of the horizon radius and near extremality, can be understood as due to the violation of the AdS$_2$ BF and that the associated near-horizon bound \eqref{scalar:qminNH} is sharp ({\it i.e.} it is attained at extremality). On the other hand, as pointed out before, the near-horizon red dashed curve diverges as $r_+/L\to 0$. However, Fig.~\ref{scalar-fig:onset} shows that $q_{onset} L$ is finite for small $r_+/L$. Actually, in this regime the numerical onset curves are well described by the superradiant bound  \eqref{scalar:qminSR} (blue dashed curve with negative slope). This suggests that for small horizon radius and near extremality the instability has a superradiant nature and the superradiant bound  \eqref{scalar:qminSR}  is sharp. For finite values of $r_+/L$, {\it i.e.} away from the $r_+/L\to 0$ and $r_+/L\to \infty$ regions, the superradiant and near-horizon instabilities are entangled. These features are not unique to the massless case. For example, the onset charge plot for a scalar mass of $mL=2$ is shown in the right panel of Fig.~\ref{scalar-fig:onset}. Again, as extremality is approached the numerical green curves increasingly approach the near-horizon onset bound \eqref{scalar:qminNH} (in this case we do not show the curve corresponding to the perturbative superradiant curve because it was not computed in \cite{Dias:2016pma} but we see that the behaviour of the onset curves for $r_+/L\ll 1$ is similar to the massless case).  

To compare with what happens in the Dirac field case, it is enlightening to do the  following exercise whose results are summarized in Fig.~\ref{scalar-fig:freqVSq}. We pick a particular AdS-RN background with chemical potential $\mu = 0.99 \mu_{\rm ext}$ and horizon radius $r_{+}/L = 0.5$. We also fix the scalar mass to be $m L=0$ and the scalar field harmonic number $\ell=1$. Then we solve the Klein-Gordon equation to find the imaginary and real  parts of the frequency $\omega L$ as a function of the dimensionless scalar field charge $q L$: these are shown in the left and right panels, respectively, of Fig.~\ref{scalar-fig:freqVSq}. From the left panel we see that, in accordance with the conclusions of Fig.~\ref{scalar-fig:onset}, for small $q L$ the system is stable (since ${\rm Im}(\omega L)<0$) but there is a critical charge  $q_\star L\sim 1.863$ (vertical dotted line) above which the system becomes unstable. Precisely at this critical onset charge one has $Re(\omega)-q\mu=0$ and this quantity is negative (positive) for $q< q_\star$ ($q> q_\star$). The inset plot of Fig.~\ref{scalar-fig:onset} zooms-in the region  $q< q_\star$. In section \ref{sec:DiracRN3} we will find that the partner plot for Dirac fields is substantially different.

\begin{figure}[t]
\centerline{
\includegraphics[width=.49\textwidth]{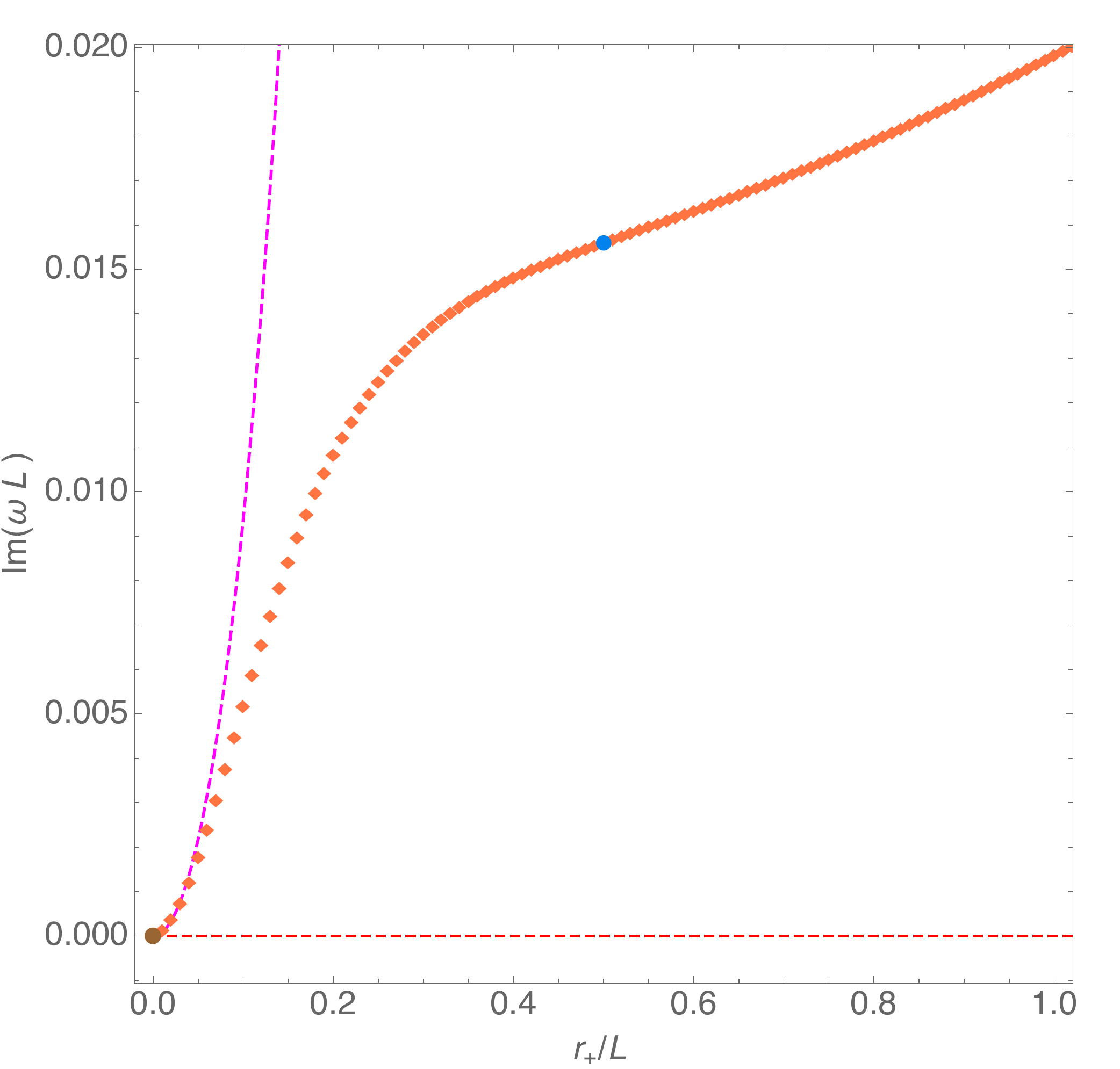}
\hspace{0.3cm}
\includegraphics[width=.455\textwidth]{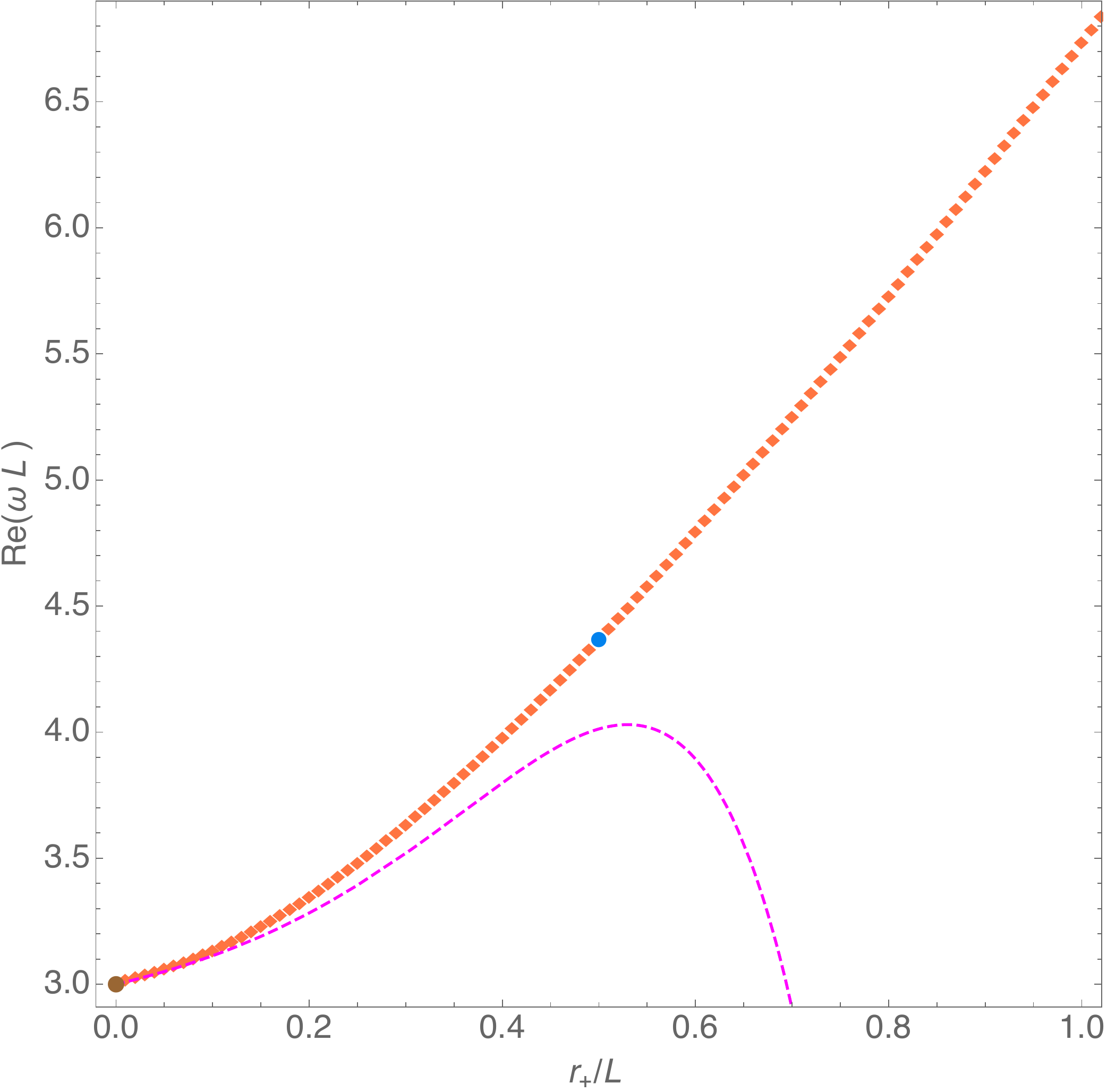}
}
\caption{Scalar field frequency as a function of the dimensionless horizon radius $r_{+}/L$  for a AdS-RN black hole with $\mu = 0.99 \mu_{\rm ext}$ and $q L= 2.5$ that is always above the near-horizon bound \eqref{scalar:qminNH} ($m L=0$ and $\ell=1$). {\it Left panel}: Imaginary part of the dimensionless frequency, ${\rm Im}(\omega L)$. {\it Right panel}: Real part of the dimensionless frequency, ${\rm Re}(\omega L)$. 
In both plots, the magenta dashed curves describe the analytic prediction of the perturbative expansion in $r_+/L$ about AdS. The unstable modes are thus connected to the AdS normal modes when $r_+\to 0$ (brown disk). For reference, the blue disk with $r_+/L=0.5$ is shown (which makes contact with Fig.~\ref{scalar-fig:freqVSq}).}
\label{scalar-fig:freqVSrp:q2.5}
\end{figure}

\begin{figure}[th]
\centerline{
\includegraphics[width=.49\textwidth]{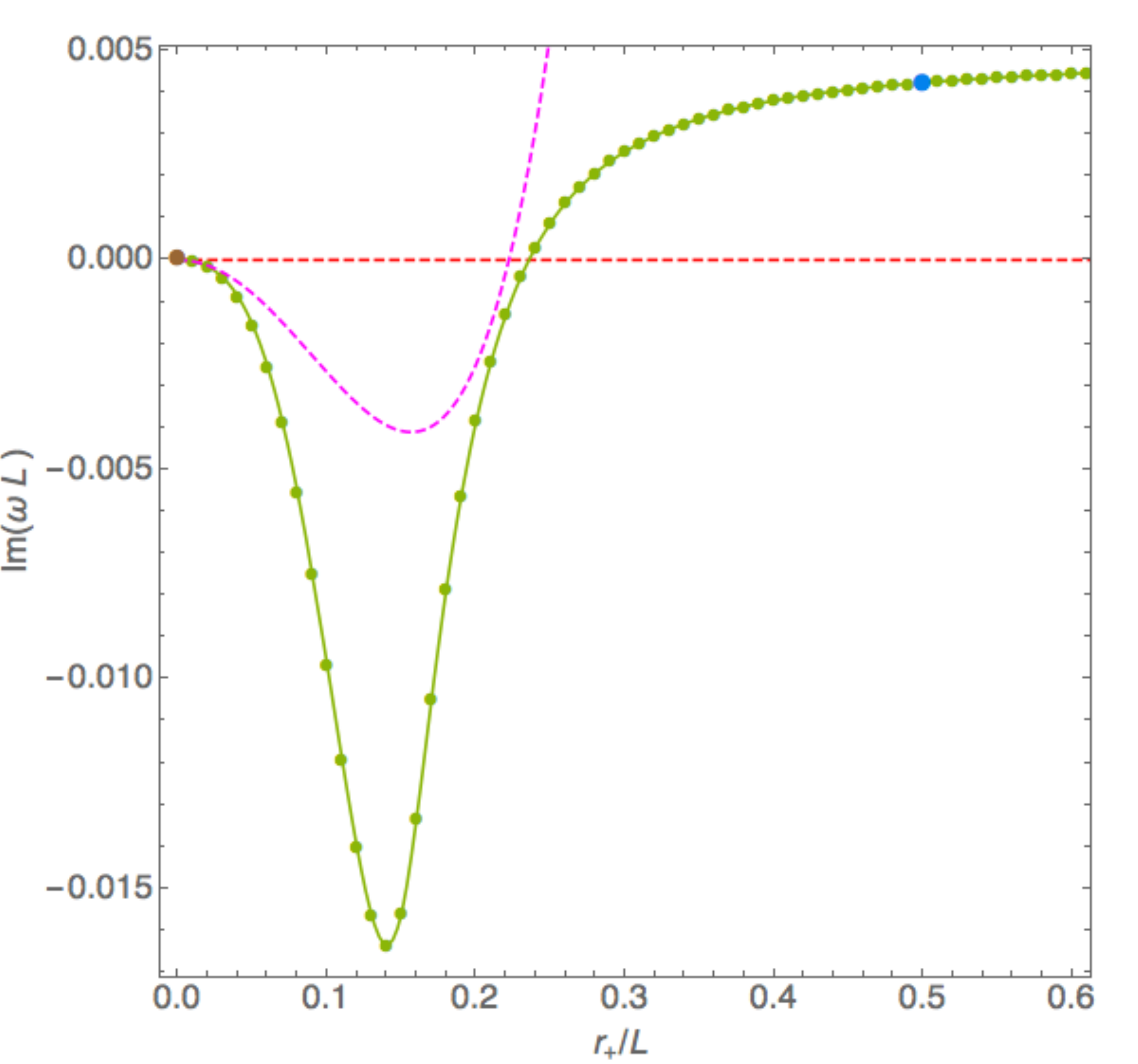}
\hspace{0.3cm}
\includegraphics[width=.445\textwidth]{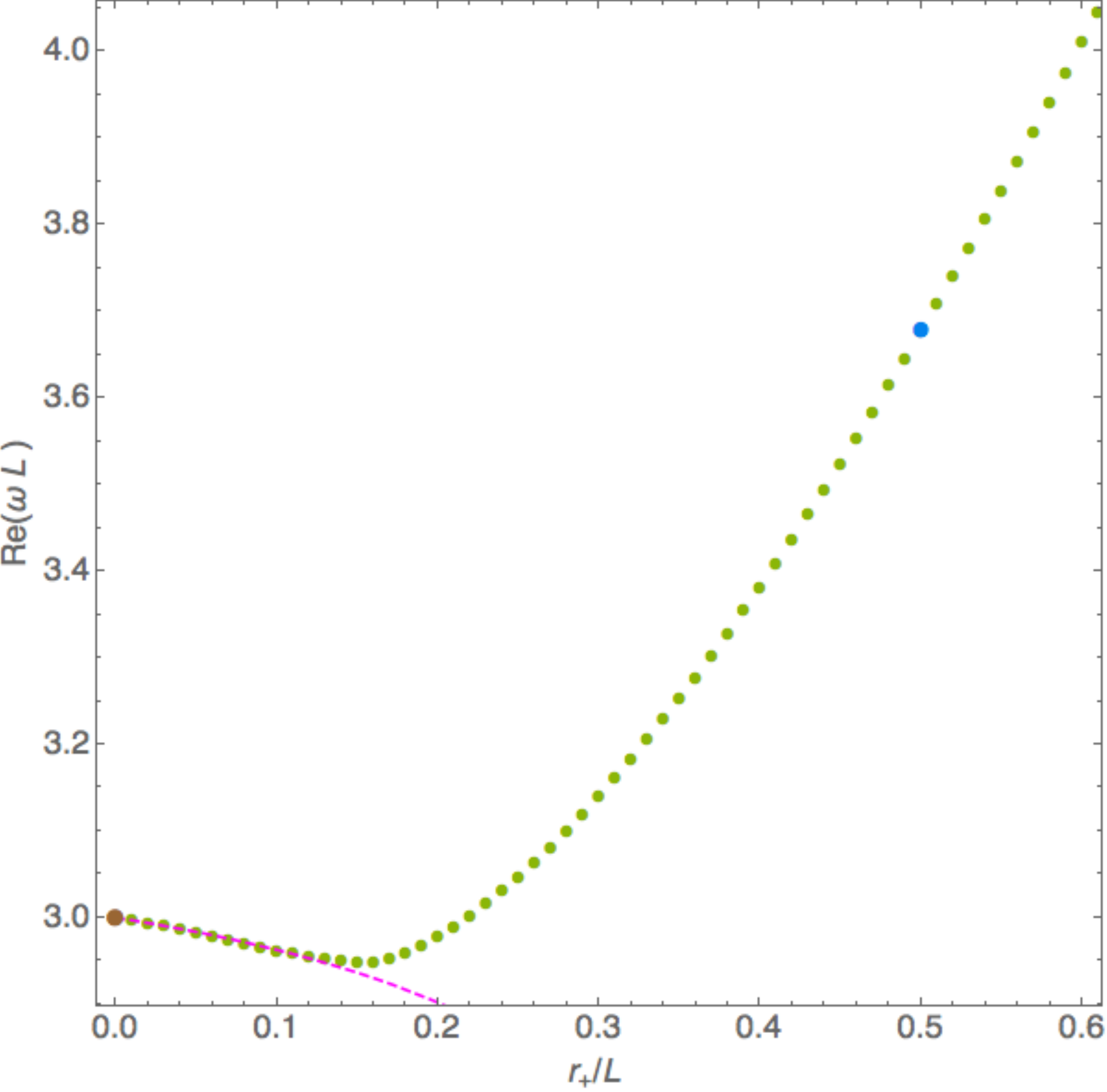}
}
\caption{Similar to Fig.~\ref{scalar-fig:freqVSrp:q2.5} but this time for a scalar field charge $q L= 2$ that is above the near-horizon bound \eqref{scalar:qminNH} only for $r_+/L$ above a certain value as seen in the left plot of Fig. \ref{scalar-fig:onset}.}
\label{scalar-fig:freqVSrp:q2}
\end{figure}

We take also the opportunity to understand better the frequency spectrum of scalar fields in AdS-RN. For global AdS RN black holes there are two quasinormal mode families \cite{Wang:2000gsa,Berti:2003ud,Uchikata:2011zz}: one whose imaginary part grows  negative without bound as the horizon radius $r_+/L$ decreases, and another whose imaginary part vanishes as $r_+/L \to 0$ and whose real part approaches the normal modes of AdS. The unstable modes are found in this second family. This could well be the complete story. However, in {\it de Sitter} black holes there is a third family of quasinormal modes $-$ called the near-extremal family $-$ whose wavefunctions are spatially peaked near the horizon and that is distinct from the de Sitter family (as the name suggests, the latter is connected to the normal modes of de Sitter when the black hole shrinks). This naturally raises the question: could it be that in AdS one also has a near-extremal family of quasinormal modes that is not connected to the AdS family? If so, do the near-horizon unstable modes with bound \eqref{scalar:qminNH} fit in this near-extremal family? We find a negative answer to these questions: the unstable modes belong to the AdS family of modes and the near-extremal family {\it coincides} with the AdS family. To arrive to this conclusion we first use  a matching asymptotic expansion similar to the one used in de Sitter \cite{Cardoso:2017soq,Dias:2018ynt,Dias:2018etb,Dias:2018ufh} to find the frequency spectrum of the near-extremal family of quasinormal modes. This is done in Appendix \ref{secA:nearExtremalQNMs} and here we just quote the final result: near-extremality and for small scalar field charge  one finds that near-extremal modes have the frequency (for the lowest radial overtone $p=0$)  
\begin{eqnarray}\label{scalar:NearExtremalFreq}
&&  \hspace{-1.5cm} \omega L \sim e \mu + \sigma \Bigg[  \frac{e  \sqrt{1+3 R_+^2}}{\sqrt{2}}
-i\, \frac{1}{4 R_+}\Bigg((1 + 6 R_+^2) (1 + 2 p) \nonumber \\
&&  \hspace{-0.8cm} +\sqrt{(1 + 6 R_+^2) 
\left[1+6 R_+^2+4m^2 R_+^2+4\ell(\ell+1)\right]-8q^2 R_+^2(1+3R_+^2)}\Bigg)\Bigg] +
\mathcal{O}(\sigma^2)
\end{eqnarray}
where $R_+=r_+/L$, $e = q L$ and $\sigma=\frac{r_+-r_-}{r_+}$ measures the distance away from extremality with $r_-(r_+,\mu,L)$ being the inner (Cauchy) horizon for which $f(r_-)=0$.
In Fig.~\ref{scalar-fig:freqVSq}, this analytical near-extremal frequency (with $\ell=1$) is described by the dashed black curve. We find that it matches quite well the numerical result for small scalar charge. This indicates that the unstable modes fit into the {\it near-extremal family} of modes. But they also fit into the {\it AdS family} of normal modes. That is to say, unlike in the de Sitter case,  in AdS the near-extremal and AdS family of modes coincide. To see this is indeed the case we pick two solutions in Fig.~\ref{scalar-fig:freqVSq} that have $qL=2.5$ (orange diamond) and $qL=2$ and (keeping $\mu=0.99\mu_{\rm ext}$ fixed) we follow this family of unstable modes as $r_+/L$ decreases to zero.\footnote{Note that $qL=2.5$ is well above the onset curves of Fig.~\ref{scalar-fig:onset} for any $r_+/L$ while the $qL=2$ line is above the onset curves only above a certain horizon radius. So, for the latter charge, the system is unstable only above a critical value of $r_+/L$, as shown in Fig.~\ref{scalar-fig:freqVSrp:q2}.} This is done in Fig.~\ref{scalar-fig:freqVSrp:q2.5} for $qL=2.5$ and Fig.~\ref{scalar-fig:freqVSrp:q2} for $qL=2$. In both cases we  find that, as $r_+/L\to 0$, ${\rm Im}(\omega L)\to 0$ and ${\rm Re}(\omega L)\to 3$, which is  indeed the normal mode frequency of AdS with $\ell=1$ (and lowest radial overtone).

In Fig.~\ref{scalar-fig:freqVSrp:q2.5} and Fig.~\ref{scalar-fig:freqVSrp:q2}  the magenta dashed lines departing from the normal mode of AdS describe the frequency that one obtains when we consider a perturbative expansion in $r_+/L$ and near-extremality about global AdS (and $\ell=1$, $m=0$). This result is taken from \cite{Dias:2016pma} (we already mentioned it to get the bound \eqref{scalar:qminSR}):  
\begin{eqnarray}\label{scalar:pertAdSFreq}
&&  \hspace{-0.2cm} \omega L\sim 3-\frac{r_+}{L}\,\frac{4 \left(6-4 e \mu +3 \mu ^2\right)}{3 \pi }+\frac{r_+^2}{L^2}\,\Bigg[ -\frac{4 \left(-4 e \mu +3 \mu ^2+6\right) \left(-52 e \mu +45 \mu ^2+90\right)}{27 \pi ^2}\nonumber \\
&&  \hspace{0.5cm}  +\frac{1}{96} \left[108+\mu  \Big(224 e^2 \mu -264 e \left(\mu ^2+2\right)+9 \mu  \left(3 \mu ^2+52\right)\Big)\right] -i\,\frac{16 (3-e \mu )}{3 \pi } \Bigg] 
+ \mathcal{O}\left(\frac{r_+^3}{L^3}\right)\nonumber \\
&&
\end{eqnarray}
where $e=q L$. So we see that not only the unstable modes approach the normal modes of AdS but they also do it at the expected rate in an expansion in $r_+/L$. The matching of our numerical results with the perturbative results \eqref{scalar:NearExtremalFreq} and \eqref{scalar:pertAdSFreq}  represents a non-trivial check of our results and illustrates the regime of validity of the perturbative results.

Now that we have highlighted the key features of the near-horizon (and superradiant) instabilities due to scalar perturbations in AdS-RN, we can proceed to the study of perturbations of Dirac fields in AdS-RN.

\section{Searching for an instability of Dirac fields in the AdS-RN background \label{sec:DiracRN}}

In section \ref{sec:Scalar} we have seen that scalar fluctuations in the AdS-RN background give rise to the near-horizon scalar condensation instability. Moreover, we have seen that this instability is closely associated to the violation of the AdS$_2$ scalar BF stability bound. So much that the associated stability bound \eqref{scalar:qminNH} for the onset of the instability is sharp. This naturally invites the questions: in the fermionic case can we also have a range of parameters where the AdS$_2$ fermionic stability bound is violated? If so what is the equivalent bound to \eqref{scalar:qminNH} for the onset of the instability?

In this section we will address these questions. We will find that a near-horizon analysis of the Dirac equation indeed indicates that the AdS$_2$ fermionic stability bound can be violated near-extremality if the charge of the fermion is above a critical value (subsection \ref{sec:DiracRN1}). Encouraged by this result we will do a numerical analysis that will search for unstable modes in the region of parameters of interest (subsection \ref{sec:DiracRN2}). However, we will find no trace of instabilities, unlike in the scalar field case. 

\subsection{Argument for a near-horizon instability of Dirac fields\label{sec:DiracRN1}}

The Dirac equation in the near-horizon geometry \eqref{NHgeometry} of the extreme AdS-RN black hole can be obtained taking the near-horizon limit of section \ref{sec:NH} directly on the Dirac equation \eqref{radial:2ndorder} for the extreme AdS-RN black hole. Concretely, applying the near-horizon coordinate transformation \eqref{NHtransf} together with the near-horizon frequency transformation $\tilde{\omega}\to \widehat{\omega} \,\varepsilon/L_{AdS_2}^2$ (so that $e^{-i\tilde{\omega} t}\to e^{-i\widehat{\omega} \tau}$) followed by the near-horizon limit $\varepsilon \to 0$ yields the Dirac equation in the near-horizon geometry \eqref{NHgeometry}:\footnote{The field $R_2$ obeys a similar near-horizon Dirac equation that is just the complex conjugate of \eqref{NH:Dirac}.}
\be\label{NH:Dirac}
 \rho  \,\frac{d}{d\rho} \left( \rho\, \frac{d R_1(\rho)}{d\rho}\right)+\left[ \frac{\left( \widehat{\omega}+ q\,\alpha \,\rho \right)^2}{\rho^2}+i\,\frac{\widehat{\omega}}{\rho}-L_{AdS_2}^2\left(m^2+\frac{\lambda^2}{r_+^2}\right)\right]R_1(\rho)=0
\ee
where the AdS$_2$ radius $L_{AdS_2}$ and the Maxwell near-horizon parameter $\alpha$ are defined in \eqref{NHgeometry} and $\lambda$ is the angular eigenvalue quantized as in \eqref{lambda:quantization}. Also, recall that $m$ and $q$ are the mass and charge of the fermionic field. 

\begin{figure}[bh]
\centerline{
\includegraphics[width=.4\textwidth]{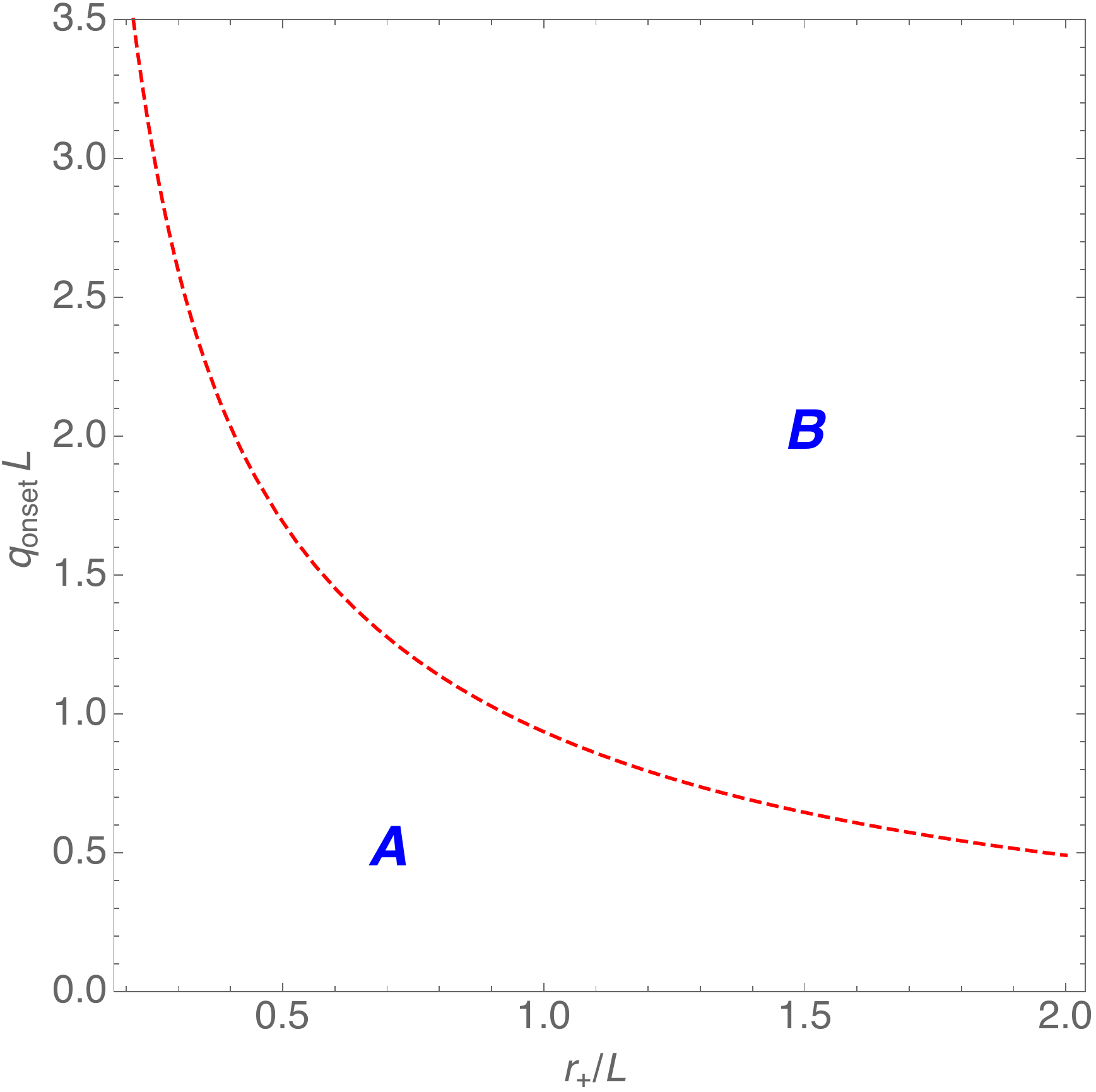}
\hspace{2cm}
\includegraphics[width=.4\textwidth]{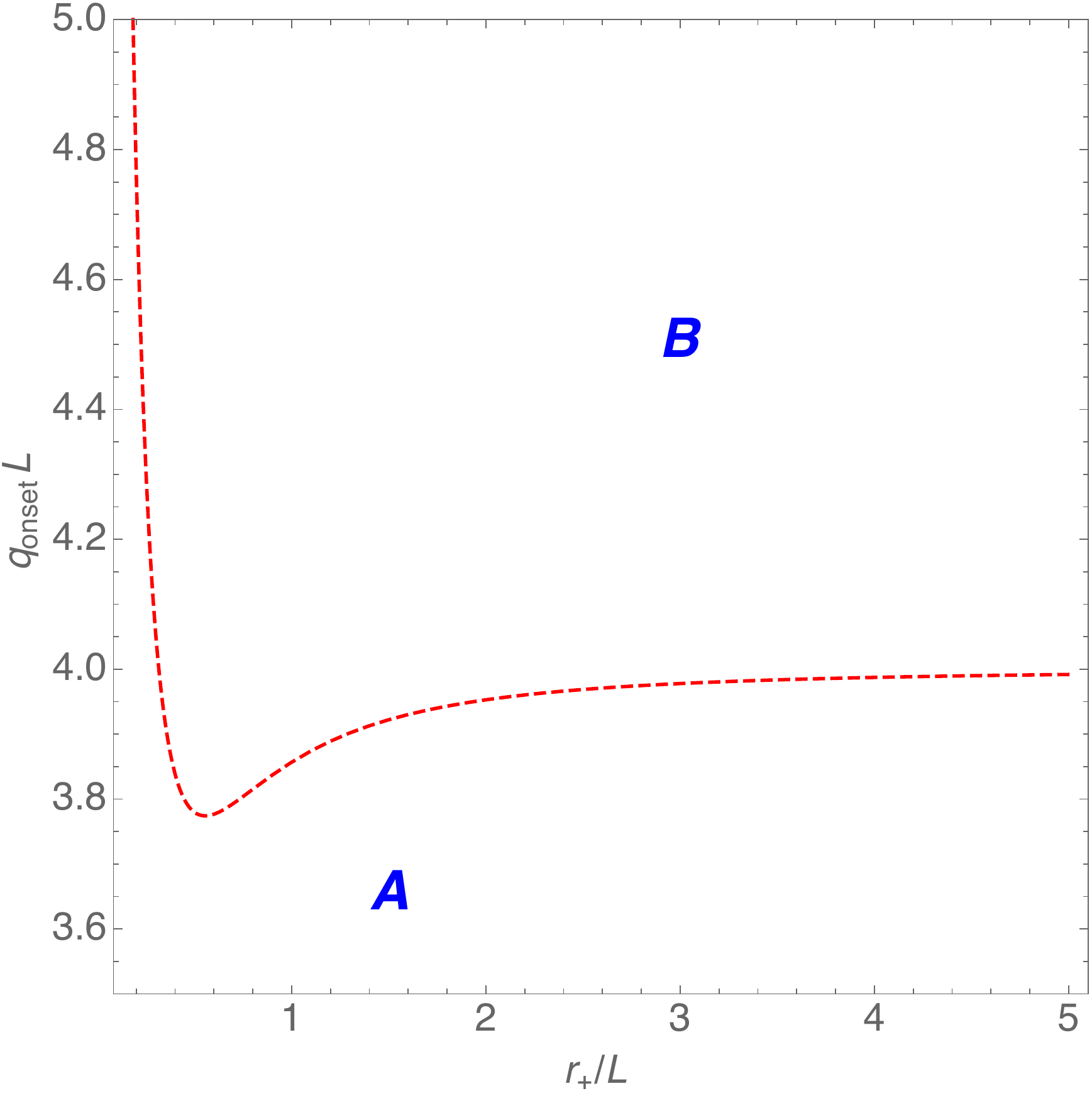}
}
\caption{{\it Predicted} Dirac field charge for the onset of an instability as a function of the horizon radius for an extremal AdS-RN black hole ($\mu=\mu_{\rm ext}$). In both plots the red dashed curve is the lower bound of \eqref{Dirac:qmin}. The plot in the {\it left} ({\it right}) panel is for fermion mass $mL=0$ ($mL=4$) and harmonic number $\ell=1/2$. The near-horizon analysis of the 2-dimensional stability bound violation leading to \eqref{Dirac:qmin} predicts that region $B$ should be unstable while region $A$ should be stable (at least with respect to the stability mass bound mechanism). Note that for small $r_+/L$ the system is not unstable because there is no superradiance for fermions and the predicted near-horizon instability is also suppressed. These Dirac figures can be (qualitatively) compared with Fig.~\ref{scalar-fig:onset} for the scalar field.}
\label{dirac-fig:onset}
\end{figure}

Asymptotically, as $\rho\to \infty$, a Frobenius analysis of \eqref{NH:Dirac} finds that the solution $R_1(\rho)$ decays as  
\begin{equation}\label{AsympDecaysR:NH}
\rho^{-\frac{1}{2}}R_1\big|_{\rho\to \infty}\sim \rho^{-\widehat{\Delta}_-}\left(\widehat{\alpha}_{1}+\cdots \right)+  \rho^{-\widehat{\Delta}_+}\left(\widehat{\beta}_{1}+\cdots \right), 
\end{equation}
where $\widehat{\alpha}_{1},\widehat{\beta}_{1}$ are two arbitrary constants and we have introduced the AdS$_2$ conformal dimensions 
\be\label{ConfDim:NH}
\widehat{\Delta}_{\pm}=\frac{1}{2}\pm m_{\rm eff} L_{AdS_2} \quad \hbox{with}\quad 
m_{\rm eff}=\sqrt{m^2+\frac{\lambda^2}{r_+^2}-\frac{q^2 \alpha^2}{L_{AdS_2}^2}}  \,.
\ee

The $s=1/2$ stability bound is {\it independent} of the spacetime dimension and still given by \eqref{BFfermionic}, $m^2\geq 0$ \cite{Amsel:2008iz,Andrade:2011dg}. Thus, the 2-dimensional fermionic stability bound is obeyed if $m_{\rm eff}^2\geq 0$ in \eqref{ConfDim:NH}. It follows that we can have situations where the Dirac field obeys the 4-dimensional fermionic stability bound \eqref{BFfermionic}, $m^2\geq 0$, but violates the 2-dimensional stability bound. When this happens, {\it i.e.} when $m_{\rm eff}^2<0$, one might expect an instability. This condition can be rewritten: the 2-dimensional stability bound is violated if the charge of the fermion is larger than
\begin{align}\label{Dirac:qmin}
q\geq \frac{1}{\sqrt{2} r_+}\sqrt{\frac{L^2+6 r_+^2}{L^2+3 r_+^2}}\left(m^2 r_+^2+\lambda ^2\right)\,.
\end{align}
The equality applies strictly to the extremal case; as we move away from extremality, by continuity the instability should still be present but a higher fermion charge is needed to trigger it.  Fig.~\ref{dirac-fig:onset} illustrates the regions where \eqref{Dirac:qmin} predicts instability/stability.

At this level, we see that the near-horizon analysis of the possible violation of the AdS$_2$ stability bound for a Dirac field parallels very much the partner analysis done for a scalar field in section \ref{sec:Scalar}, with the minimum value for the charge \eqref{scalar:qminNH} for the scalar case just replaced by the fermionic minimum value \eqref{Dirac:qmin}. In the scalar field case, we found (through a numerical study of linear perturbations in the AdS-RN background) that the violation of the 2-dimensional stability bound translates into the existence of a linear scalar condensation instability. Moreover, the near-horizon scalar bound \eqref{scalar:qminNH} turns out to be very sharp, as best illustrated in Fig.~\ref{scalar-fig:onset}. This scalar condensation linear instability indicates that non-linearly the AdS-RN black hole, when perturbed by a scalar field evolves towards a new configuration $-$ a hairy black hole  (with a scalar condensate floating above the horizon) $-$ with the same UV asymptotics (since the 4-dimensional stability bound is satisfied) but with a different near-horizon geometry where the 2-dimensional stability bound is no longer violated \cite{Basu:2010uz,dias2012hairyBHs,Arias:2016aig,Dias:2016pma}. 

These considerations motivate the study done in this manuscript for a Dirac field. In this case the AdS$_2$ stability bound can also be violated: at extremality this occurs for a fermion charge that saturates \eqref{Dirac:qmin}. From the lessons learned in the scalar field case one might well expect that the AdS-RN black hole, when perturbed by a Dirac field, is linearly unstable. To confirm whether this is the case, in the rest of this section we will solve numerically the Dirac equation in the AdS-RN background to hunt for a signature of the near-horizon linear instability. However, unlike the scalar field case, we will not find any evidence of a {\it linear} instability.  

\subsection{Dirac normal modes of global AdS \label{sec:DiracNormalModes}}

Before looking for potential instabilities (or frequency spectrum of damped oscillations) of Dirac modes in the global AdS-RN black hole it is convenient to first compute the normal mode spectrum of Dirac fields in global AdS. Indeed, some families of AdS-RN perturbations must reduce to these in the limit where the horizon shrinks to zero. 
Massive (section \ref{sec:DiracNormalModes2}) and massless (section \ref{sec:DiracNormalModes1}) Dirac fields require a distinct analysis.

\subsubsection{Massive normal modes \label{sec:DiracNormalModes2}}

For  massive fermions in global AdS, it is not easy to solve directly the Dirac equations to get the radial functions $R_{1,2}$. There is however an appropriate combination of  $R_{1,2}$ that yields equations of motion that are explicit hypergeometric equations. The linear combination for  $R_{1,2}$ that we use below is motivated by a similar analysis done to compute the massive normal modes of fermions for {\it de Sitter} in \cite{lopez2006Dirac_dS}.

For $m\neq 0$ and in global AdS, we introduce the new radial variable  $y=-i r/L$ and make the following field redefinitions 
\bea  \label{massive1}
&&R_{1}(y) = (1-y^2)^{-\frac{1}{4}} (1-y)^{\frac{1}{2}} \Big[f_{1}(y) - f_{2}(y) \Big], \nn \\
&& R_{2}(y) = (1-y^2)^{-\frac{1}{4}} (1+y)^{\frac{1}{2}} \Big[f_{1}(y) + f_{2}(y) \Big],
\eea
where $f_{1,2}(y)$ are functions to be determined. In these conditions, the coupled system of Dirac equations \eqref{radial:1storder} yields 
\begin{eqnarray}
&&\left(1-y^2\right) \left(f_{1}^{\prime} - f_{2}^{\prime} \right)   + \left(\omega L - \frac{1}{2} \right) \left(f_{1} - f_{2} \right)  + \left[m L(1+y) - \frac{1+y}{y} \left(\ell +\frac{1}{2} \right) \right] \left(f_{1}+ f_{2} \right)=0\,,  \label{eq1}  \nonumber\\
&&\left(1-y^2\right) \left(f_{1}^{\prime} + f_{2}^{\prime} \right) - \left(\omega L- \frac{1}{2} \right) (f_{1} + f_{2} ) - \left[m L (1-y) + \frac{1-y}{y} \left(\ell +\frac{1}{2} \right) \right] \left(f_{1}- f_{2} \right)=0\,.\label{eq2} \nonumber
\end{eqnarray}
Adding and subtracting these two ODEs yields
\begin{eqnarray}
&&\left(1-y^2\right) f_{1}^{\prime}(y)+ \left[m L \,y - \frac{1}{y}\left(\ell +\frac{1}{2}\right) \right] f_{1}(y) =\big( \omega L -m L+ \ell   \big) f_{2}(y)\,, \nn \\
&& \left(1-y^2\right) f_{2}^{\prime}(y)- \left[m L \,y - \frac{1}{y}\left(\ell +\frac{1}{2}\right) \right] f_{2}(y) = \big(\omega L +  m L  -\ell -1  \big) f_{1}(y)\,. 
\end{eqnarray}
This pair of coupled first order ODEs can be straightforwardly rewritten as a decoupled pair of second order ODEs for $f_1$ and $f_2$. Moreover, if we introduce the new radial coordinate $z=y^2$ and the field redefinitions 
\bea \label{massive2}
&&f_{1}(z) = z^{\frac{\ell + 1/2}{2}} \left(1-z\right)^{\frac{1}{4} (1-2\omega L)} F_{1}(z), \nn \\
&&f_{2}(z) = z^{\frac{\ell + 3/2}{2}} \left(1-z\right)^{\frac{1}{4} (1-2\omega L)}  F_{2}(z),
\eea
each of the ODEs becomes a  hypergeometric ODE with the standard form 
\be\label{hyperMassive}
z(1-z)  F_{i}^{\prime\prime}(z)+\big[ c_{i} - (a_{i} +b_i +1) z \big]  F_{i}^{\prime}(z) - a_{i} b_{i} \,F_i(z) = 0\,,\qquad \hbox{for $i=1,2$}
\ee
with parameters $a_{i}, b_{i}$ and $c_{i}$ given by
\bea
&& \hspace{-1.5cm} a_1 = \frac{1}{2} \left(1+\ell -\omega L - mL \right), \qquad b_1 = \frac{1}{2} \left(2+\ell -\omega L  +  mL  \right), \qquad  c_1 = 1 + \ell\,; \\
&& \hspace{-1.5cm}  a_2 = \frac{1}{2} \left(3+\ell -\omega L - mL  \right), \qquad b_2 = \frac{1}{2} \left(2+\ell -\omega L  +  mL\right),\qquad c_2 = 2 + \ell.
\eea
The most general solutions of \eqref{hyperMassive} are  \cite{1965handbook}
\bea \label{hyperMassiveSoln}
F_1(z)&=&A_1 \, _2F_1\left(\frac{1}{2} (\ell-m L -\omega L +1),\frac{1}{2} (\ell+m L -L
   \omega +2);\ell +1;z \right)  \nonumber\\
   && +A_2 z^{-\ell} \, _2F_1\left(\frac{1}{2} (1-m L-\ell -\omega L),\frac{1}{2} (m L-\ell -\omega L +2);1-\ell ;z\right),
    \nonumber\\
 F_2(z)&=&B_1 \, _2F_1\left(\frac{1}{2} (\ell-m L -\omega L +3),\frac{1}{2} (\ell+m L -\omega L +2);\ell
   +2;z\right) \nonumber\\
   && +B_2 z^{-\ell -1} \, _2F_1\left(\frac{1}{2} (1-m L-\ell -\omega L),\frac{1}{2} (m L-\ell
   -\omega L );-\ell ;z\right),
\eea
where $\,_2F_1(a,b,c;z)$ is the Gaussian (ordinary) hypergeometric function and $A_{1,2}$, 
$B_{1,2}$ are arbitrary amplitudes.
We can now plug \eqref{hyperMassiveSoln} into \eqref{massive2} and then into \eqref{massive1} to get the most general solution for $R_1(r)$ and $R_2(r)$. Finally,  we can insert this most general solution for $R_{1,2}(r)$ into \eqref{Radial-psi} to get the most general solution for the Dirac fields $\psi_\pm(r)$. These are the physical fields that have to be regular everywhere and this constrains some of the amplitudes $A_{1,2}$ and $B_{1,2}$ and the frequencies. Namely, at the origin, $r=0$, one finds that both $\psi_\pm$ have two divergent terms of the form $B_2/r^{\ell+3/2}$ and $(2A_2-B_2)/r^{\ell+1/2}$. Regularity at the origin thus requires that one sets $A_2=0$ and $B_2=0$ and the other two amplitudes $A_1$ and $B_1$ are left arbitrary.
It follows that the regular normal eigenmodes are  
\bea\label{MassiveNormalEigenmodes}
\hspace{-3cm} R_1&=&\sqrt{\frac{r}{L} \left(\frac{r}{L}-i\right)} \left(\frac{r^2}{L^2}+1\right)^{-\frac{\omega L }{2}} \left(-\frac{i r}{L}\right)^{\ell} \nonumber\\
&& \Bigg[A_1  \, _2F_1\left(\frac{1}{2} (\ell-m L -\omega L +1),\frac{1}{2} (\ell+m L -\omega L +2);\ell +1;-\frac{r^2}{L^2}\right) \nonumber\\
&& \quad+i B_1 \frac{r}{L} \, _2F_1\left(\frac{1}{2} (\ell-m L -\omega L +3),\frac{1}{2} (\ell+m L -\omega L +2);\ell
   +2;-\frac{r^2}{L^2}\right)
\Bigg],
   \nonumber\\
\hspace{-3cm} R_2&=&\sqrt{1-\frac{i r}{L}} \left(\frac{r^2}{L^2}+1\right)^{-\frac{\omega L }{2}} \left(-\frac{i
   r}{L}\right)^{\ell +\frac{1}{2}}\nonumber\\
&& \Bigg[ A_1  \, _2F_1\left(\frac{1}{2} (\ell-m L -\omega L +1),\frac{1}{2} (\ell+m L -\omega L +2);\ell
   +1;-\frac{r^2}{L^2}\right)\nonumber\\
&& \quad-i B_1 \frac{r}{L} \, _2F_1\left(\frac{1}{2} (\ell-m L -\omega L +3),\frac{1}{2} (\ell+m L -\omega L +2);\ell
   +2;-\frac{r^2}{L^2}\right)
\Bigg].
\eea  
We have not yet imposed the asymptotic boundary condition. A Frobenius analysis of \eqref{MassiveNormalEigenmodes} near the conformal boundary together with the use of \eqref{Radial-psi} finds that $\psi_\pm$ behaves as \eqref{AsympDecays:psi+-1} or \eqref{AsympDecays:psi+-2} with
\bea
&& \alpha_1= \frac{A_1 (-i)^{\ell } L^{-m L} \Gamma (\ell +1) \Gamma \left(m L+\frac{1}{2}\right)}{\Gamma
   \left[\frac{1}{2} (\ell+2+m L -\omega L)\right] \Gamma \left[\frac{1}{2} (\ell+1+m L +\omega L)\right]}
   \nonumber\\
&& \beta_1 = \frac{i B_1 (-i)^{\ell } L^{m L} \Gamma (\ell +2) \Gamma \left(\frac{1}{2}-m L \right)}{\Gamma \left[\frac{1}{2} (\ell+3-m L -\omega L)\right] \Gamma \left[\frac{1}{2} (\ell+2-m L +\omega L)\right]}
\eea 
For $m>0$ ($m\neq 1/2$) the no-source standard boundary condition \eqref{StandardQuant} requires $\alpha_1=0$. Using $\Gamma[-p]=\infty$ for $p=0,1,2,\cdots$ this  quantizes the frequency as
\bea\label{StandardMassiveNormalModes}
&& \omega L=\ell +2+m L+2 p \quad \hbox{or} \quad  \omega L= -(\ell +1+m L+2 p)\,, \qquad \hbox{(standard quantization)}\nonumber\\
&&
\eea
For $0<m L<\frac{1}{2}$ we can also impose the alternative quantization \eqref{AlternativeQuant}, {\it i.e.} $\beta_1=0$. This quantizes the frequency spectrum as (also with radial overtone $p=0,1,2,\cdots$)
\bea\label{AlternativeMassiveNormalModes}
&& \omega L=\ell +3-m L+2 p  \quad \hbox{or} \quad \omega L= -(\ell +2-m L+2 p)\,, \qquad \hbox{(alternative quantization)}\nonumber\\
&&
\eea

\subsubsection{Massless normal modes \label{sec:DiracNormalModes1}}

In this section we find the normal modes in global AdS for a massless fermionic field. These have been previously discussed in \cite{Cotaescu:1998ts,Wang:2017fie} but these references have not identified the full spectra of frequencies.  

For $m=0$ in global AdS, introducing the change of coordinates and field redefinition
\bea\label{massless1}
&& z = \frac{2 r}{r+i L }\,,\qquad 0 \leq z \leq 2; \nonumber\\
&& R_{1}(z) = z^{\ell+\frac{1}{2} } (1 - z)^{\frac{1}{2} \omega L } F(z),
\eea
the radial equation \eqref{radial:2ndorder} can be rewritten as a hypergeometric ODE in the standard form $z(1 - z) F''(z) + (c - (a+b+1)z)F'(z) - a b F(z) = 0$, with 
\be
a = \ell+\frac{1}{2} \,, \qquad b = \ell+1 + \omega L \,, \qquad c = 2 \left( \ell+1 \right).
\ee
Its most general solution is \cite{1965handbook}
\bea\label{farSol}
&&F(z) = C_1\, _2F_1\Big(\frac{1}{2} + \ell, \omega L+\ell+ 1  ,2(1+ \ell),z\Big) + C_2 \, z^{-1-2 \ell} \, _2F_1\Big(-\frac{1}{2} - \ell, \omega L- \ell ,-2 \ell, z\Big).\nn\\
&&
\eea
Introducing this into \eqref{massless1} one gets $R_{1,2}(r)$  (note that $R_2=R_1^*$ as discussed in the next section). Plugging this  into \eqref{Radial-psi} one finds the most general solution for the Dirac fields $\psi_\pm(r)$. At the origin, $r=0$, these $\psi_\pm$ have a divergent term proportional to $C_2 r^{-\ell-3/2}$. Regularity at the origin thus requires that we set $C_2=0$.
Now we need to impose the asymptotic boundary condition. One finds that asymptotically $\psi_{\pm}$ decays as \eqref{AsympDecays:psi+-m0} with
\bea
 \alpha_1&=& C_1 \,i^{\omega L} \,2^{\ell +\frac{1}{2}} \, _2F_1\left(\ell +\frac{1}{2},\ell +\omega L +1;2 \ell  +2;2\right), \nonumber\\
\beta_1&=& -C_1\, 2^{\ell -\frac{1}{2}}\, i^{\omega L +1} \Bigg[(2 \omega L +2 \ell +1) \, _2F_1\left(\ell
   +\frac{1}{2},\ell +\omega L +1;2 \ell +2;2\right)\nonumber\\
&&   +\frac{(2 \ell +1) (\omega L +\ell +1)}{\ell +1}\, _2F_1\left(\ell
   +\frac{3}{2},\ell +\omega L +2;2 \ell +3;2\right)\Bigg].
\eea
As explained previously, for $m=0$ we can impose either the standard or alternative boundary conditions.
The no-source standard boundary condition \eqref{StandardQuant}, $\alpha_1 (\lambda +\omega L)-i \beta_1=0$, quantizes  the frequency spectrum as 
\be \label{normal_mode_StandardBCp}
\omega  L= \ell+2+2 p   \quad \hbox{and} \quad  \omega L =- \left(\ell+ 1+2 p \right) , \quad p = 0,1,2,\dots, \quad\hbox{(Standard quantization)}
\ee 
On the other hand, for the no-source alternative quantization \eqref{AlternativeQuant}, $\alpha_1 (\lambda -\omega L)+i \beta_1=0$, the normal mode frequencies of a massless Dirac field in global AdS are:
\be \label{normal_mode_AlternativeBCm}
\omega L = \ell+ 1+2 p \quad \hbox{and} \quad \omega L= -  \left(\ell+2+2 p \right), \quad p = 0,1,2,\dots, \quad\hbox{(Alternative quantization)}
\ee 
The positive frequencies in \eqref{normal_mode_StandardBCp} and \eqref{normal_mode_AlternativeBCm} were computed in \cite{Wang:2017fie} using vanishing flux boundary conditions that, as explained in the end of section \ref{sec:BCsDiracAdS-RN}, are exactly the AdS/CFT standard and alternative boundary conditions. However, \cite{Wang:2017fie} missed the existence of half of the normal mode spectrum, namely the half part that has negative frequencies. The relevance of the full spectrum (and associated relations between standard/alternative quantizations) is further analysed in the discussion of Fig. \ref{fig-D:Schw}. Further note that in RN, the four families of modes that reduce to \eqref{normal_mode_StandardBCp}-\eqref{normal_mode_AlternativeBCm} in the AdS limit become completely independent (i.e. they are not related by complex conjugation and the ``degeneracy" is broken). This is further discussed in the next subsection.

\subsection{Setup of the numerical problem\label{sec:DiracRN2}}

In this section we solve numerically the Dirac equation and search for 
linear instabilities of the Dirac solution in the AdS-RN background.
Before proceeding it is important to note that: 1) the Dirac radial equation \eqref{radial:2ndorder}  for $R_2(r)$ is just the complex conjugate of the radial equation for $R_1(r)$ so if $R_1(r)$ is a solution one automatically has $R_2(r)=R_1(r)^*$, and 2) the Dirac angular equations for $S_{1,2}$ are related by the symmetry $\theta\to\pi-\theta$ so if $S_1(\theta)$ is a solution then $S_2(\theta)=S_1(\pi-\theta)$. Therefore, we just need to find the solutions $R_1(r)$ ($S_1(\theta)$ are just the spin-weighted $s=1/2$ spherical harmonics with quantum number $\ell$). 

Further note that if $R_1$ has charge $q$ then $R_2=R_1^*$ has charge $-q$, and complex conjugation maps quasinormal modes to quasinormal modes. It follows that if $\omega=\omega_r + i \omega_i$ is a linear mode frequency of $R_1$ then $-\omega^* = -\omega_r + i \omega_i$ is a linear mode frequency of $R_2=R_1^*$. Thus, if we compute the frequency spectrum of $R_1$, we have the spectrum of $R_2$ too.
It also follows that there is no loss of generality in assuming that $qQ>0$ in our analysis: results for $qQ<0$ are obtained simply by reversing the sign of the real part of the frequencies. Finally note that when we compute $\omega$ we have to allow both positive and negative values of $\omega_r$, {\it i.e.} if $\omega_r={\rm Re}(\omega)$ is a frequency of $R_1$ there is no symmetry in the system that requires $-\omega_r$ to be also a frequency of $R_1$. The only exception is if  $\mu$=0 (or $e=0$) and $m=0$ {\it i.e.} a massless Dirac field in Schwarzschild-AdS. In this case if $\omega=\omega_r + i \omega_i$ is an eigenvalue of $R_1$ so is $-\omega^* = -\omega_r + i \omega_i$, although with the opposite quantization: see \eqref{radial:2ndorder} and \eqref{StandardQuant}-\eqref{AlternativeQuant} or \eqref{normal_mode_StandardBCp}-\eqref{normal_mode_AlternativeBCm}.

For concreteness, we will set the mass of the fermion to zero, {\it i.e.} we solve the Dirac equation \eqref{radial:2ndorder} for $R_1$ with $m=0$ (and in the gauge $A |_{\infty}=0$ where the frequency of the fermionic wave is $\omega$; see footnote \ref{foot:gauge}) subject to the physically relevant boundary conditions. The asymptotic decay of $R_1(r)$ is given in \eqref{AsympDecaysRm0}. For reasons discussed previously, we impose the asymptotic boundary condition \eqref{StandardQuant} (standard quantization, $\psi_+^{(0)}=0$) or \eqref{AlternativeQuant} (alternative quantization, $\psi_-^{(0)}=0$). 
At the horizon, for a non-extreme black hole, a Frobenius analysis finds that the two pairs of independent solutions are
\begin{equation}\label{HorizonSolns}
R_1\big|_{r=r_+}=A_{in}\left(r-r_+\right)^{\frac{1}{2}-i\,\frac{\omega-q \mu}{4\pi T_H}}\Big(1+{\cal O}(r-r_+) \Big)+B_{out}\left(r-r_+\right)^{i\,\frac{\omega-q \mu}{4\pi T_H}}\Big(1+{\cal O}(r-r_+) \Big).
\end{equation}
Rewriting this  in ingoing Eddington-Finkelstein coordinates $(v,r,\theta,\phi)$, with $v=t+\int f^{-1}\mathrm dr$, which are smooth across the future event horizon ${\cal H}^+$, we find that regularity of $R_{1}(r)$ at ${\cal H}^+$ requires that we impose the boundary condition $B_{out}=0$.
\footnote{For scalar fields, \cite{BVR2009real} used the real-time holography formalism \cite{skenderisBVR2009realholog} to show that imposing  ingoing boundary conditions in the bulk horizon translates on the CFT side of the AdS/CFT correspondence to study {\it retarded} two-point functions.}

For the numerical solution it is convenient to redefine
\begin{equation}\label{redefR}
R_1(r)=\left(1-\frac{r_+}{r}\right)^{\frac{1}{2}-i\,\frac{\omega-q \mu}{4\pi T_H}}q(r)\,,
\end{equation}
and to work with the compact radial coordinate
\be \label{def:z}
z=\sqrt{1-\frac{r_+}{r}}
\ee
such that the horizon is now at $z=0$ and the asymptotic infinity at $z=1$. This has the advantage that analytical solutions that are smooth at the horizon simply have to obey the horizon Neumann boundary condition $q'(z=0)=0$. The asymptotic boundary condition for $q(z)$ follows straightforwardly from \eqref{StandardQuant}-\eqref{AlternativeQuant} and \eqref{redefR}-\eqref{def:z}.

The numerical methods that we use are very well tested \cite{Dias:2009iu,Dias:2010maa,Dias:2010eu,Dias:2010gk,Dias:2011jg,Dias:2014eua,Dias:2010ma,Dias:2011tj,Dias:2013sdc,Cardoso:2013pza,Dias:2015wqa} and reviewed in  \cite{Dias:2015nua}.
To discretize the field equations we use a pseudospectral collocation grid on Gauss-Chebyshev-Lobbato points. The eigenfrequencies and associated eigenvectors are found using {\it Mathematica}'s built-in routine {\it Eigensystem}. For a given $\ell$, this method has the advantage of finding several modes (i.e., from distinct families and with distinct radial overtones) simultaneously. However, to increase the accuracy of our results at a much lower computational cost we use a powerful numerical procedure which uses the Newton-Raphson root-finding algorithm discussed in detail in section III.C of the review \cite{Dias:2015nua}. All our results have the exponential convergence on the number of gridpoints, as expected for a code that uses pseudospectral collocation. In particular, all the results that we present are accurate at least up to the 10th decimal digit.

The Dirac equation in AdS-RN also has the scaling symmetry that determines that the physical  dimensionless quantities are those listed in \eqref{dimensionless}.

\subsection{Main results}\label{sec:DiracRN3}

As discussed in section \ref{sec:DiracRN2}, for fermion mass $m=0$, we can have two independent homogeneous boundary conditions that yield normalizable modes: the standard ($\psi_+^{(0)}=0$) and alternative ($\psi_-^{(0)}=0$) boundary conditions. Moreover, for each of these boundary conditions, the eigenvector $R_1$ can have negative or positive real part of the frequency. It follows that, for a given harmonic ${\ell,m_\phi}$ (and $m=0$) we  have  a total of two frequency spectra to discuss for each one of the two possible boundary conditions.

Before proceeding to the actual physical analysis of the frequency spectrum and instabilities of the system, in Appendix~\ref{sec:DiracRN3a} we first test our numerical code by comparing the associated numerical results with some analytical perturbative expansions that are derived in Appendix \ref{secA:Perturbative_results}.   
This confirms that our numerical code is generating physical data and we can now proceed and discuss our main physical findings.

Our aim is {\it not} to present the full spectrum of frequencies of a Dirac field in AdS-RN black hole. Instead, we are motivated to search for unstable modes, {\it i.e.} on eventually finding modes that, in some range of parameters, have ${\rm Im}(\omega L)>0$. There is a wide window of parameters to explore although the instability, if it exists, should appear near extremality  for fermion charges $q$ above a critical value. Thus one needs a good strategy to hunt efficiently for unstable modes. We proceed as follows. From the near-horizon bound \eqref{Dirac:qmin} arguing for the existence of an instability, we see that this bound is lower if we set $m=0$ and $\ell=1/2$. So, first, we either: 1) fixed $m=0,\ell=1/2$ and $\mu$ close to $\mu_{\rm ext}$, and varied $\{r_+/L,qL\}$, or 2) fixed $m=0,\ell=1/2$ and $qL$ and varied  $\{r_+/L, \mu\}$. In both cases, as described in the end of section \ref{sec:DiracRN2}, we solved our system as an eigenvalue problem for $\omega L$. This finds ``all" the solutions of the system (as long as the hierarchies do not grow large, {\it e.g.} $|\omega L|\gg 1$, which makes the numerical problem hard). This allows to eventually identify unstable modes with ${\rm Im}(\omega L)>0$ or, in the worst case, to identify modes with ${\rm Im}(\omega L)<0$ that are closest to the marginal case for instability (${\rm Im}\,\omega =0$). Once these interesting modes are identified we then used a Newton-Raphson root-finding algorithm to follow efficiently the modes to other values of the parameter space.

In spite of our efforts, we have found no sign of an instability. Recall that for $mL=0$ both the standard  \eqref{StandardQuant} and the alternative \eqref{AlternativeQuant} boundary conditions yield normalizable modes. In general, we do find that the stable modes with smallest  $|{\rm Im}(\omega L)|$ are those that reduce to the alternative normal modes of AdS \eqref{normal_mode_AlternativeBCm}  or to the standard AdS normal modes \eqref{normal_mode_StandardBCp}, when the horizon radius shrinks to zero. Among these, we further find that the modes with smallest $|{\rm Im}(\omega L)|$ are, for both quantizations, the ones that reduce to the {\it positive} normal mode frequencies when $r_+/L\to 0$, {\it i.e.} $\omega L=3/2$ (alternative boundary condition) and $\omega L=5/2$ (standard quantization). Therefore, to avoid distraction from the main point, in the rest of this manuscript we only discuss these two families of modes.

\begin{figure}[th]
\centerline{
\includegraphics[width=.5\textwidth]{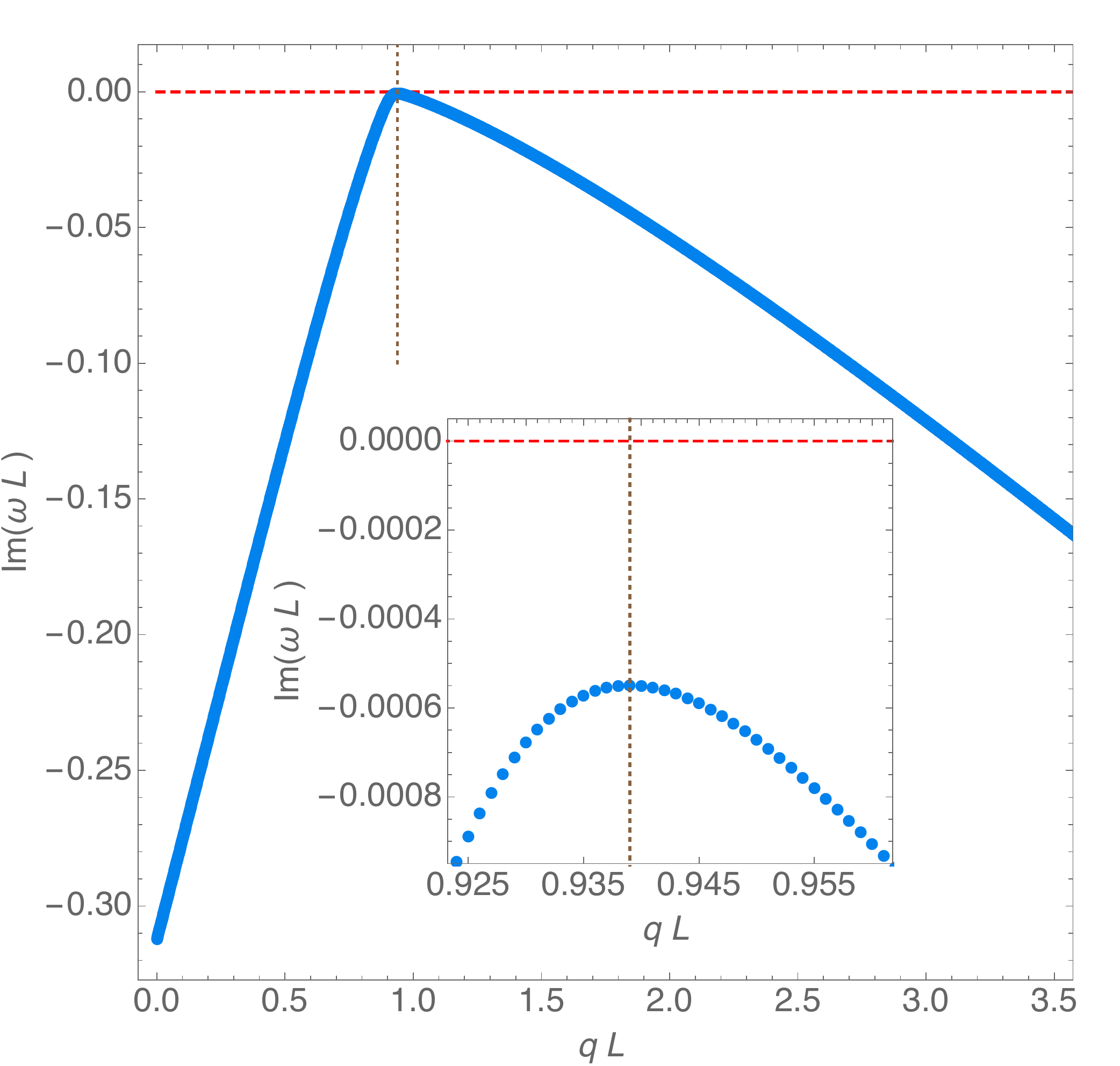}
\hspace{0.3cm}
\includegraphics[width=.46\textwidth]{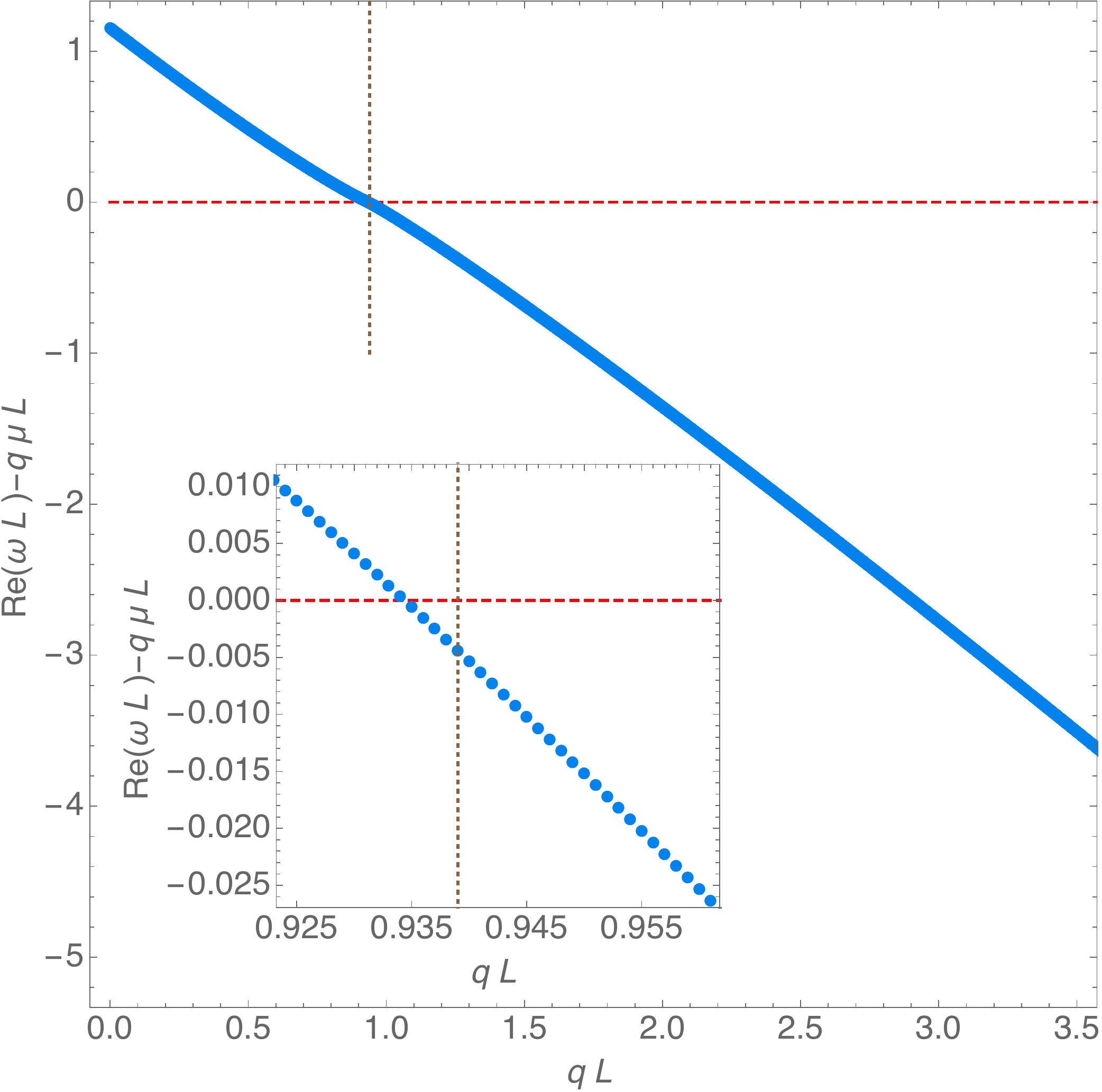}
}
\caption{Dirac field frequency with alternative quantization \eqref{AlternativeQuant} as a function of the dimensionless scalar charge $q L$ for a AdS-RN black hole with $\mu = 0.99 \mu_{\rm ext}$ and $r_{+}/L = 0.5$ (also, $m L=0$ and $\ell=1/2$). {\it Left panel}: Imaginary part of the dimensionless frequency, ${\rm Im}(\omega L)$ which attains a maximum for $q =q_{\rm max} \sim 0.9390/L$ (vertical brown dashed line). The inset plot zooms-in around this maximum and shows that  ${\rm Im}(\omega L)<0$ for any $q L$. {\it Right panel}: Real part of the dimensionless frequency, ${\rm Re}(\omega L)$, measured with respect to $q \mu L$. This quantity changes sign at $q=q_\star\sim 0.9344/L$ with $q_\star<q_{\rm max}$, {\it i.e.} for a smaller $qL$ than the one where ${\rm Im}(\omega L)$ attains its maximum value  (vertical brown dashed line): this is better seen in the inset plot which zooms-in the relevant region.}
\label{Dirac-fig:AlternativeFreqVSq}
\end{figure}

\begin{figure}[b]
\centerline{
\includegraphics[width=.49\textwidth]{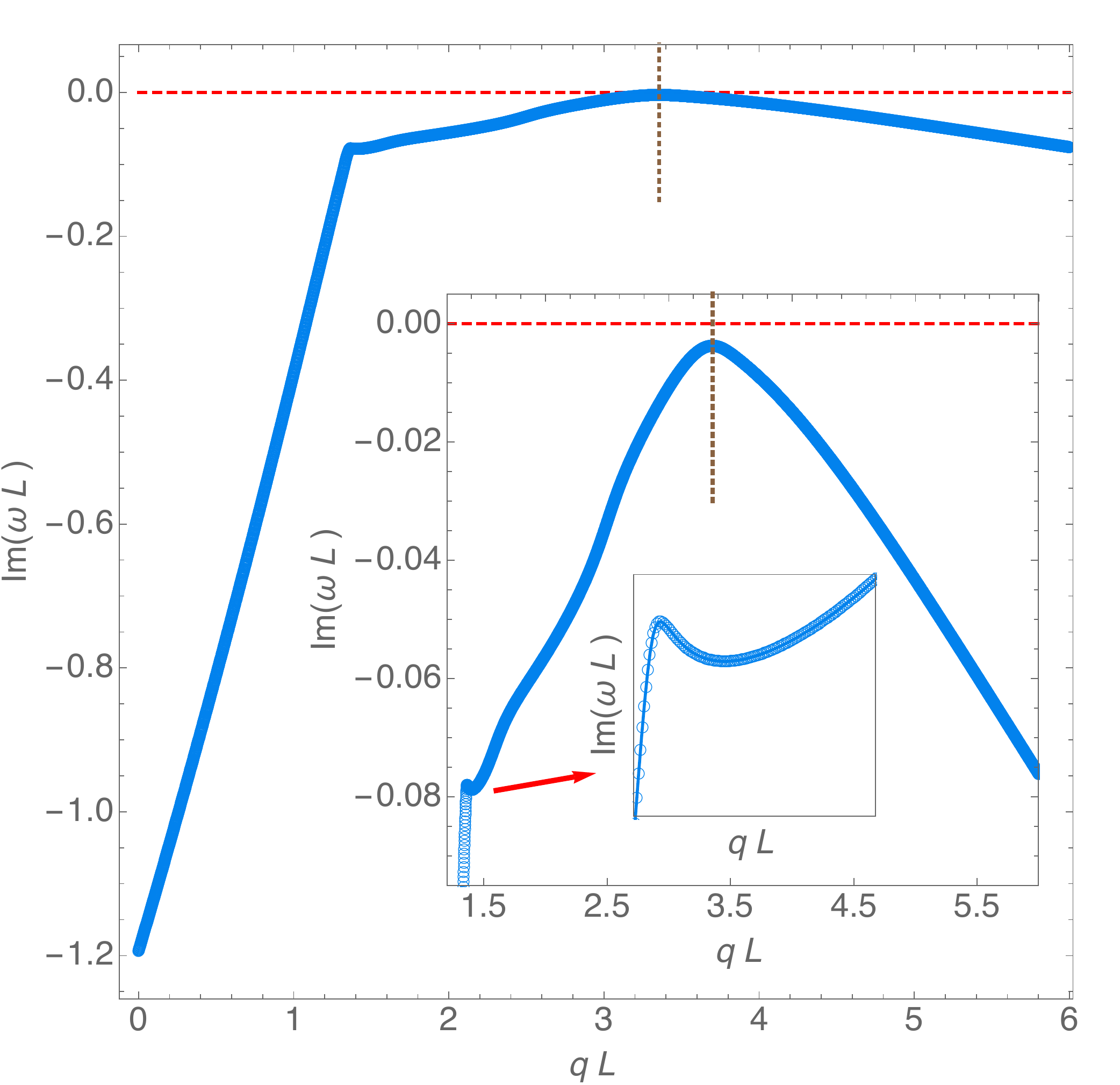}
\hspace{0.3cm}
\includegraphics[width=.47\textwidth]{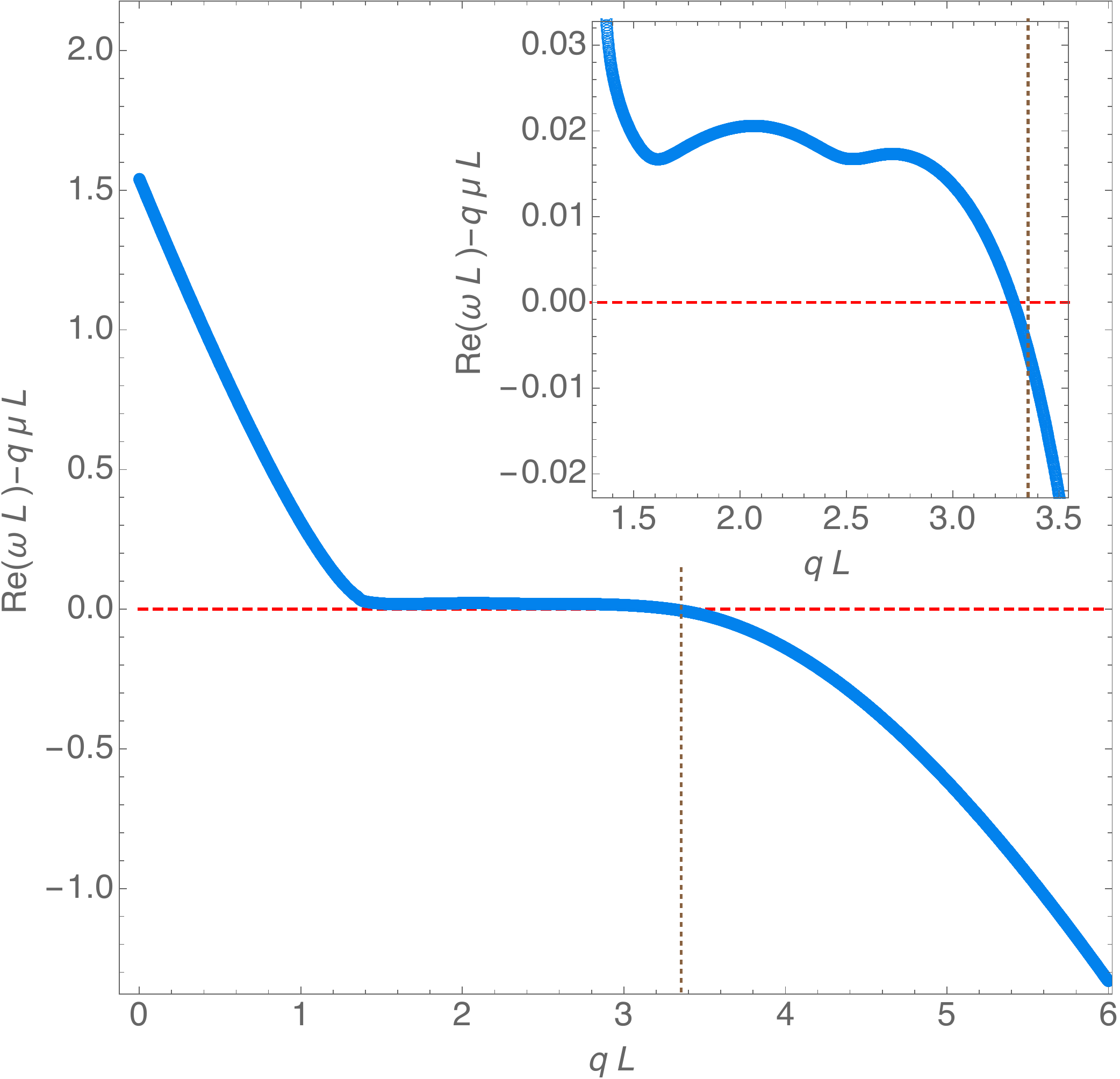}
}
\caption{Dirac field frequency with standard quantization \eqref{StandardQuant} as a function of the dimensionless scalar charge $q L$ for a AdS-RN black hole with $\mu = 0.99 \mu_{\rm ext}$ and $r_{+}/L = 0.5$ (also, $m L=0$ and $\ell=1/2$). {\it Left panel}: Imaginary part of the dimensionless frequency, ${\rm Im}(\omega L)$ which attains a maximum of ${\rm Im}(\omega L)\sim -0.0037491$ for $q =q_{\rm max} \sim 3.3555/L$ (vertical brown dashed line). The main inset plot zooms-in around this maximum and shows that  ${\rm Im}(\omega L)<0$ for any $q L$. The secondary inset plot shows the detail of the curve around $q L\sim 1.45$ to show that the apparent cusp  in the main plot is smooth. {\it Right panel}: Real part of the dimensionless frequency, ${\rm Re}(\omega L)$, measured with respect to $q \mu L$. This quantity changes sign at $q=q_\star\sim 3.2873/L$ with $q_\star<q_{\rm max}$, {\it i.e.} for a smaller $qL$ than the one where ${\rm Im}(\omega L)$ attains its maximum value  (vertical brown dashed line): this is better seen in the inset plot which zooms-in the relevant region.}
\label{Dirac-fig:StandardFreqVSq}
\end{figure}

Probably the plots that best illustrate the main conclusions of our Dirac study are those of Fig.~\ref{Dirac-fig:AlternativeFreqVSq} (for alternative quantization) and of Fig.~\ref{Dirac-fig:StandardFreqVSq}  (for standard quantization). Recall that in the best case scenario the expectation is that, close to extremality, modes should become unstable above a fermion charge $q$ that should be higher than the near-horizon bound \eqref{Dirac:qmin}. Thus, in these figures we fix the black hole horizon to be $r_+/L=0.5$ and choose a chemical potential close to extremality, $\mu = 0.99 \mu_{\rm ext}$. Starting from $qL=0$, where ${\rm Im}(\omega L)<0$,  we then increase this charge to see if there is a critical value above which ${\rm Im}(\omega L)$ becomes positive. (That is to say, we adopt a similar strategy as the one followed in the scalar field case to get Fig.~\ref{scalar-fig:freqVSq}).

For the alternative quantization, the left panel of Fig.~\ref{Dirac-fig:AlternativeFreqVSq} shows that, starting from $q=0$, as $qL$ grows, ${\rm Im}(\omega L)<0$ increases and approaches ${\rm Im}(\omega L)=0$ very closely. However, no matter how large $qL$ is we never reach a situation where ${\rm Im}(\omega L)\geq 0$. Interestingly, there is a critical value of $q$, namely  $q L=q_{\rm max} L\sim 0.9390$ (vertical brown dashed line) where ${\rm Im}(\omega L)$ reaches a maximum value of ${\rm Im}(\omega L)\sim -0.000548$ (see the inset plot which zooms-in around this maximum). But increasing $q L$ further, ${\rm Im}(\omega L)$ becomes again increasingly more negative (instead of becoming positive). The Dirac field system behaves therefore substantially distinctly from the scalar field case of Fig.~\ref{scalar-fig:freqVSq} (left panel) where there was a critical $q L$ above which ${\rm Im}(\omega L)$ becomes positive. To complete the analysis, in the right panel of Fig.~\ref{Dirac-fig:AlternativeFreqVSq}, we plot  ${\rm Re}(\omega L)-q \mu L$. We find that for small $qL$ this quantity is positive but becomes negative above $q=q_\star\sim 0.9344/L$. Interestingly, this occurs at a charge that is smaller than $q_{\rm max}$ where the maximum of ${\rm Im}(\omega L)$ is reached (vertical brown dashed line): this is better seen in the inset plot which zooms-in the relevant region. Again we note the difference to the scalar field case displayed in the right panel of Fig.~\ref{scalar-fig:freqVSq} where  ${\rm Re}(\omega L)-q \mu L$ changes sign precisely at the critical value of $q L$ where ${\rm Im}(\omega L)=0$. Further note that these plots also show that for a Dirac field we do not have a value of $q L$ for which we simultaneously have  ${\rm Re}(\omega L)-q \mu L=0$ and ${\rm Im}(\omega L)=0$. Therefore, we cannot set $\omega L=q \mu L$ in the equations of motion and solve these as an eigenvalue problem for the instability onset charge. That is to say, unlike the scalar field case, we do not have an onset charge that would produce the partner plots of the scalar field onset plots of Fig.~\ref{scalar-fig:onset}. The predictions of Fig.~\ref{dirac-fig:onset} do not hold (at least at the linear mode level). 

We have done similar experiments as those of Fig.~\ref{Dirac-fig:AlternativeFreqVSq}  for other black hole parameter values $\mu$ and $r_+/L$. 
Keeping $\mu$ fixed, black holes with distinct $r_+/L$ have plots similar to Fig.~\ref{Dirac-fig:AlternativeFreqVSq} with the feature that larger values of $r_+/L$ reach the maximum of ${\rm Im}(\omega L)$ (but remaining negative)  at smaller critical values of $q=q_{\rm max}$. On the other hand, keeping $r_+/L$ fixed, black holes with distinct $\mu $ also have similar plots to Fig.~\ref{Dirac-fig:AlternativeFreqVSq} with the property that larger values of $\mu$ reach the maximum of ${\rm Im}(\omega L)$ (but still negative)  at smaller critical values of $q=q_{\rm max}$ and this maximum of ${\rm Im}(\omega L)$ is increasingly closer to zero as $\mu$ approaches the extremal value $\mu_{\rm ext}$. 
\begin{figure}[th]
\centerline{
\includegraphics[width=.46\textwidth]{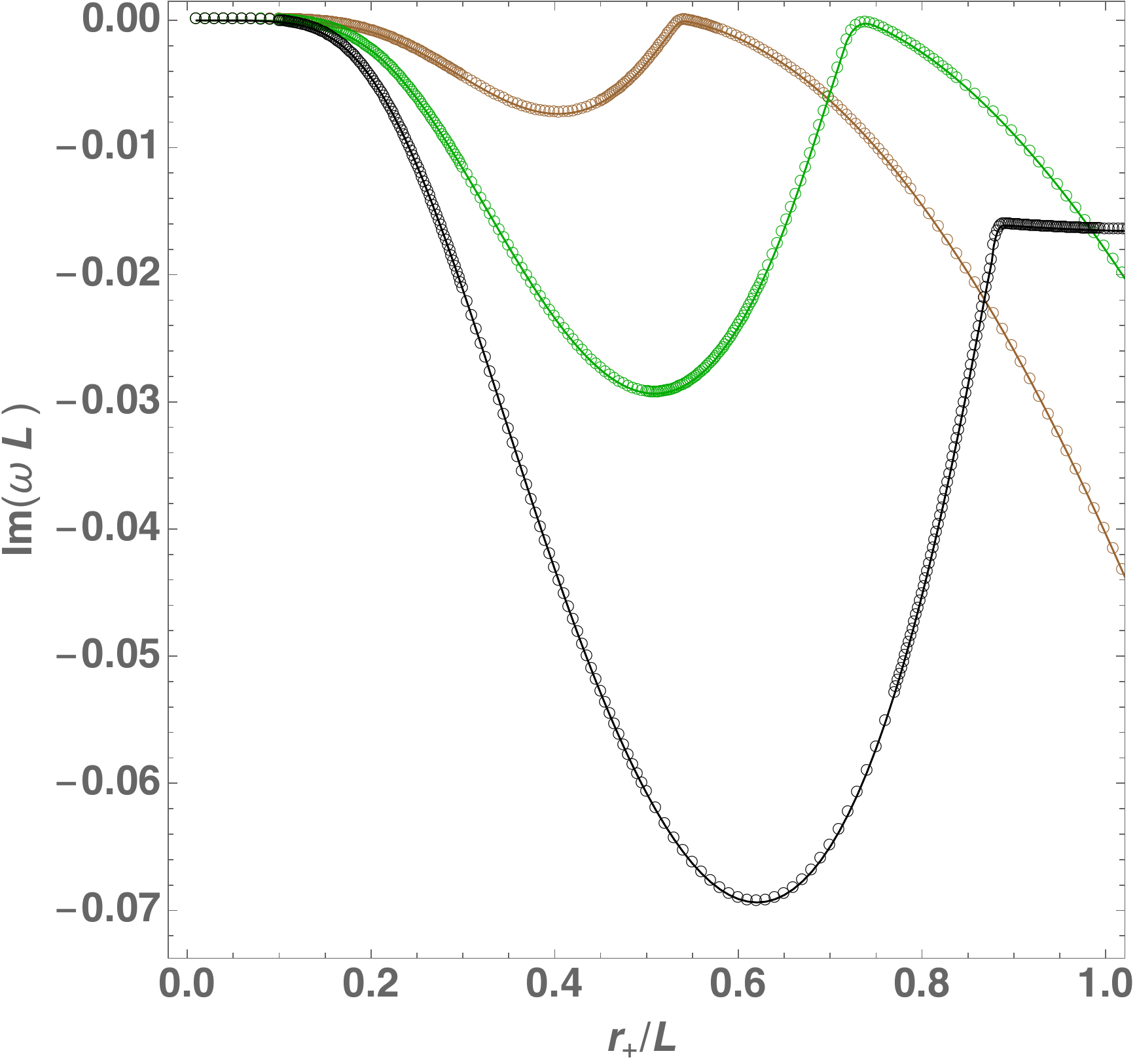}
\hspace{0.3cm}
\includegraphics[width=.49\textwidth]{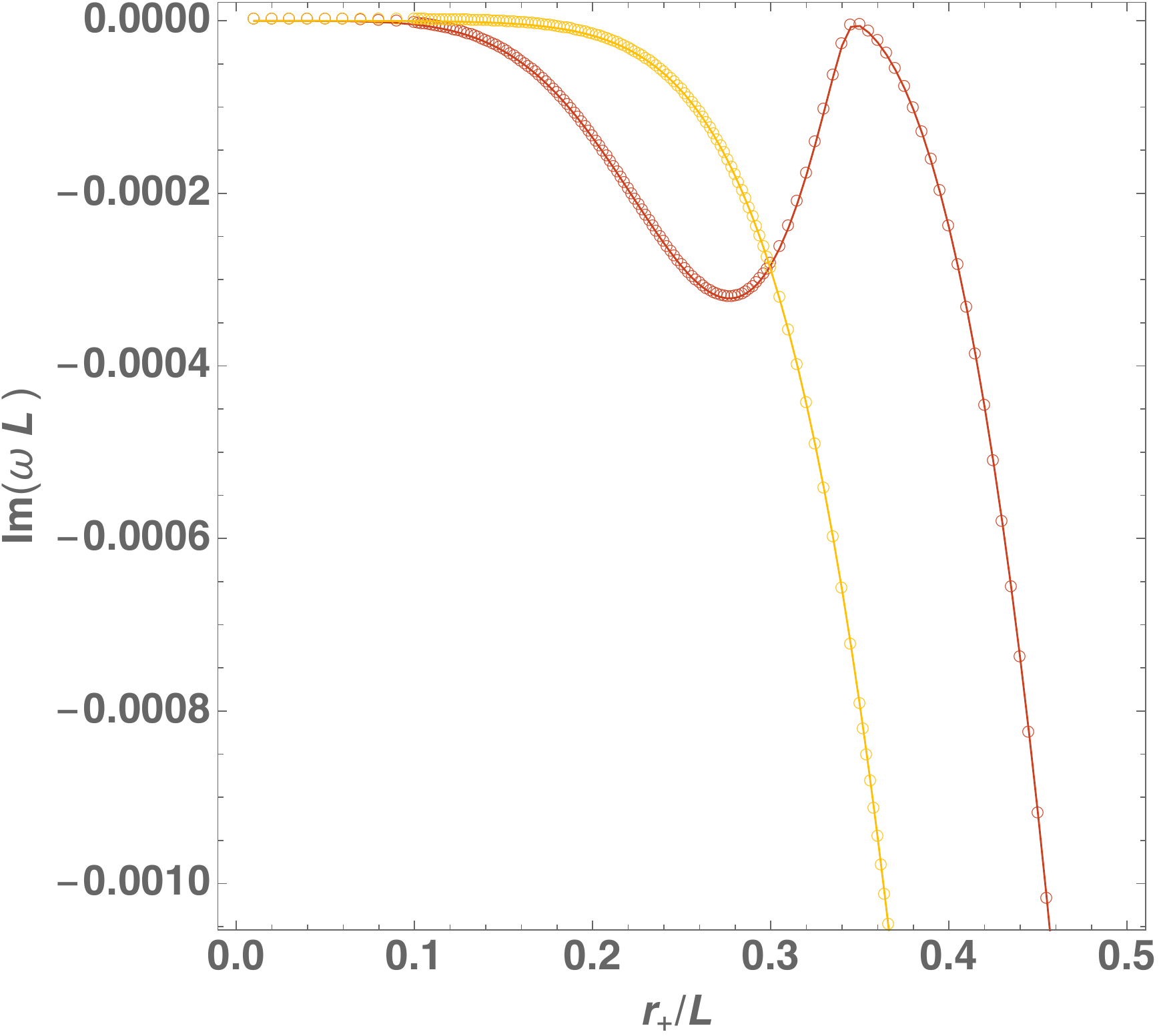}
}
\caption{Imaginary part of the frequency as a function of the horizon radius for chemical potential $\mu=0.999 \mu_{\rm ext}$ for different values of the fermion field charge. In the left panel the charges are (from bottom bump to top) $q L=0.7$ (black), $q L=0.8$ (green), $q L=0.9$ (brown). In the right panel the charges are $q L=1$ (red) and $q L=1.1$ (yellow). Notice the different regions scanned by the axes in the two plots.}
\label{Dirac-fig:AlternativeFixCharge}
\end{figure} 
\begin{figure}[ht]
\centerline{
\includegraphics[width=.47\textwidth]{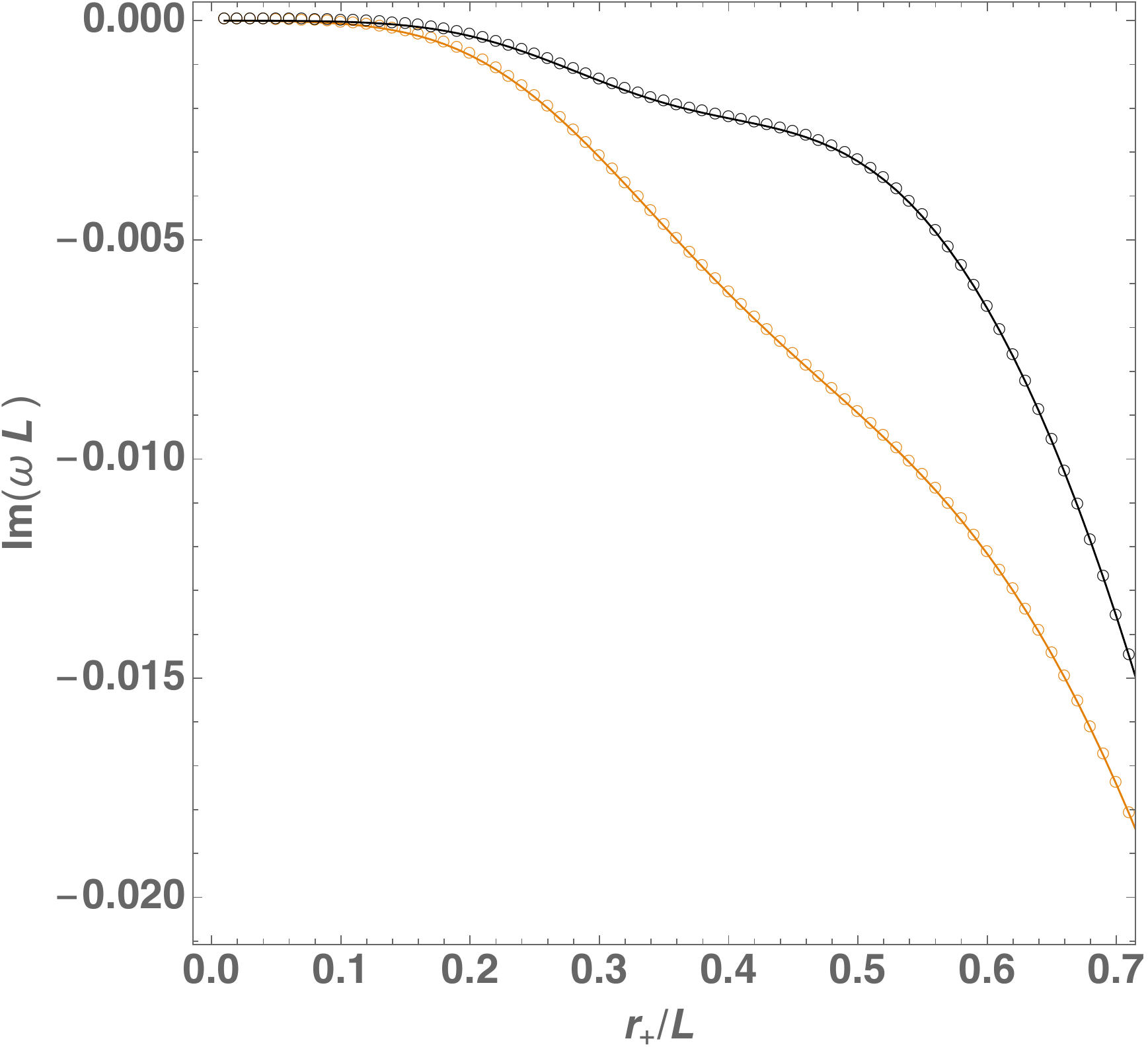}
\hspace{0.3cm}
\includegraphics[width=.48\textwidth]{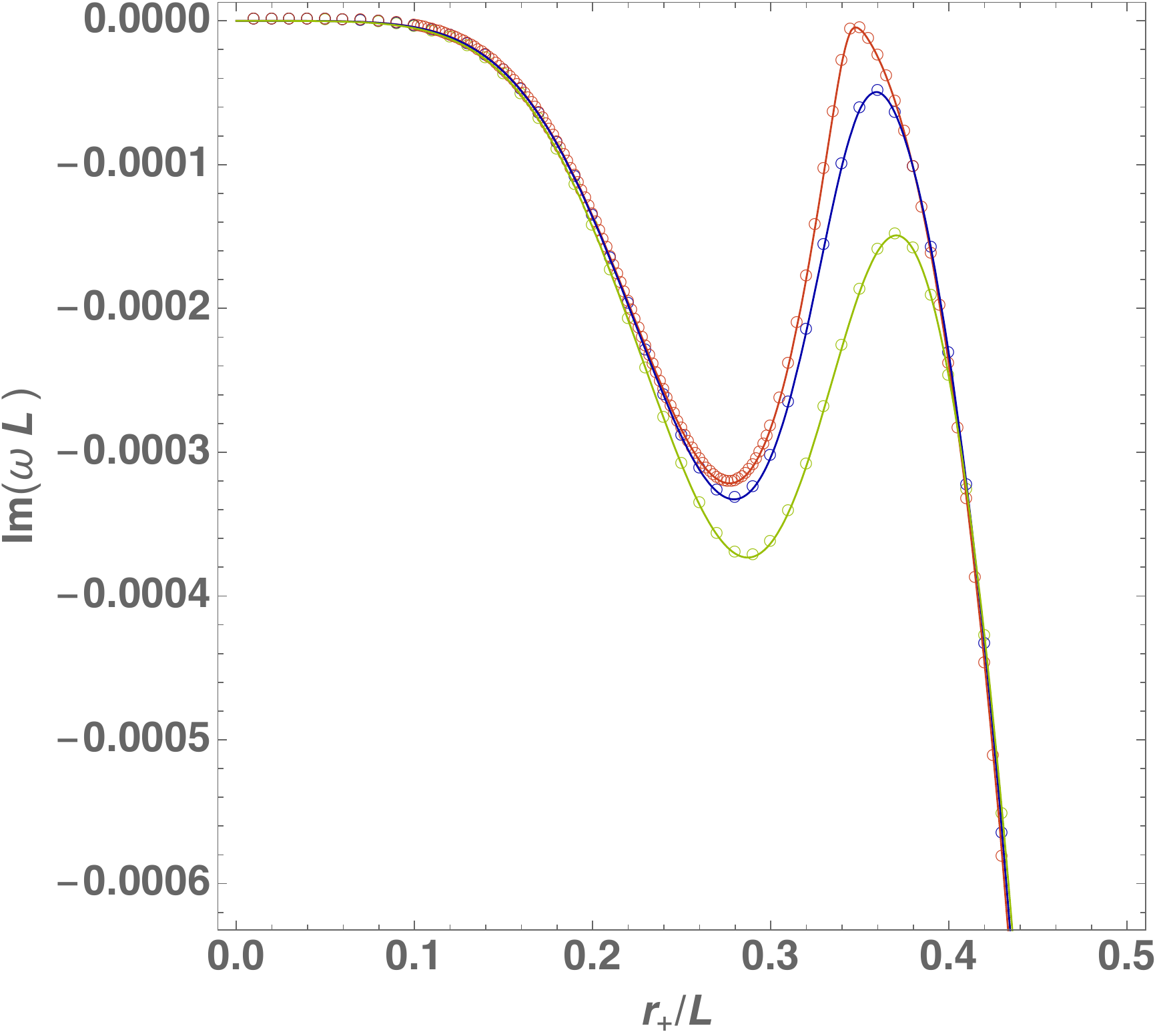}
}
\caption{Imaginary part of the frequency as a function of the horizon radius for fermion field charge $q L=1$ and different values of the chemical potential. In the left panel the chemical potentials are (from bottom to top) $\mu=0.9\mu_{\rm ext}$ (orange) and $\mu=0.95\mu_{\rm ext}$ (black). In the right panel the chemical potentials are closer to extremality, namely (from bottom to top curves): $\mu=0.99\mu_{\rm ext}$ (green), $\mu=0.995\mu_{\rm ext}$ (blue) and $\mu=0.999\mu_{\rm ext}$ (red). Notice the different regions scanned by the axes in the two plots. }
\label{Dirac-fig:AlternativeFixchp}
\end{figure}

To have a complementary perspective of the system's properties, in Fig.~\ref{Dirac-fig:AlternativeFixCharge}  and in Fig.~\ref{Dirac-fig:AlternativeFixchp} we illustrate other attempts we have made to find an instability. In Fig.~\ref{Dirac-fig:AlternativeFixCharge}, we keep the alternative quantization and fix the chemical potential at $\mu=0.999 \mu_{\rm ext}$, and plot ${\rm Im}(\omega L)$ as a function of $r_+/L$ for five different values of $qL$, namely, $q L=0.7,\,0.8,\,0.9$ (from bottom to top in the left panel) and $q L=1,\, 1.1$ (right panel). (The two plots are needed for the presentation of the results because the relevant $qL=1$ case in the right panel reaches a maximum that is approximately two orders of magnitude higher than the first three cases on the left panel). 
The main feature in these plots is the typical presence of a local minimum and local maximum (bump).
As we increase the fermion charge from zero to a value slightly above 1, the relative minimum and relative maximum of ${\rm Im}(\omega L)$ raise and shift to lower values of $r_+/L$. But the local maximum always has  ${\rm Im}(\omega L)<0$, {\it i.e.} there is no instability. However, for charges $qL$ above a value that is in between 1 and 1.1, the local minimum and maximum are no longer present and  ${\rm Im}(\omega L)$ decreases monotonically with $r_+/L$  (see {\it e.g.} $qL=1.1$ displayed as the yellow curve in the right panel; higher values, $qL\geq 1.1$, have a similar monotonic behaviour). 

As yet another illustration of experiments we made, in Fig.~\ref{Dirac-fig:AlternativeFixchp} we fix the fermion charge to be $qL=1$ (which was already analysed in Fig.~\ref{Dirac-fig:AlternativeFixCharge} for $\mu=0.999 \mu_{\rm ext}$) and we study the effect that changing the chemical potential has by considering a total of 5 curves with 5 different values of $\mu$. Namely, in the left plot we consider the cases $\mu=0.9\mu_{\rm ext}$ and $\mu=0.95\mu_{\rm ext}$. These cases have no bump (no local maximum) and illustrate that it only appears close to extremality. In the right panel we show three more cases where we fix $\mu=0.99 \mu_{\rm ext}$, $\mu=0.995\mu_{\rm ext}$ and $\mu=0.999\mu_{\rm ext}$ (from bottom to top). 
The bump is now present and the local maximum increases as one approaches extremality but never becomes positive. For the case  $\mu=0.999\mu_{\rm ext}$ this local maximum is at  ${\rm Im}(\omega L) \sim -5.93\times 10^{-6}$.

So far we have focused our discussion of the results for the alternative quantization case because, typically, for the same values of black hole parameters this is the case where ${\rm Im}(\omega L)$ approaches ${\rm Im}(\omega L)=0$ the most. Nevertheless, we have also tried hard to find an instability in the standard boundary condition \eqref{StandardQuant} case. Again without success. To illustrate briefly this conclusion, in Fig.~\ref{Dirac-fig:StandardFreqVSq} we give the partner plot of  Fig.~\ref{Dirac-fig:AlternativeFreqVSq} but this time for the standard quantization. Although the features of Fig.~\ref{Dirac-fig:StandardFreqVSq} are clearly more elaborated than those of  Fig.~\ref{Dirac-fig:AlternativeFreqVSq} ({\it e.g.} there are several local maxima and minima), the main conclusions are still the same: {\it i}) one always has ${\rm Im}(\omega L)<0$; {\it ii}) there is a $q=q_{\rm max}$ where the solution approaches ${\rm Im}(\omega L)=0$ the most (vertical brown dashed line); {\it iii}) $Re(\omega L)-q L \mu$ changes sign at $q=q_\star<q_{\rm max}$. It follows that we find no sign of an instability and the standard boundary condition case, much like the alternative quantization case, also gives results that are very different from the scalar field case of  Fig.~\ref{scalar-fig:freqVSq}.

Altogether, all our attempts $-$ best illustrated in Figs. \ref{Dirac-fig:AlternativeFreqVSq}-\ref{Dirac-fig:AlternativeFixchp}$-$ to find an instability due to Dirac field perturbations with alternative boundary condition \eqref{AlternativeQuant} failed miserably. The outcome is similar when we consider the standard boundary condition \eqref{StandardQuant}. This is best summarized in Fig.~\ref{Dirac-fig:StandardFreqVSq} where we show the partner plots of the alternative quantization Fig.~\ref{Dirac-fig:AlternativeFreqVSq} but this time for the standard quantization. Again, and essentially, we find that ${\rm Im}(\omega L)$ reaches a maximum at a critical $q_{\rm max}$ but never crosses the borderline    ${\rm Im}(\omega L)=0$ that would signal an instability.

\section{Discussion and conclusions}\label{sec:Conc}

A scalar field in an asymptotically AdS$_4$ Reissner-Nordstr\"om black hole can satisfy the asymptotically AdS$_4$ UV Breitenl\"ohner-Freedman (BF) stability bound but violate the infrared 2-dimensional BF stability bound associated to the AdS$_2\times S^2$ near-horizon geometry of the extremal black hole of the system, as reviewed in section \ref{sec:Scalar}. When this is the case, the AdS-RN black hole is unstable to scalar condensation and the system evolves to a new configuration in the phase diagram of solutions that preserves both the UV BF bound and the near-horizon 2-dimensional stability bound. Such a solution is a hairy black hole with a charged scalar field floating above the horizon \cite{hartnoll2008building,Faulkner:2009wj,Murata:2010dx,Basu:2010uz,Dias:2011tj,Bosch:2016vcp,Arias:2016aig}. Coulomb repulsion balances the gravitational force and the system is static. There is no doubt that the violation of the AdS$_2$ stability bound is the physical mechanism responsible for the near-horizon scalar condensation instability since the associated minimum bound \eqref{scalar:qminNH} on the scalar field charge that triggers the instability is sharp (at extremality) as best demonstrated by Fig.~\ref{scalar-fig:onset}.

Given these considerations, the study done in this manuscript for  Dirac field perturbations in the global AdS$_4$ RN black hole was motivated by the following observation. Dirac fields in AdS-RN can also preserve the UV fermionic stability bound \eqref{BFfermionic} \cite{Amsel:2008iz,Andrade:2011dg} but violate the near-horizon infrared fermionic stability bound, as seen in section \ref{sec:DiracRN1}. From the scalar field case lessons, this suggests that the system might be unstable to fermion condensation.  However, in spite of our efforts to scan the  relevant parameter space near extremality, we found no sign of a {\it linear mode} instability. The sharp distinction between the scalar and Dirac field cases is best illustrated comparing the scalar Fig.~\ref{scalar-fig:freqVSq} with the Dirac  Fig.~\ref{Dirac-fig:AlternativeFreqVSq} (for alternative quantization) or Fig.~\ref{Dirac-fig:StandardFreqVSq} (for standard quantization).
Of course our numerical study does not prove linear stability but we did such a detailed scan that we are very confident that no linear instability is present. Our stability results are also consistent with the stability study of fermions in {\it planar} AdS, where no instability was found \cite{Faulkner:2009wj,Iqbal:2009fd} (see also \cite{Lee:2008xf,Cubrovic:2009ye,Liu:2009dm,Guarrera:2011my,Iqbal:2011ae,Hartnoll:2016apf}).\footnote{This sharp difference between the stability conclusion of scalar and Dirac fields in RN-AdS is probably due to the fact that the Dirac equation is originally a first order PDE. We acknowledge Don Marolf for this observation (private communication).} Indeed, the planar AdS case is the $r_+/L \to \infty$ limit of the global AdS system. 

So, the planar AdS studies \cite{Faulkner:2009wj,Iqbal:2009fd,Lee:2008xf,Cubrovic:2009ye,Liu:2009dm,Guarrera:2011my,Iqbal:2011ae,Hartnoll:2016apf} and our present study in global AdS establish that the violation of the 2-dimensional stability bound of a Dirac field in AdS-RN does not lead to a {\it linear} mode instability. However, such solutions correspond to negative energy Schr\"odinger states: without a positive self-adjoint extension for the Schr\"odinger operator the dynamical evolution of the system should develop an instability... In particular, the system might indeed still be unstable if {\it non}-linear effects play a role in the discussion. That is to say, if we perturb a AdS-RN black hole with a Dirac field in a region of parameter space where the infrared stability bound is violated, it could still be the case that the system evolves non-linearly to a new configuration that has a charged Dirac field floating above the horizon and that preserves both the UV and IR stability bounds. How difficult would it be to prove whether this scenario is correct? 
      
One must proceed with caution. To begin with one needs to first formulate more precisely the setup of the problem. It is certainly much harder to find, if they exist, the proposed Dirac hairy black holes than it was to construct the scalar hairy black holes \cite{hartnoll2008building,Faulkner:2009wj,Murata:2010dx,Basu:2010uz,Dias:2011tj,Bosch:2016vcp,Arias:2016aig}. There is a fundamental difference between fermionic and bosonic fields. The fermion has no classical limit: Planck's constant $\hbar$ is present in the stress tensor and associated equations of motion for a fermion. As discussed in detail in section 14.3 of Wald's textbook \cite{Wald:106274}, the absence of a classical limit means that in Einstein's equation we have to promote the differential Einstein and energy-momentum operators $G$ and $T$ to quantum operators and the quantum version of Einstein's that gives the back-reaction of a fermion on the gravitational field is $\langle G_{\mu\nu}\rangle=8\pi (\langle T_{\mu\nu}^{\rm Max}\rangle+\langle T_{\mu\nu}\rangle)$, where $T_{\mu\nu}^{\rm Max}$ and $T_{\mu\nu}$  stands for the Maxwell and Dirac stress tensor contributions and  $\langle \cdots \rangle$ stands for the expectation value of the corresponding operator.

 Thus, to find the backreaction that fermions induce on the gravitoelectromagnetic background one needs to first compute the expectation value of the fermion energy momentum tensor $\langle T_{\mu\nu}\rangle$. This is a highly non-trivial task. Even worse, once we consider the quantum backreaction of fermions one also needs to consider the quantum backreaction of gravitons and photons, {\it i.e.} one also needs to compute $\langle G_{\mu\nu}\rangle$ and $\langle T_{\mu\nu}^{\rm Max}\rangle$  \cite{Wald:106274}. In a best case scenario, where we have a large number $N$ of Dirac fields, one might be able to assume that, roughly speaking, the effects of $N$ Dirac fields are $N$ times as relevant as that of the gravitons and photons  \cite{Wald:106274}. For a `fermionic hairy black hole', the fermionic condensate should be made of a large number of fermions. In these conditions, for large $N$, one might be able to neglect the quantum backreaction of gravitons and photons and work in the semi-classical limit whereby the backreaction of the Dirac field on the gravitoelectromagnetic background is simply governed by $G_{\mu\nu}=8\pi T_{\mu\nu}^{\rm Max}+ 8\pi N \langle T_{\mu\nu}\rangle$.  This semi-classical system should be viewed as the leading term of a $1/N$ expansion of the full theory  \cite{Wald:106274}.  But this semi-classical computation still requires that one computes $\langle T_{\mu\nu}\rangle$. And this is still a remarkable task. An overview on the physical and technical tools required to accomplish this task can be found in \cite{Hartnoll:2010gu,Allais:2013lha,Hartnoll:2016apf} (and references there-in) where  asymptotically  planar AdS quantum electron stars are discussed as semi-classical solutions of Einstein-Maxwell theory. 

Finally note that in the present manuscript we focused our attention on modes that could eventually 
become unstable. We have not studied in detail the full spectrum of quasinormal mode frequencies of the Dirac field in AdS-RN. Moreover, we focused on the case of a massless fermion because, as explained previously, this was enough for our purposes.
However, the equations of motion and relevant boundary conditions for any fermion mass and any sector of perturbations are given in section \ref{sec:WarmUp}. We have also computed the normal modes of massive fermions in AdS (previously only the massless spectrum was computed). It might be useful to have a more complete frequency spectra study for future studies/applications. It might also be interesting to look for perturbations of spin $3/2$ Rarita-Schwinger fields about AdS-RN  black holes. In this case, there are also normalizable solutions that become unstable for negative square masses \cite{Amsel:2009rr}. Probably there will be no linear near-horizon instabilities when the effective 2-dimensional mass violates the AdS$_2$ stability bound but, as far as we know, this was never checked.

\vskip .5cm
\centerline{\bf Acknowledgements}
\vskip .2cm
We wish to thank Nabil Iqbal, Jorge Santos, Kostas Skenderis, Marika Taylor and David Vegh for useful discussions and, in particular, Don Marolf for detailed discussions of \cite{Amsel:2008iz}.
The authors acknowledge financial support from the STFC Ernest Rutherford grants ST/K005391/1 and ST/M004147/1 and from the STFC ``Particle Physics Grants Panel (PPGP) 2016" Grant No. ST/P000711/1.


\begin{appendix}

\section{Near-extremal modes in AdS \label{secA:nearExtremalQNMs}}

For a Reissner-Nordstr\"om de Sitter ($\Lambda>0$) background, in \cite{Cardoso:2017soq,Dias:2018ynt,Dias:2018etb} it was found that there is a family of quasinormal modes $-$ denoted as the  `near-extremal' family of modes $-$ that is distinct from the 'de Sitter' family quasinormal mode, where the latter connects to the normal modes of de Sitter when the horizon radius shrinks to zero size, $r_{+} \rightarrow 0$.  

This naturally raises the question of whether there is also such a `near-extremal' family of modes in AdS and, if so, wether they do or not coincide with the AdS family of modes. In this appendix, we address this question in the simplest case, namely in the case of a (charged) scalar field that obeys the Klein-Gordon equation  \eqref{ScalarEqn}. More concretely, we arrive to the near-extremal frequency \eqref{scalar:NearExtremalFreq} which is used in the main text (see section \ref{sec:Scalar} and the discussion there-in of the dashed curves of Fig.~\ref{scalar-fig:freqVSq}) to show that in AdS the `near-extremal' and AdS families of modes coincide (unlike in the de Sitter case).

The  `near-extremal'  modes we seek obey \eqref{ScalarEqn} in the background \eqref{RNAdS} (we will work with the gauge choice $C=0$) and, at least in the near extremal limit, are expect to be highly peaked near the horizon. So we want to simultaneously zoom into the horizon and approach extremality. For that we first introduce the dimensionless quantities 
\be\label{def:sigma}
x = 1- \frac{r}{r_{+}}, \qquad \sigma = 1 - \frac{r_{-}}{r_{+}}\,.
\ee
For $x\ll 1$ one is close to the outer horizon and for $\sigma\ll 1$ the inner and outer horizon are very close, {\it i.e.} one is close to extremality. 
Next, we take the limit $\sigma \to 0$ whilst keeping $z = \frac{x}{\sigma}$ fixed. From \cite{Cardoso:2017soq,Dias:2018ynt,Dias:2018etb} the `near-extremal' modes are expected to saturate the superradiant bound $\omega=q \mu$ at extremality so onwards we measure the frequency difference $\delta\omega$ with respect to this bound via the redefinition
\be\label{freqRedef}
\omega = \frac{e \,Q}{r_+} + \sigma \,\delta\omega \,.
\ee 
Using the condition $f(r_-)=0$ for the location of the inner horizon one can find $r_-=r_-(r_+,Q,L)$ which is then inserted into \eqref{def:sigma} to express $Q$ as a function of $(r_+,\sigma,L)$.

In these near-extremality conditions, we are ready to find the near-horizon solution of the Klein-Gordon equation. Concretely, introducing the above redefinitions into the Klein-Gordon equation \eqref{ScalarEqn}, to {\it leading order in $\sigma$}, we obtain:
\be \label{eqNExt}
\left(1-z\right) z \,\de_{z}^{\,2} \phi(z) +\left( 1- 2 z \right) \de_{z} \phi(z) + \left[\frac{\varphi^2 - \hat{\lambda}\, z}{z(1-z)} + \eta \right] \phi(z) = 0\,,
\ee
where 
\begin{eqnarray}
&& \varphi = \frac{R_{+} \delta \tilde{\omega}}{1+ 6 R_{+}^2}\,,
\nonumber \\
&& \hat{\lambda} =  \frac{2 e\, R_{+}^2 \left[ \,\delta \tilde{\omega} \sqrt{2} \sqrt{1+3 R_{+}^2}- e (1+ 3\,R_{+}^2)  \right]}{\left(1+ 6 R_{+}^2\right)^2}\,,
\nonumber \\
&& \eta = \frac{1}{6 R_+^2+1}\left( \tilde{m}^2 R_+^2+\ell  (\ell +1)  -\frac{2 e^2 \left(1+ 3 R_{+}^2\right) R_+^2}{1+ 6 R_{+}^2}\right)\,,
\end{eqnarray}
where we have introduced the dimensionless quantities $R_+=r_+/L, e=q L, \tilde{m}=m L, \delta \tilde{\omega}=L\, \delta \omega$.

With the field redefinition 
\be
\phi(z) = z^{- i \varphi} \left(1-z\right)^{i \sqrt{\varphi^2 - \hat{\lambda}}} \hat{f}(z)\,,
\ee
\eqref{eqNExt} is rewriten as a standard hypergeometric ODE 
\bea
&&\left(1-z\right) z \,\hat{f}''(z) +\left[1-2 i \varphi -2 i z \left(-i -\varphi + \sqrt{\varphi^2 - \hat{\lambda}} \right) \right] \hat{f}'(z) \nn \\
&&\hspace{2.5cm}+ \left[\eta - \hat{\lambda} - \left(i + 2 \varphi \right)\left(-\varphi+\sqrt{\varphi^2 - \hat{\lambda}} \right)\right] \hat{f}(z) = 0 \,.
\eea
The regular ({\it i.e.} ingoing) solution at the future event horizon is given by 
\be\label{nearRegion}
\phi(z) = z^{- i \varphi} \left(1-z\right)^{i \sqrt{\varphi^2 - \hat{\lambda}}} {} _{1}F_{2} \left( a_{-} , a_{+} , 1- 2 i \varphi, z \right),
\ee
where ${} _{1}F_{2}(a,b,c;z)$ is the standard hypergeometric function and $a_{-} , a_{+}$ are defined by:
\be
 a_{\pm} = \frac{1}{2} \left(1 \pm \sqrt{1+ 4 \eta} - 2 i \varphi + 2 i \sqrt{\varphi^2 - \hat{\lambda}} \right).
\ee

In the context of a matched asymptotic expansion, the near-region (near-horizon) solution \eqref{nearRegion} must now be matched with the far-region solution of \eqref{ScalarEqn} (in near-extremality conditions). As explained above we expect the `near-extremal' modes we are looking into to have wavefunctions that die-off very quickly away from the black hole horizon (at least near-extremality). Therefore, as a first rude approximation we take the far-region to be described by a vanishing wavefunction. That is to say, in the overlapping region, we match the near-region solution \eqref{nearRegion} with $\phi=0$. In the end of the day, this approximation turns out to be quite good because the analytical approximation for the `near-extremal' frequency that we obtain $-$see \eqref{scalar:NearExtremalFreq2} $-$ matches remarkably well the numerical solution of \eqref{ScalarEqn}. This is best seen comparing the  black dashed analytical curve of our expansion in Fig.~\ref{scalar-fig:freqVSq} with the numerical blue dot results. For this reason, we do not try to improve further our matching asymptotic approximation. 

Proceeding in these conditions, the leading order behaviour of the large $R=r/L$ series expansion ($z \rightarrow -\infty$) of $\phi$, namely $\phi \approx$ $(-z)^{\pm \sqrt{1 + 4 \eta}}$, needs to be matched with the far-region solution $\phi=0$. 
Before we can do it, we still need to distinguish the cases $1 + 4 \eta \geq 0$ and $1 + 4 \eta < 0$. For our proposes (comparing with the numerical results of section \ref{sec:Scalar}), we want to consider the small scalar field charge case for which one finds that $1 + 4 \eta \geq 0$ holds as long as e $\leq e_{c}$ where 
\be
e_{c}^{2} =  \frac{1+6 R_{+}^2}{8 R_{+}^2 (1+ 3 R_{+}^2)}\left[ \left(6+ 4 m^2\right)R_{+}^2 + \left(1+2 \ell \right)^2 \right]
\ee
(the reader also interested in the case $1 + 4 \eta < 0$ can follow the steps detailed in  \cite{Dias:2018etb}).
In these conditions, from the matching condition one finds that 
\bea \label{NExt_modes}
&&  \hspace{-1.5cm}   \delta \tilde{\omega} = \frac{\sqrt{2}}{2} e \sqrt{1+3 R_{+}^2} - \frac{i}{4 R_{+}} \Bigg((1 + 6 R_+^2) (1 + 2 p) \nonumber \\
&&   +\sqrt{(1 + 6 R_+^2) 
\left[1+6 R_+^2+4m^2 R_+^2+4\ell(\ell+1)\right]-8e^2R_+^2(1+3R_+^2)}\Bigg),
\eea
where $p=0,1,2,\cdots$ is the radial overtone of the mode. 
Replacing this into \eqref{freqRedef} one finally finds that `near-extremal' modes have a frequency given by:
\begin{eqnarray}\label{scalar:NearExtremalFreq2}
 \omega L \simeq e \mu + \sigma\, \delta \tilde{\omega}  +
\mathcal{O}(\sigma^2).
\end{eqnarray}
This is \eqref{scalar:NearExtremalFreq} in the main text when we set the radial overtone $p=0$.

\section{Perturbative results}\label{secA:Perturbative_results}
Although we solve the Dirac equation numerically in the main text, it is very good practice to testify the numerical results against analytical predictions that can be obtained within perturbation theory in some region of the parameter space. Therefore, in this appendix we find some useful analytical perturbative approximations for the Dirac frequencies.   More concretely, this Appendix is divided in two parts. In Appendix~\ref{secB1:matching} we use a matching asymptotic expansion approach to find the frequency approximations \eqref{deltam}-\eqref{deltap} that, in Fig.~\ref{s:Numeric:fig:ImAnal} of Appendix~\ref{sec:DiracRN3a}, are compared against our numerical results. Then, in Appendix~\ref{secB2:perturbative} we use a systematic perturbative expansion in the dimensionless horizon radius $r_+/L\ll 1$ (with no further approximations) to find the analytical frequency approximations \eqref{D:perturbative:eq:freqm}-\eqref{D:perturbative:eq:freqp} which, in Fig.~\ref{s:Numeric:fig:ReAnal} of Appendix~\ref{sec:DiracRN3a}, are also compared with our numerical results. In both cases, there is agreement between the analytical approximation predictions and the numerical results in the regime of parameter space where the former are valid.

In this appendix, as in the main text, we solve the Dirac equation  \eqref{radial:2ndorder} in a AdS-RN background \eqref{RNAdS} (gauge choice $C=0$) with regular (ingoing) boundary conditions at the future event horizon and the standard \eqref{StandardQuant} or alternative \eqref{AlternativeQuant} boundary conditions at the conformal boundary. We will work exclusively with vanishing fermion mass, $m=0$.

\subsection{Matched asymptotic expansion}\label{secB1:matching}

In this section we derive an analytical expression for the imaginary part of the Dirac frequency  using the method of matched asymptotic expansion introduced in \cite{Starobinsky:1973aij,PhysRevUnruh}  (see also e.g. \cite{maldacena1997matching,Cardoso:2004nk,Cardoso:2004hs}). We assume $r_{+} \ll L$ and split our spacetime into two regions; an asymptotic globally AdS {\it far region} where the effects of the black hole can be neglected and a {\it near region} about  the black hole outer horizon where the effects of the cosmological constant can be neglected. In each region the associated perturbation equation can be solved analytically, then matching the near and far region solutions in their {\it overlapping region} will fix the integration constants as well as the imaginary part of the perturbation frequency $\omega$. More concretely, the near region is defined by $r - r_{+}\ll 1/ \omega$ and the far region is defined by $r - r_{+} \gg  r_{+}$. It  follows that the overlapping region exists for $\omega r_{+} \ll 1$. A further assumption we must make is that the Coulomb interaction is weak, $Q q \ll 1$, where $q$ ($Q$) is the fermion (black hole) charge. 

\subsubsection{Far region solution \label{appB1:far}}

Since in the far region the effect of the black hole (BH) is assumed to be negligible, we effectively have a fermion field in the global AdS background. Thus, the general solution for the massless fermionic field $R_1$ is given by \eqref{farSol} that we reproduce here:
\bea\label{farSol:appendix}
F(r) &=& C_1\, _2F_1\Big(\frac{1}{2} + \ell, \omega L+\ell+ 1  ,2(1+ \ell),\frac{2 r}{r+i L }\Big) \nn \\
&&+ C_2 \, \Big(\frac{2 r}{r+i L }\Big)^{-1-2 \ell} \, _2F_1\Big(-\frac{1}{2} - \ell, \omega L- \ell ,-2 \ell, \frac{2 r}{r+i L }\Big),
\eea
where $C_{1,2}$ are two arbitrary amplitudes to be determined below. 
Asymptotically this solution decays as \eqref{AsympDecaysRm0}, namely  $R_1\big|_{r\to \infty}\sim   \alpha_1+\beta_1\,\frac{L}{r} +{\cal O}(r^{-2})$ with 
\bea
 \alpha_1&=& \,i^{\omega L} \,2^{\ell +\frac{1}{2}} \left( C_{1} \,_2F_1\left(\ell +\frac{1}{2},\ell +\omega L +1;2 \ell  +2;2\right)+ C_{2} \,_2F_1\left(-\ell -\frac{1}{2},-\ell +\omega L ;-2 \ell  ;2\right) \right) , \nonumber\\
\beta_1&=& C_2 i^{1+L \omega} 2^{-\frac{3}{2}-\ell}\Bigg[ \left(1+2\ell-2 L \omega \right) \, _2F_1\left(-\ell
   -\frac{1}{2},-\ell +\omega L ;-2 \ell ;2\right) + \nn\\
  && \frac{1}{\ell}\left(1+2\ell\right)\left(\ell - L \omega \right) \, _2F_1\left(-\ell
   +\frac{1}{2},-\ell +\omega L +1;-2 \ell +1;2\right) \Bigg] - \nn\\ 
  &&C_{1}\, 2^{\ell -\frac{1}{2}}\, i^{\omega L +1} \Bigg[(2 \omega L +2 \ell +1) \, _2F_1\left(\ell
   +\frac{1}{2},\ell +\omega L +1;2 \ell +2;2\right)\nonumber\\
&&   +\frac{(2 \ell +1) (\omega L +\ell +1)}{\ell +1}\, _2F_1\left(\ell
   +\frac{3}{2},\ell +\omega L +2;2 \ell +3;2\right)\Bigg].
\eea
This solution has to satisfy the asymptotic boundary condition (for a massless fermion). 
For the standard quantization this is \eqref{StandardQuant} while for the alternative quantization the boundary condition is \eqref{AlternativeQuant} which fix  $\beta_1$  as a function of $\alpha_1$ or, equivalently, $C_1$ as a function of $C_2$. 
This yields
\bea \label{B1:asymptoticBC}
 &&\alpha_1 (\lambda \pm \omega L) \mp i \beta_1=0 \quad \Leftrightarrow \quad C_1=C_2 (-1)^{-2 \ell} 2^{-1-2 \ell} \gamma_{\pm} \,, \qquad  \hbox{where} \nn \\ 
 \\
 &&\mbox{{\normalsize $\gamma_{\pm} = $} \Large $
 \frac{(1+2 \ell \pm 4 L \omega) _2F_1\Big(-\frac{1}{2} - \ell,-\ell+ \omega L  ,-2 \ell, 2 \Big) \mp (1+2 \ell) _2F_1\Big(\frac{1}{2} - \ell, \omega L-\ell  ,-2 \ell, 2\Big)}{(1+2 \ell \pm 4 L \omega) _2F_1\Big(\frac{1}{2} + \ell,1+\ell+ \omega L  ,2(1+\ell), 2 \Big) \pm (1+2 \ell) _2F_1\Big(\frac{3}{2} + \ell, \omega L+\ell+1  ,2(1+\ell), 2\Big)}\,,
$} \nn
\eea
where the upper sign refers to the {\it standard quantisation} \eqref{StandardQuant} and the lower sign to the {\it alternate quantisation} \eqref{AlternativeQuant}.

Note that we do not impose any boundary condition at a inner boundary since this far region solution does not extend till there.

\subsubsection{Near region solution \label{appB1:near}}
In the near region we can approximate $\Delta(r)=r^2f(r)$ by:
\be
\Delta(r) \approx (r - r_{+})(r - r_{-})
\ee
where $r_{-} \approx \frac{Q^{2}}{2 r_{+}}$. This follows from the assumption that $r_{+} \ll L$ and therefore in the near region we have $r\sim{\cal O}(r_+) \ll L$, so we can neglect the $r^{2}/L^{2}$ term in $f(r)$. Further applying the near region assumptions to the Dirac equation \eqref{radial:2ndorder} we can neglect terms of order $\o r_{+}$ or higher powers. Other terms appear which are dominated by a $1/ \Delta$ term in the small black hole (BH) approximation near the horizon; therefore we can evaluate the numerators of these terms at $r \approx  r_{+}$.

With these approximations, and the coordinate transformation
\be
z = \frac{r - r_{+}}{r - r_{-}}\,, \qquad 0 \leq z \leq 1,
\ee
(the horizon $r=r_{+}$ is at $z=0$) the Dirac equation is approximately given by the near region equation, 
\be
\Big( \hat{\omega} \frac{1-z}{z} - \frac{\lambda^{2}}{1-z}\Big) R_{1}(z) +\frac{1}{2} (1- 3 z) R_{1}'(z) + (1-z) z R_{1}''(z) = 0,
\ee
where  
\be \label{B1:omegaHat}
\hat{\omega} = \frac{r_{+}^{2} \omega (r_{+}(i + 2 r_{+} \omega) - i r_{-})}{2 (r_{+} - r_{-})^{2}}.
\ee
Making the field redefinition 
\be
R_{1}(z) = z^{ \tilde{\alpha}} (1-z)^{\tilde{\beta}} R(z)
\ee
where 
\be \label{B1:sigmaTilde}
 \tilde{\alpha} = \frac{1}{4} + i \tilde{\sigma}, \quad \tilde{\beta} = \frac{1}{2} + \ell,  \quad
\tilde{\sigma} = \frac{1}{4} \sqrt{16 \hat{\omega} - 1}\,,
\ee 
the near horizon equation is rewritten in the standard hypergeometric form $-$ see \eqref{hyperMassive} $-$ with parameters $a, b$ and $c$ given by 
\be
a = 1+ \ell,\quad b = 1+ \ell + 2 i \tilde{\sigma}, \quad c = 1 + 2 i \tilde{\sigma}\,.
\ee
The most general near horizon solution is
\bea \label{B1:near}
R_{1}^{\rm near}(z) = &\alpha z^{\frac{1}{4} - i \tilde{\sigma}} (1-z)^{\ell + \frac{1}{2}} \,_2F_1(1+\ell, 1+ \ell - 2 i \tilde{\sigma}, 1- 2 i \tilde{\sigma}, z) \nn \\ 
&\quad + \beta z^{\frac{1}{4} + i \tilde{\sigma}} (1-z)^{\ell + \frac{1}{2}} \,_2F_1( 1+ \ell, 1 + \ell + 2 i \tilde{\sigma}, 1 + 2 i \tilde{\sigma}, z)
\eea
Using the property of the hypergeometric function $ \,_2F_1(a, b, c, 0) = 1$ we find that the 
 $z \to 0$ behaviour of the near region solution is $R_{1}^{\rm near}(z) \approx z^{\frac{1}{4}}( \alpha z^{- i \tilde{\sigma}} + \beta z^{ i \tilde{\sigma}} )$. Requiring regularity (only ingoing modes) at the horizon implies that we must set $\beta = 0$.

\subsubsection{Matching}
To find the large $r$ ($z \to 1$) behaviour of the near region solution \eqref{B1:near} with $\beta=0$, we use the $z \rightarrow 1 - z$ transformation law of the hypergeometric function \cite{1965handbook}:
\bea
&& \,_2F_1(a, b, c, z) = \frac{\Gamma(c) \Gamma(c - a - b)}{\Gamma(c - a) \Gamma(c - b)} \,_2F_1(a, b, a+b-c+1,1-z) \nn \\
 &&+ (1- z)^{c - a - b} \frac{\Gamma(c) \Gamma(a+ b - c)}{\Gamma(a) \Gamma(b)} \,_2F_1(c - a, c - b, c- a - b + 1,1 - z).
\eea
We again use that $ \,_2F_1(a, b, c, 0) = 1$  as well as $1- z \approx \frac{r_{+} - r_{-}}{r}$ (when $r \to \infty$), to obtain:
\be \label{B1:nearLarge}
 R_{1}^{\rm near}\Big|_{{\rm large}\, r} \approx \Gamma(1 - 2 i \tilde{\sigma}) \Big[ \frac{ (r_{+} - r_{-})^{- \ell - \frac{1}{2}} \Gamma(1+ 2 \ell )}{\Gamma(1+\ell) \Gamma(1 + \ell - 2 i \tilde{\sigma})} r^{\ell + \frac{1}{2}} + \frac{(r_{+} - r_{-})^{ \ell + \frac{1}{2}} \Gamma(-1- 2 \ell )}{ \Gamma(- \ell ) \Gamma(- \ell - 2 i \tilde{\sigma})} r^{-\ell - \frac{1}{2}} \Big] .
\ee
This needs to be matched (in the overlapping region) with the small  $r$ behaviour of the far region solution \eqref{farSol:appendix}  subject to the asymptotic boundary conditions \eqref{B1:asymptoticBC}.\be \label{B1:farSmall}
 R_{1}^{\rm far}\Big|_{{\rm small}\, r} \approx { C_1}  \, 2^{\frac{1}{2} + \ell} (i L)^{-\frac{1}{2} - \ell} r^{\frac{1}{2} + \ell}+  C_2 \, 2^{-\frac{1}{2} + \ell} (i L)^{\frac{1}{2} + \ell} r^{-\frac{1}{2} - \ell}.
\ee

In the overlapping  region, one must have $R_{1}^{\rm near}\big|_{{\rm large}\, r} =  R_{1}^{\rm far}\big|_{{\rm small}\, r}$. That is to say, we must match independently the $r^{\frac{1}{2} + \ell}$ and $r^{-\frac{1}{2} - \ell}$ terms of \eqref{B1:nearLarge} with those of \eqref{B1:farSmall}.
This matching yields:
\be \label{matchingeq}
{\frac{C_2}{C_1}} \Big(\frac{i}{2}\Big)^{2 \ell + 1} = \frac{ \Gamma\left[1+\ell \right] \Gamma\left[-1- 2 \ell \right] }{\Gamma\left[1+2 \ell \right] \Gamma\left[-\ell \right]} \frac{\Gamma\left[\frac{3}{2} + \ell - \frac{2 i r_{+}^2}{r_{+} - r_{-}} \omega \right]}{\Gamma\left[\frac{1}{2}-\ell - \frac{2 i r_{+}^2}{r_{+} - r_{-}} \omega \right]}  \Big(\frac{r_{+} - r_{-}}{L}\Big)^{2 \ell + 1}
\ee
where we have used  \eqref{B1:sigmaTilde} and \eqref{B1:omegaHat} for $\tilde{\sigma}$ to restore the explicit dependence on the frequency $\omega$. The ratio $\frac{C_2}{C_1}$ follows straightforwardly from \eqref{B1:asymptoticBC}.
We want to solve the transcendental equation \eqref{matchingeq} to get an analytical expression for the frequency. There is no closed form solution, unless we do some educated approximations that we now discuss. Since we are working with very small and weakly charged RN-AdS black hole, one expects that the mode frequencies in such a background are close to the massless Dirac normal modes frequencies of $AdS_4$, already computed in  \eqref{normal_mode_StandardBCp} (standard quantization) or \eqref{normal_mode_AlternativeBCm}  (alternative quantization). Denote this normal mode frequency by $\omega_{AdS_4}$. However, since the background now has a  horizon, the system becomes dissipative and with respect to the normal modes of $AdS_4$, the frequency of the system should  acquire a small imaginary contribution. Denote it by $i\,\delta$. In \eqref{matchingeq}, it is thus a good approximation to replace the frequency $\omega$ by $\omega = \omega_{AdS_4} + i \, \delta$ with $|\delta| \ll  \omega_{AdS_4} $. Our target now is to solve \eqref{matchingeq} at {\it leading order} for $\delta\ll \omega_{AdS_4}\sim \mathcal{O}(1)$.
 {\it A posteriori}, we compare the prediction of our analytical computation with the numerical result to confirm that this approximation is valid.  
 
Equation  \eqref{matchingeq} has the additional challenges that: 1) the frequency appears in the argument of the Gamma functions and 2) $\Gamma\left[-1- 2 \ell \right]$ in the numerator diverges for the allowed values \eqref{lambda:quantization} of the harmonic number $\ell$ (recall that $\Gamma[-p]=\infty$ for non-negative $p$). To deal with these obstacles, we use the Gamma function property $\Gamma[z+1] = z \Gamma[z]$ and the assumptions of our problem, $\omega r_{+} \ll 1$ and $\delta\ll \omega_{AdS_4}\sim \mathcal{O}(1)$. This allows to expand the Gamma functions whose argument depends on $\omega$ (and thus on $\delta$) to extract $\delta$ out of the argument of the Gamma functions. In particular, this permits to find that the divergence of $\Gamma\left[-1- 2 \ell \right]$ in the numerator is cancelled by the Gamma function in the denominator that depends on $\delta$.
 
In these conditions, one finds that the leading order solution for $\delta$ is
\be \label{delta_approx}
\tilde{K} \delta \approx \frac{ i (-1)^{2 \ell +3/2} 2^{-4 \ell - 2} \ell ! }{\sqrt{\pi} \left(\frac{2 \ell -1}{2}\right)! \left(\ell + 1/2\right)} \prod_{k=0}^{\ell - 1/2} \frac{\ell + 1/2 - k}{\ell - k} \Big(\frac{r_{+} - r_{-}}{L}\Big)^{2 \ell + 1}.
\ee
where  $ \tilde{K} $ is a positive real number for each $\ell$, $n$. For the alternative boundary condition \eqref{AlternativeQuant} it is given by (for overtone $p=0$)
\be
\tilde{K} = -\frac{  e^{3 i \pi  \ell } \left(\,_2F_1^{(0,1,0,0)}\left(\ell +\frac{1}{2},2 \ell +2,2 \ell +2,2\right)+\,_2F_1^{(0,1,0,0)}\left(\ell +\frac{3}{2},2 \ell +2,2 \ell +2,2\right)\right)}{\, _2F_1\left(1,-\ell -\frac{1}{2};-2 \ell ;2\right)-\, _2F_1\left(1,\frac{1}{2}-\ell ;-2 \ell ;2\right)},
\ee
whereas for the standard boundary condition \eqref{StandardQuant} we have (for $p=0$)
\be
\tilde{K} = \frac{  e^{3 i \pi  \ell } \left(-\,_2F_1^{(0,1,0,0)}\left(\ell +\frac{1}{2},2 \ell +3,2 \ell +2,2\right)+\,_2F_1^{(0,1,0,0)}\left(\ell +\frac{3}{2},2 \ell +3,2 \ell +2,2\right)\right)}{\, _2F_1\left(2,-\ell -\frac{1}{2};-2 \ell ;2\right)+\, _2F_1\left(2,\frac{1}{2}-\ell ;-2 \ell ;2\right)}.
\ee
Thus the two boundary conditions \eqref{AlternativeQuant}, \eqref{StandardQuant} yielddifferent values of $\delta$. As a concrete example (that we use to compare our numerics with), for $\ell = 1/2$, $p=0$ we have for the alternative boundary condition \eqref{AlternativeQuant},
\be\label{deltamA}
\delta \approx -\frac{1}{4 \pi} \Big(\frac{r_{+} - r_{-}}{L}\Big)^{2}\,,
\ee
while for standard boundary condition \eqref{StandardQuant},
\be\label{deltapA}
\delta \approx -\frac{3}{4 \pi} \Big(\frac{r_{+} - r_{-}}{L}\Big)^{2}.
\ee
These are the analytical predictions we use in \eqref{deltam}-\eqref{deltap} of Appendix~\ref{sec:DiracRN3a} to compare against the numerical results: see Fig.~\ref{s:Numeric:fig:ImAnal}. 
We expect these analytical predictions  to be valid only for small horizon radius and away from extremality and for $Qq\ll 1$, which we are able to confirm numerically in section \ref{sec:DiracRN3}.  An added bonus for this method is that we have an expression for general $\ell$. The method we present in the next Appendix  below has to be done for each $\ell$ individually, but it is more systematic than this one, since it only requires an expansion in $r_+/L\ll 1$.

\subsection{Perturbative expansion in $R_{+}$}\label{secB2:perturbative}

In this section we find an analytical prediction for the frequency using a systematic perturbative expansion in $r_+/L$. Unlike in the previous subsection, the only approximation that will be made is that the expansion parameter of this expansion is small, $r_+/L\ll 1$. We will do this expansion up to the order that finds the first correction (in the real part of the frequency) to the global AdS normal mode frequency. Should we wish, we could go one order higher in the analysis and find also the correction to the imaginary part of the frequency (although this is computationally more demanding). For our purposes of comparing with the numerical results, it is enough to have the correction to the real part of the frequency (the results of appendix \ref{secB1:matching} already allow us to test independently the imaginary part). 

The systematic perturbative expansion in $r_+/L\ll 1$ used in this Appendix was first introduced in 
 \cite{Basu:2010uz} and further explored in \cite{Bhattacharyya:2010yg,Dias:2011at,dias2012hairyBHs,Dias:2016pma,Dias:2018zjg} where the reader can find full details of the method (we will be very succinct in our exposition). In short, we split our spacetime into a near and far regions.
We expand the frequency $\Omega=\omega L$ and the field $R_{1}$ in each region in a power series in $R_{+}=r_+/L$:
\be
\Omega = \sum_{k=0}^{\infty} \Omega_{(k)} R_{+}^k\,; \qquad R_{1}^{near} = \sum_{k=0}^{\infty} \psi_{(k)}^{near} R_{+}^k\,, \qquad R_{1}^{far} = \sum_{k=0}^{\infty} \psi_{(k)}^{far} R_{+}^k \,.
\ee
We now series expand the Dirac equation \eqref{radial:2ndorder} in small $R_{+}$. The leading, zeroth order equation is simply the Dirac equation in global $AdS_4$ for a massless fermion \eqref{farSol:appendix}  (that we already studied in sections \ref{sec:DiracNormalModes1} and \ref{appB1:far}). Not surprisingly, the small $R=r/L$ expansion of this leading order solution breaks down at order $R_{+}/R$. This motivates splitting our spacetime into a far region, $R\gg R_+$, and a near region, $R_+ \leq R\ll 1$. In the far region we work with the radial coordinate $R$ but in the  near region we work instead with the radial coordinate $y =R/R_{+}$ (since the far region small $R$ expansion breaks down at order $R_{+}/R$).  

In the far region, at each order in $R_{+}$,  we impose  the {\it standard} boundary condition \eqref{StandardQuant} or the {\it alternative} boundary condition \eqref{AlternativeQuant}. In the near region we impose boundary conditions that only allow for ingoing waves at the horizon. We then perform a matching procedure at each order in $R_{+}$, in the region where the far and near region overlap, to determine the frequency coefficients $\Omega_{(k)}$, as well as amplitudes that were not fixed by the two boundary conditions. 
At leading (zeroth) order, we fix $\Omega_{(0)}$ to be the normal mode frequency for a massless fermion already obtained in \eqref{normal_mode_AlternativeBCm} or \eqref{normal_mode_StandardBCp} for the boundary conditions \eqref{AlternativeQuant} or \eqref{StandardQuant}, respectively. We will do this for the mode with harmonic number $\ell = 1/2$, and  radial overtone to be $p=0$. 
Our aim is then to find the first frequency correction $\Omega_{(1)}$ due to the presence of the black hole. 
For concreteness, in most of our discussion below we only explicitly present  details of the case where we impose the alternative boundary condition \eqref{AlternativeQuant}. We then present the final result also for the standard boundary condition \eqref{StandardQuant}.   

In the far region the leading order $R_{+}^0$ solution is \eqref{farSol:appendix} and imposing the asymptotic boundary conditions amounts to repeat {\it mutatis mutandis} the analysis done in \eqref{farSol:appendix}-\eqref{B1:asymptoticBC}. With our choice of $\ell=1/2$ and $p=0$ this fixes the frequency at order zero to be $\Omega_{(0)} = \frac{3}{2}$; see  \eqref{normal_mode_AlternativeBCm}. 
To fix the normalization, we set the amplitude of the Dirac field at infinity to be $1$ at all orders in $R_{+}$: $R_{1}^{far}|_{R\to \infty} = 1 + \mathcal{O}(1/R)$.  

Introducing the near region radial coordinate $y =R/R_{+}$, still at leading order $R_{+}^0$, the near region Dirac equation (for $\ell = 1/2$) reads 
\be
\frac{1}{2} \left(y-1\right) \left(2 y - \mu^2 \right) \de_{y}^2 \psi_{(0)}^{near} + \left( y-\frac{1}{2} - \frac{\mu^2}{4} \right) \de_y \psi_{(0)}^{near}  - \psi_{(0)}^{near}  = 0.
\ee
The solution which is regular at the horizon ($y=1$) is 
\bea
\psi_{(0)}^{near}=&\alpha_{(0)} \cosh \left( 2 \log \left(2 \sqrt{y-1} + \sqrt{4 y - 2 \mu^2} \right)\right) \nn \\
&+ i \beta_{(0)} \sinh\left(2 \log\left(2 \sqrt{y-1} + \sqrt{4 y - 2 \mu^2} \right)\right),
\eea
with
\be
\beta_{(0)} = -i \alpha_{(0)} \frac{1+4\left(\mu^2-2\right)^2}{1-4\left(\mu^2-2\right)^2}.
\ee

We must now match the far and near regions solutions at order $R_{+}^0$ in their overlapping region $R_+ \ll R\ll 1$. This procedure, typically fixes all other constants of the problem that were not fixed by the boundary conditions. The large $R$ expansion of $\psi_{(0)}^{near}$ is  $\psi_{(0)}^{near}|_{{\rm large} \: R} =\alpha_{(0)}$ + $\beta_{(0)} R+\cdots$  whereas the small $R$ expansion of $\psi_{(0)}^{far}$ is $\psi_{(0)}^{far} |_{{\rm small} \: R}= (-1)^{3/4} \,R+ \cdots$. Therefore, matching $\psi_{(0)}^{near}|_{{\rm large} \: R} = \psi_{(0)}^{far} |_{{\rm small} \:R}$ requires that we set $\alpha_{(0)} = 0$ and  $\beta_{(0)}= (-1)^{3/4} $. Collecting the results at order 0 for  the alternative quantization  \eqref{AlternativeQuant} we have: 
\bea
\begin{split}
&\psi_{(0)}^{near} = 0\, ,\quad
&\psi_{(0)}^{far} =  \frac{(-1)^{3/4}R\sqrt{1+ \mathrm{i} R}}{ (1+R^2)^{3/4}} \,;
& \qquad \Omega_{(0)} = \frac{3}{2}.
\end{split}
\eea
For the standard quantisation one has a similar result with $\Omega_{(0)} =5/2$.

We can now proceed to the first order $R_{+}^1$ contribution that enables us to find $\Omega_{(1)}$.
We repeat the procedure outlined above for the far and near regions, imposing the alternative boundary condition in the far region and ingoing boundary condition at the horizon in the near region. The near region equation for $\psi_{(1)}^{near} $ is the same as at order $R_{+}^{0}$ (i.e. there is no source). After imposing the horizon boundary condition, we find 
that the large $R$ expansion of the near region solution gives 
\be \label{RnearLarge}
R^{near}\big|_{{\rm large}\, R}  \approx  (-1)^{3/4} R +  \beta_{(1)} R_+ +\cdots.
\ee
The far region equation at order $R_{+}^{1}$ can also be solved analytically for $\psi_{(1)}^{far}$. As usual, we find a solution with two integration constants, say $C_1$ and $C_2$. These constants will in general depend on $\mu$, $e$ and $\Omega_{(1)}$. Firstly by imposing the alternative quantisation we find an expression for $C_2$ in terms of $\mu$, $e$ and $\Omega_{(1)}$. Then we again impose that our field has amplitude 1 at infinity, yielding an expression for $C_1$ in terms of $\mu$, $e$ and $\Omega_{(1)}$. After doing this we find the small $R$ behaviour of the far region solution to be 
\be\label{RfarSmall}
\begin{split}
&R^{far}\big|_{{\rm small}\, R} \approx (-1)^{3/4} R  +  R_{+}\left(-14 + 2 e \mu - 7 \mu^2 - 6 \pi \Omega_{(1)} \right) \\
 & \hspace{10em}- \frac{R_{+}}{R} \left( 2 - 4 e \mu +\mu^2 + 2 \pi \Omega_{(1)} \right)+\cdots, 
\end{split}
\ee

In the overlapping region $R_+ \ll R\ll 1$, \eqref{RfarSmall} must match \eqref{RnearLarge}.
A straightforward matching of the terms $R_+ R^0$ fixes the constant $\beta_{(1)}$ in \eqref{RnearLarge}. On the other hand, there is no $R_{+}/R$ term in the large $R$ series expansion of the near region solution. It follows that $\Omega_{(1)}$ must be such that it eliminates the corresponding $R_{+}/R$ term in the small $R$ expansion of the far region solution \eqref{RfarSmall}. This fixes  $\Omega_{(1)}$ to be
\be
\Omega_{(1)} = - \frac{\mu^2-4 \mu e+2}{2 \pi }.
\ee
Thus for the alternative boundary condition \eqref{AlternativeQuant}, the frequency up to ${\cal O}(R_+)$ is 
\begin{align}\label{D:perturbative:eq:freqmA}
\omega L=\frac{3}{2}-R_+\frac{\mu^2-4 \mu e+2}{2 \pi }+\mathcal O(R_+^2).
\end{align}

On the other hand, repeating the above analysis the standard boundary condition \eqref{StandardQuant} yields the frequency:
\begin{align}\label{D:perturbative:eq:freqpA}
\omega L=\frac{5}{2}+R_+\frac{11 \mu^2-20 \mu e+22}{6 \pi }+\mathcal O(R_+^2).
\end{align}

The frequencies  \eqref{D:perturbative:eq:freqmA} and \eqref{D:perturbative:eq:freqpA} are the analytical frequency approximations \eqref{D:perturbative:eq:freqm} and \eqref{D:perturbative:eq:freqp} that we reproduce in the main text and that we compare against the numerical data in Fig.~\ref{s:Numeric:fig:ReAnal} of section \ref{sec:DiracRN3a}.

\section{Comparing numerical results with analytical expansions}\label{sec:DiracRN3a}

\begin{figure}[th]
\centerline{
\includegraphics[width=.46\textwidth]{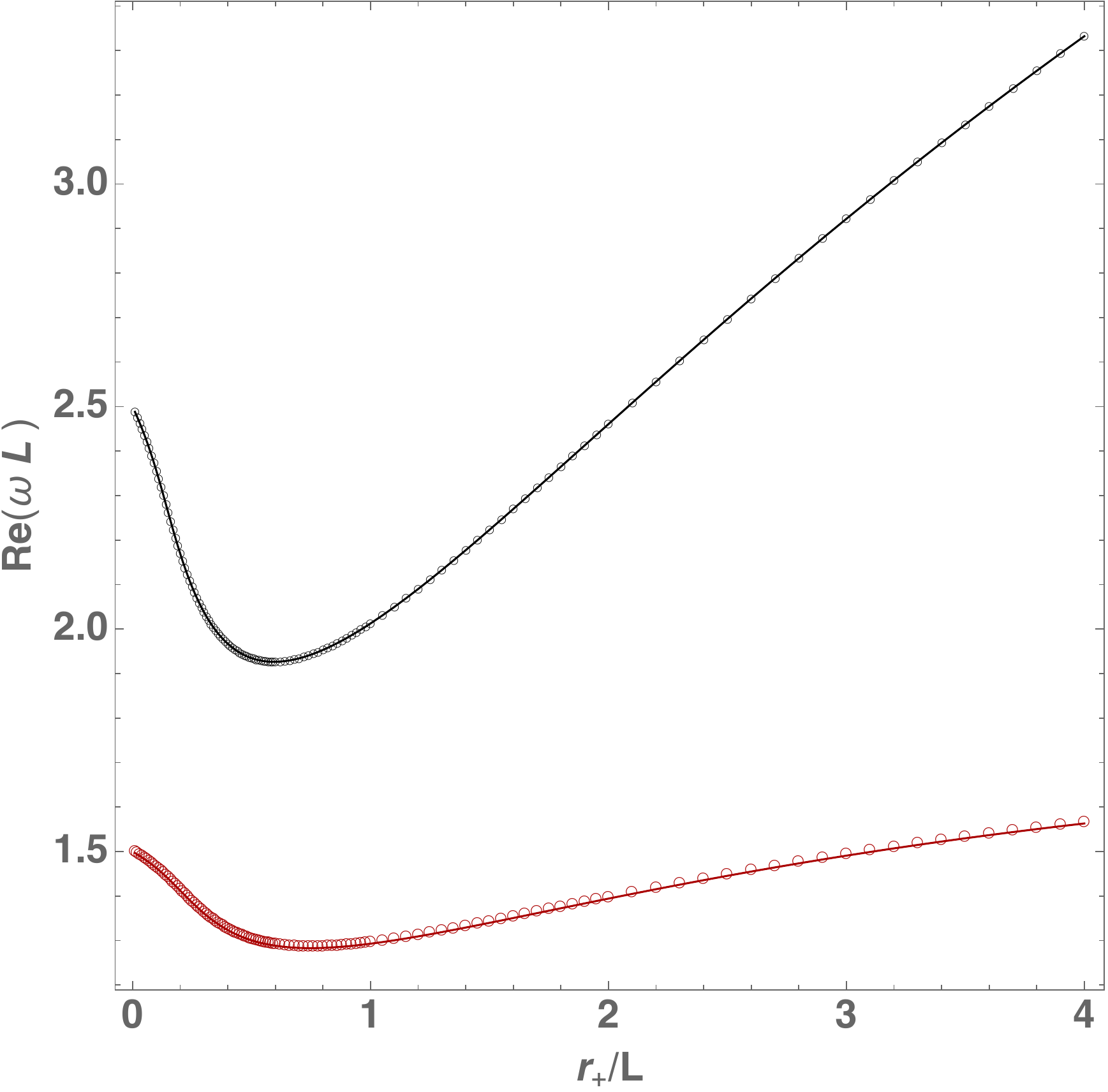}
\hspace{0.5cm}
\includegraphics[width=.46\textwidth]{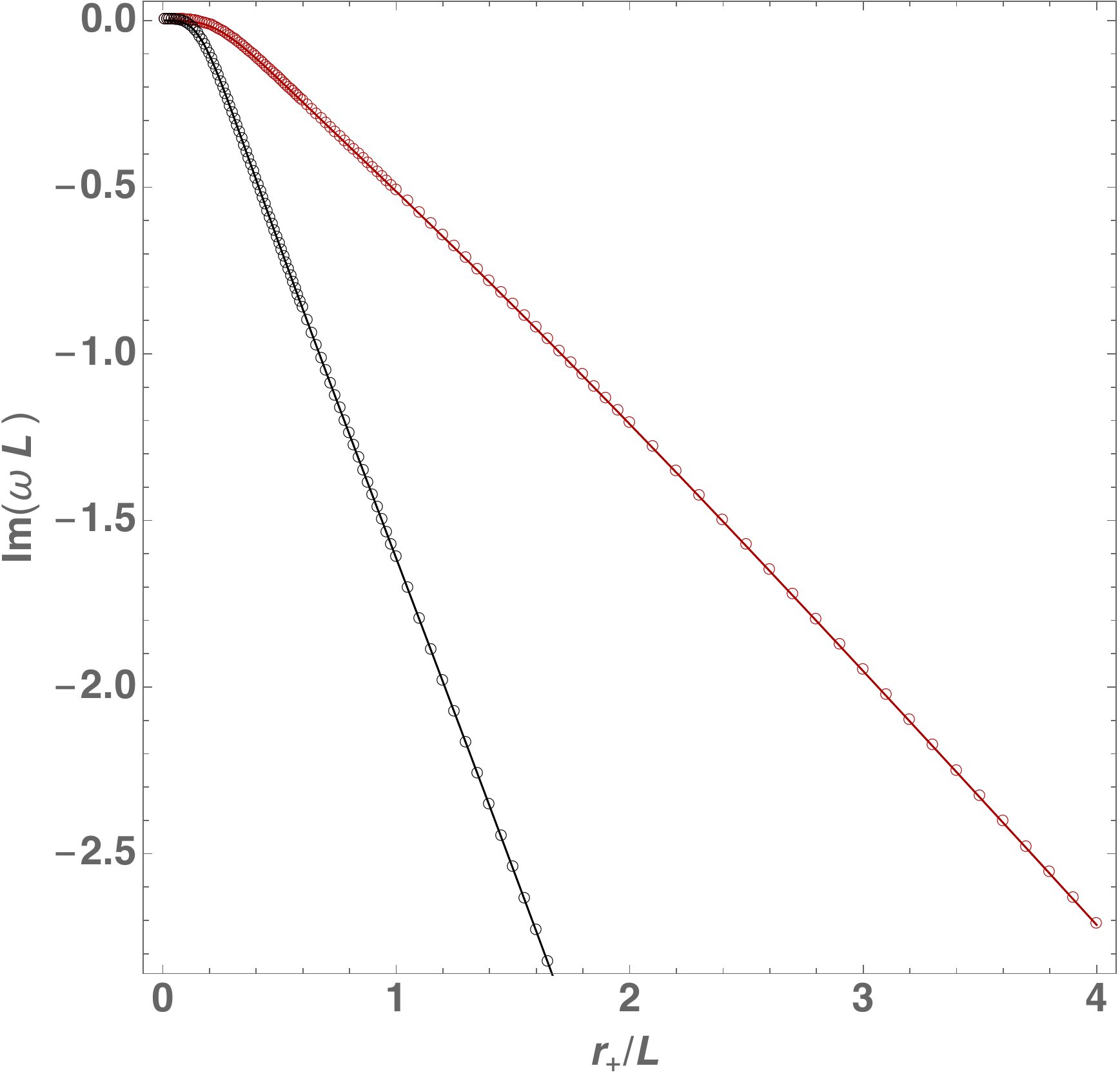}
}
\caption{Real (left) and imaginary (right) part of the QNM frequencies as a function of the horizon radius for the two families of boundary conditions in Schwarzschild-AdS background  ($m=0$, $\ell=1/2$). The black curve that reduces to the normal mode of AdS $\omega L=5/2$ has the standard quantization \eqref{StandardQuant} while the red curve (that reduces to the normal mode of AdS $\omega L=3/2$) has the alternative quantization \eqref{AlternativeQuant}. Not plotted, if we take the black curve data then $-{\rm Re}(\omega)+i\,{\rm Im}(\omega)$ is also a solution with alternative quantization \eqref{AlternativeQuant} (which reduces to the AdS normal mode $\omega L=-5/2$ when the horizon shrinks). Similarly, if we take the red curve data then $-{\rm Re}(\omega)+i\,{\rm Im}(\omega)$ is also a solution with standard quantization \eqref{StandardQuant}.}
\label{fig-D:Schw}
\end{figure}

To test our numerical code we first compute the quasinormal mode frequencies for a massless Dirac field in global AdS and in Schwarzschild-AdS: see Fig.~\ref{fig-D:Schw} for the standard \eqref{StandardQuant} and alternative \eqref{AlternativeQuant} quantizations. When $r_+=0$ our frequencies reduce to the normal mode frequencies of global AdS \eqref{normal_mode_StandardBCp} and \eqref{normal_mode_AlternativeBCm} for the standard and alternative quantizations, respectively. Recall that  $\ell$ is the spin-weighted harmonic number and $n$ is the radial overtone  (related to the number of nodes of the wavefunction along the radial direction).
So the smallest $|\omega|$ is obtained for $\ell=1/2$ and $n=0$. For the alternative boundary condition these are  $\omega L=3/2$ and $\omega L=-5/2$, while for the standard quantization these are  $\omega L=5/2$ and $\omega L=-3/2$). 
All the numerical results we will present describe solutions with $\ell=1/2$ and $n=0$. If there is an instability it should already be present in this sector of perturbations (see the argument in section~\ref{sec:DiracRN1}). 

On the other hand, for finite $r_+/L$ our numerical curves reproduce the values first computed in  \cite{Wang:2017fie}. As explained in the end of section~\ref{sec:BCsDiracAdS-RN},  this is because the AdS/CFT standard  \eqref{StandardQuant} and alternative  \eqref{AlternativeQuant} quantizations have vanishing energy flux and correspond precisely to the boundary conditions imposed in \cite{Wang:2017fie}. To complete the spectrum of Schwarzschild-AdS (and global AdS) note that frequencies that are the negative of the complex conjugate of the frequencies ($-\omega^*$) plotted in  Fig.~\ref{fig-D:Schw} are also eigenvalues of the system (which was missed in \cite{Wang:2017fie}). 

\begin{figure}[t]
\centering{
\includegraphics[width=.48\textwidth]{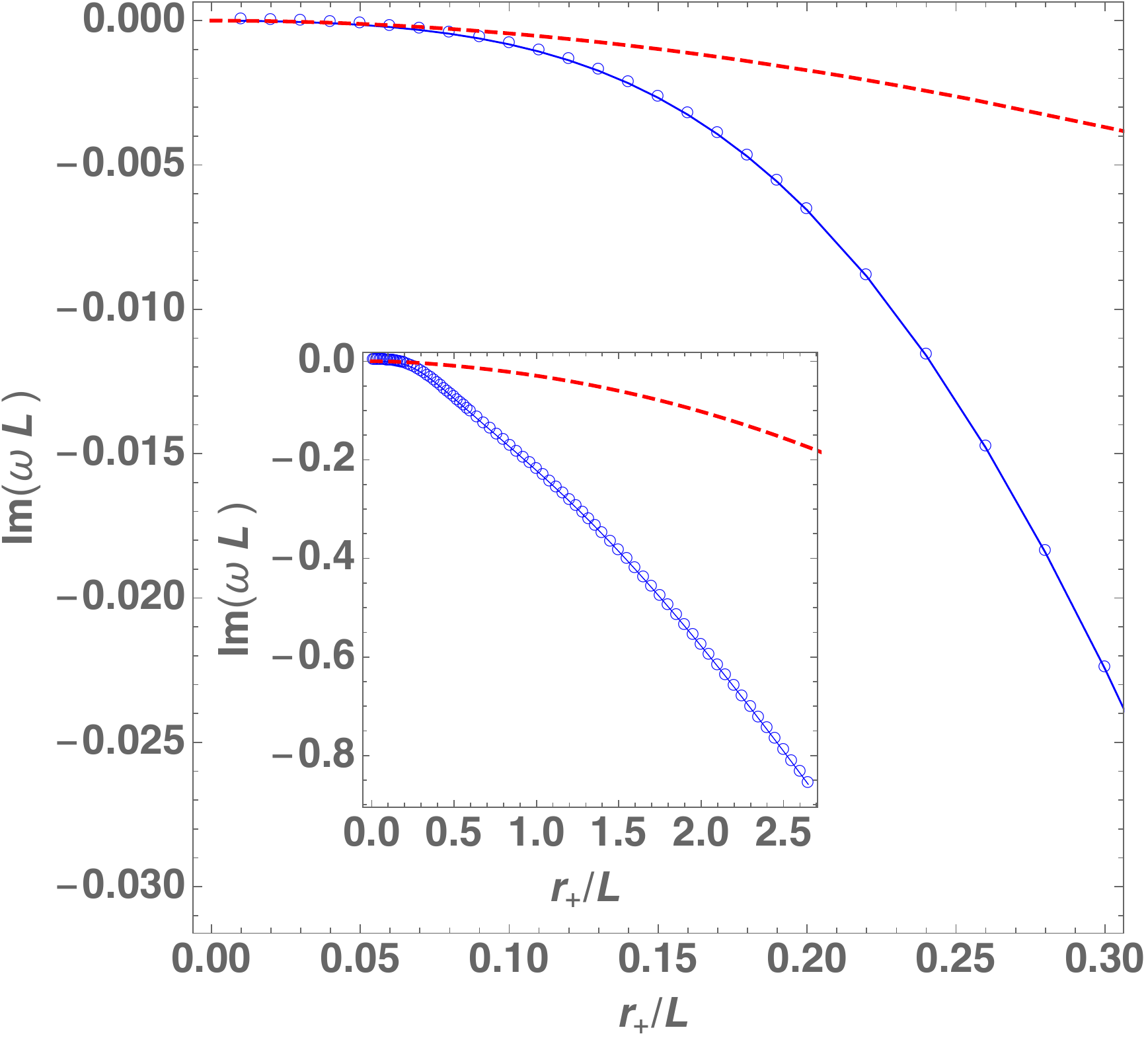}\hspace{0.5cm}}
\caption{Imaginary part of the frequency as a function of the horizon radius for fixed chemical potential $\mu=\frac{\mu_{\rm ext}}{2}$ ($m=0,\ell=1/2$ and alternative quantization). The red dashed line is the analytical approximation \eqref{deltam}. The inset plot zooms-out the main plot.}
\label{s:Numeric:fig:ImAnal}
\end{figure}

Next we test our numerical code for AdS-RN. As a first test, in Appendix \ref{secB1:matching}
we use a matching asymptotic expansion method to find that for $\omega r_+\ll 1$ and $q Q\ll 1$ (and $m=0,\ell=1/2,n=0$) the imaginary part of the frequency is approximately given by
\begin{align}
& {\rm Im}(\omega L) \approx -\frac{1}{4 \pi} \Big(\frac{r_{+} - r_{-}}{L}\Big)^{2}+{\cal O}\left( \Big(\frac{r_{+} - r_{-}}{L}\Big)^{3} \right), \qquad \hbox{{Alternative quantization}}; \label{deltam}  \\
&  {\rm Im}(\omega L) \approx -\frac{3}{4 \pi} \Big(\frac{r_{+} - r_{-}}{L}\Big)^{2}+{\cal O}\left( \Big(\frac{r_{+} - r_{-}}{L}\Big)^{3} \right), \qquad \hbox{{Standard quantization}}.\label{deltap}
\end{align}
In Fig.~\ref{s:Numeric:fig:ImAnal} we plot ${\rm Im}(\omega L)$ as a function of $r_+/L$, for $\mu=\frac{1}{2}\mu_{\rm ext}$ and a fermion charge $q L=1$ (blue dots). We compare these numerical results with the analytical result \eqref{deltam}. Both agree for small $r_+/L$, {\it i.e.} in the regime were the matching expansion \eqref{deltam} is valid.

\begin{figure}[h]
\centering{
\includegraphics[width=.45\textwidth]{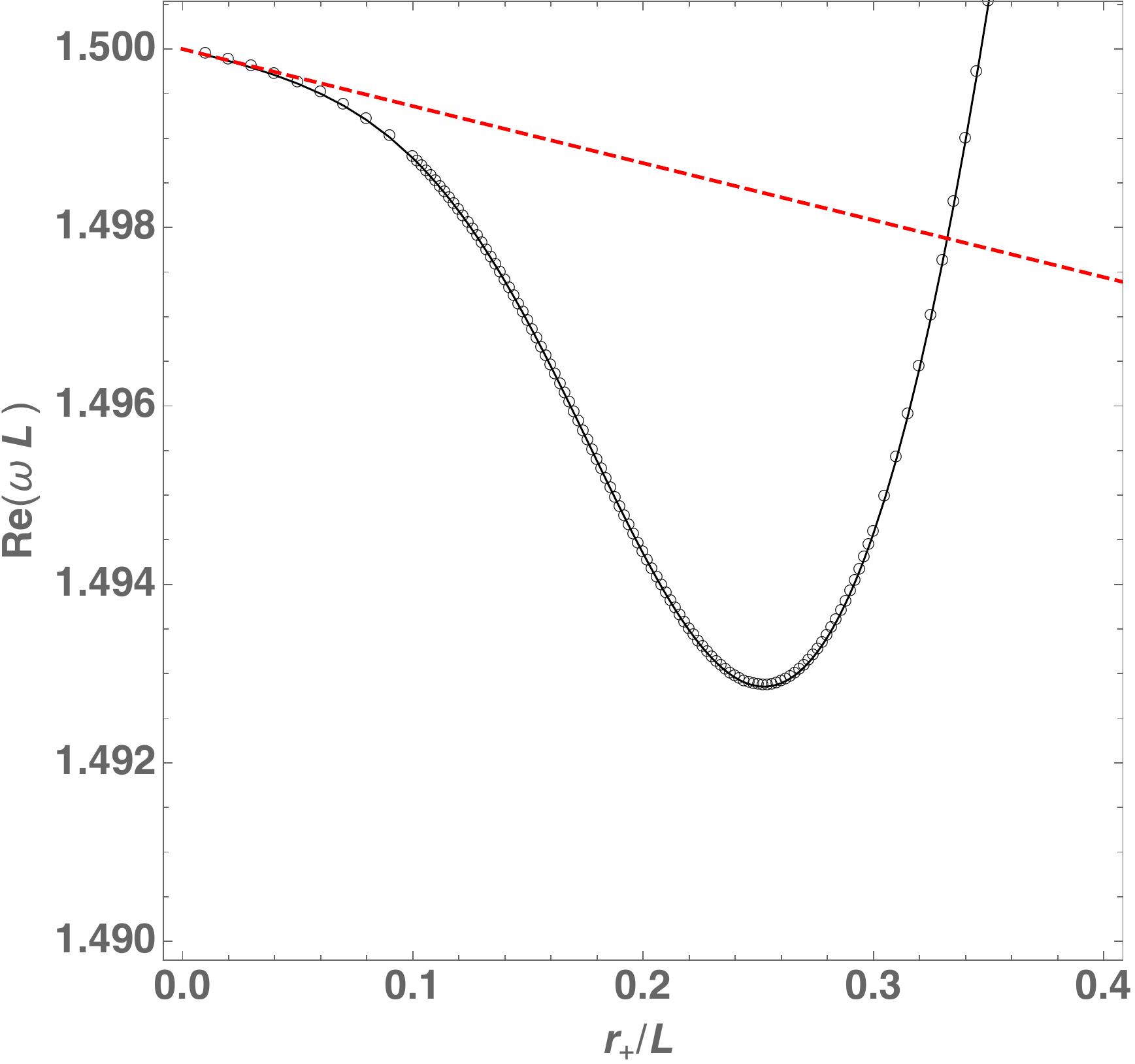}\hspace{0.3cm}
\includegraphics[width=.45 \textwidth]{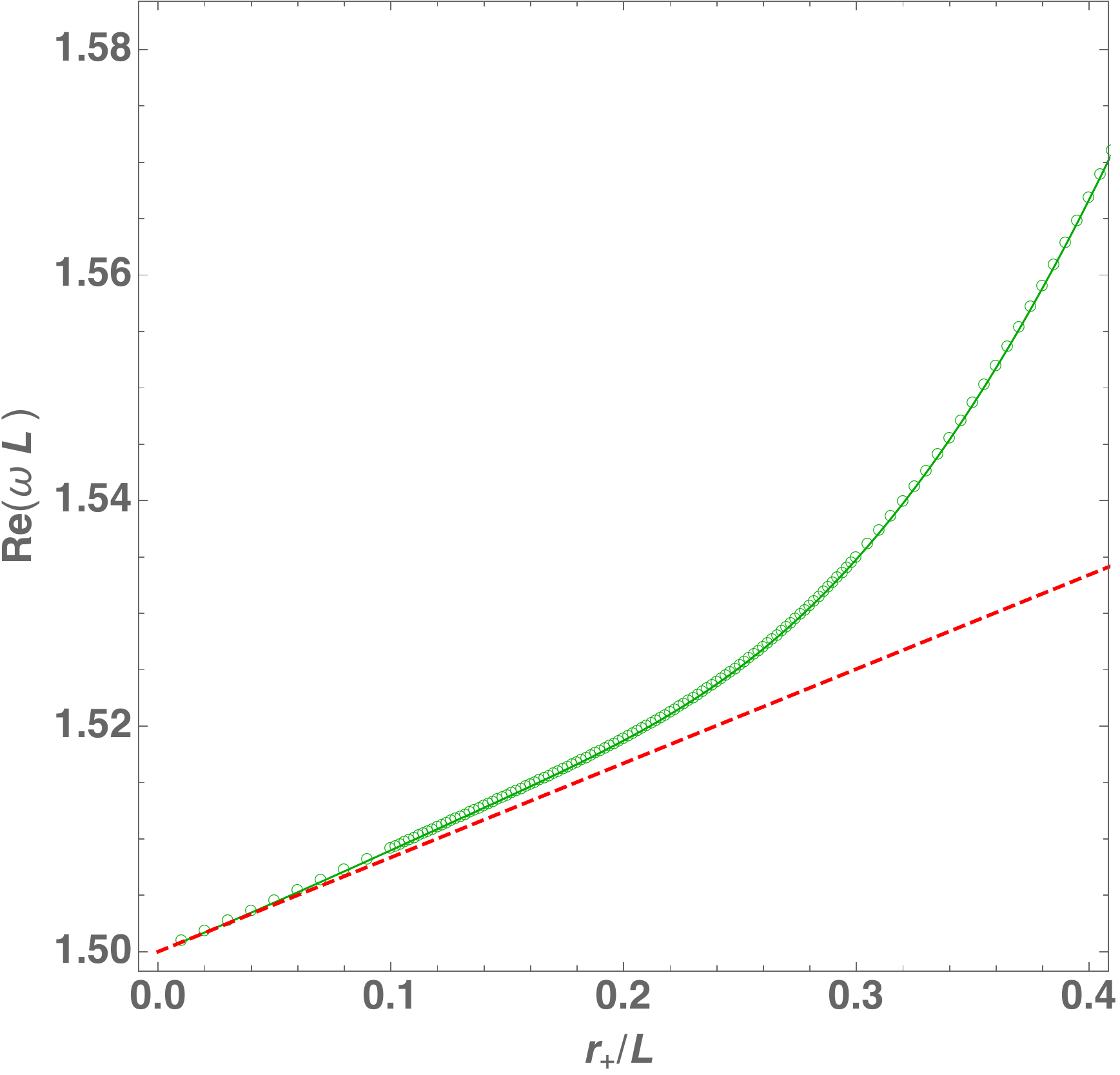}}
\centering{
\includegraphics[width=.45 \textwidth]{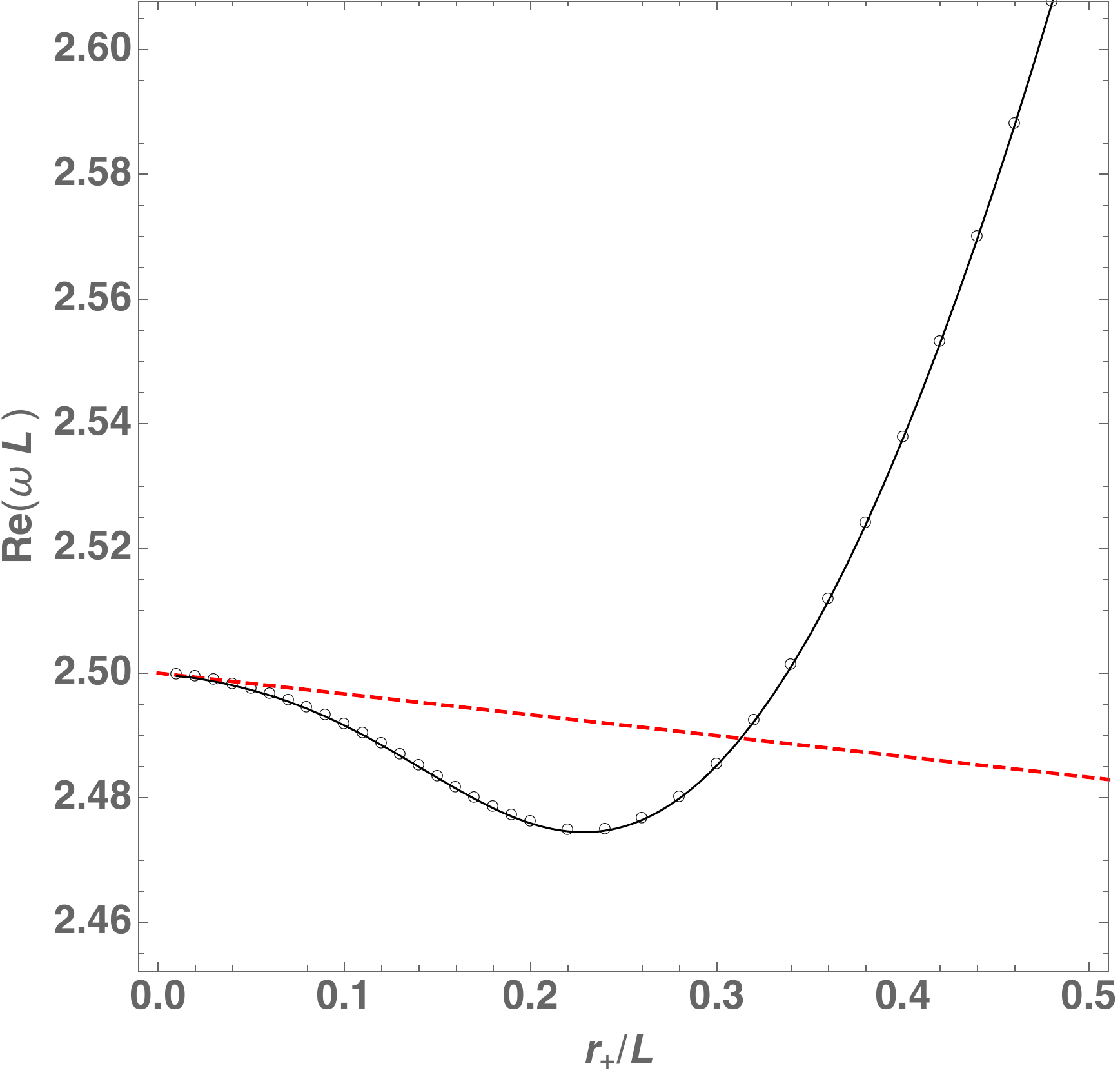}\hspace{0.3cm}
\includegraphics[width=.45 \textwidth]{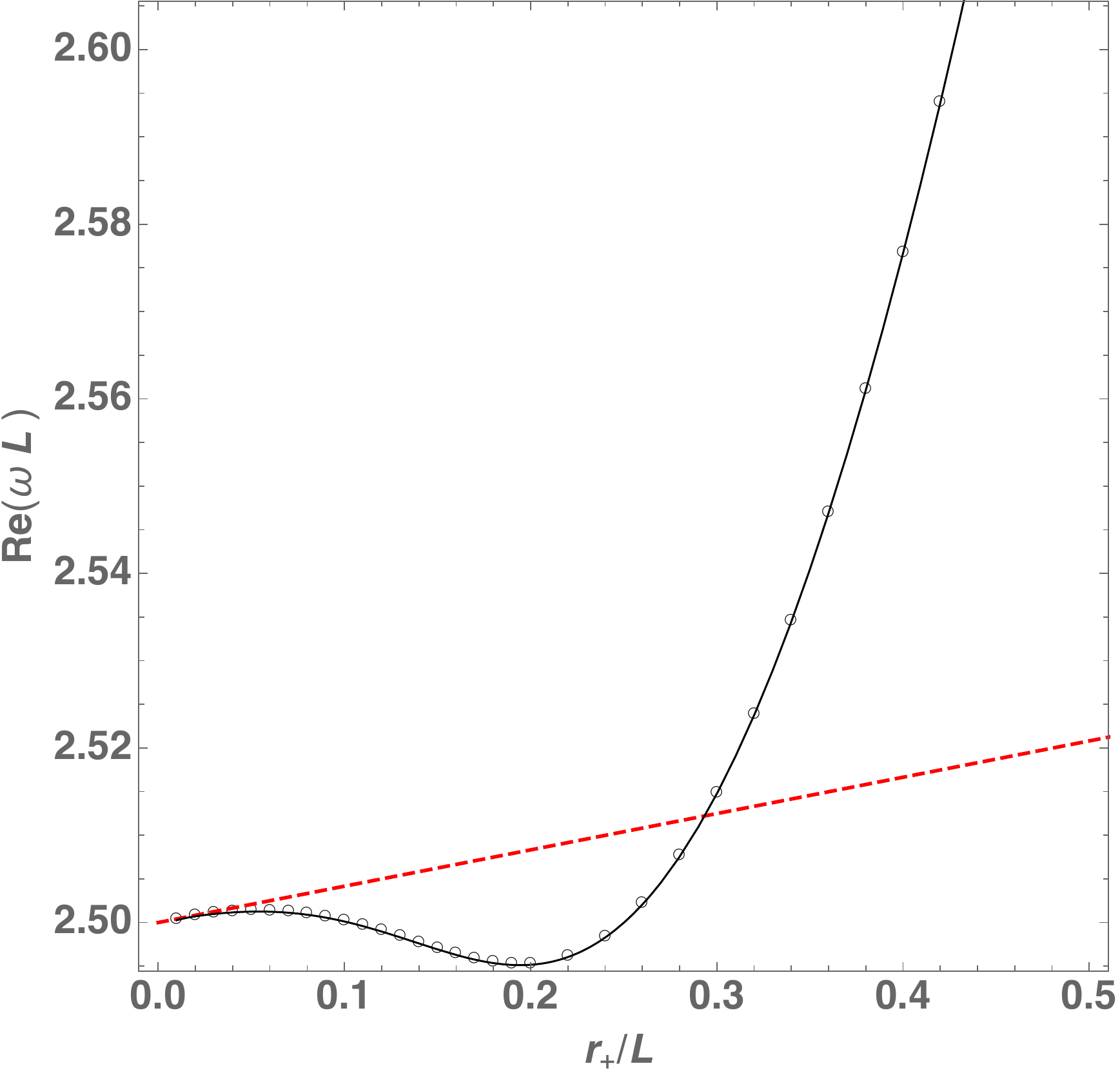}}
\caption{Real part of the frequencies as a function of the horizon radius. {\emph {Top panel}}: Alternative quantization \eqref{AlternativeQuant} and fixed chemical potential $\mu=\mu_{\rm ext}(1-10^{-2})$ and charges $q L=0.7$ (left panel) and $q L=0.8$ (right panel). The red dashed line describes the perturbative result \eqref{D:perturbative:eq:freqp}.  {\emph {Bottom}}: Standard quantization \eqref{StandardQuant} and fixed chemical potential $\mu=\frac{\mu_{\rm ext}}{2}$ and charges $q L=1.9$ (left panel) and $q L=2$ (right panel). The red dashed line describes the perturbative result \eqref{D:perturbative:eq:freqm}.}
\label{s:Numeric:fig:ReAnal}
\end{figure}

As a second test, in Appendix~\ref{secB2:perturbative} we use a perturbative expansion in $r_+/L$ about global AdS to find that the real part of the frequency behaves as
\begin{align}
& {\rm Re}(\omega L)=\frac{3}{2}-\frac{r_+}{L}\,\frac{\mu^2-4 \mu e+2}{2 \pi }+\mathcal O\left(\frac{r_+^2}{L^2}\right), \qquad \hbox{{Alternative quantization}};\label{D:perturbative:eq:freqm} \\
&  {\rm Re}(\omega L)=\frac{5}{2}+\frac{r_+}{L}\,\frac{11 \mu^2-20 \mu e+22}{6 \pi }+\mathcal O\left(\frac{r_+^2}{L^2}\right), \qquad \hbox{{Standard quantization}}.\label{D:perturbative:eq:freqp}
\end{align}

The plots for the real part of the frequency are displayed in figure \ref{s:Numeric:fig:ReAnal}. The analytical perturbative results are \eqref{D:perturbative:eq:freqm} and \eqref{D:perturbative:eq:freqp}. An important feature of this result is that for a fixed chemical potential, the slope of the real part of the frequency (the term proportional to $R_+$) changes sign for a certain electric charge. This coincides with our numerical findings. In the top panels we show the real part of the frequency for the boundary condition \eqref{AlternativeQuant} with $q L=0.7,\ 0.8$ and $\mu=\mu_{\rm ext}(1-10^{-3})$. In the bottom panels of figure \ref{s:Numeric:fig:ReAnal} we show the real part of the frequency  for the boundary condition \eqref{StandardQuant} as a function of the horizon radius for fixed chemical potential $\mu=\frac{\mu_{\rm ext}}{2}$ and fermion charges $q L=1.9,\ 2$ from left to right.  The dashed red lines are the analytical results \eqref{D:perturbative:eq:freqp} and \eqref{D:perturbative:eq:freqm} respectively. Note that for these parameters the change of sign occurs for charges $q L\sim 1.95$ and $q L\sim 0.71$ respectively.

As a general comment, it is perhaps worthy to comment that we find that matching asymptotic and perturbative analysis like the ones provided in Appendices  \ref {secB1:matching} \ref{secB2:perturbative} are valid for smaller windows of $r_+/L$ in the Dirac field case when compared with similar analysis done for the scalar field.

We have confirmed that our numerical code is generating physical data. Our main physical results are reported in section~\ref{sec:DiracRN3} of the main text.


\end{appendix}

\bibliography{refs_Spinor}{}

\providecommand{\href}[2]{#2}\begingroup\raggedright\begin{thebibliography}{10}

\bibitem{Brito:2015oca}
R.~Brito, V.~Cardoso and P.~Pani, \emph{{Superradiance}},
  \href{http://dx.doi.org/10.1007/978-3-319-19000-6}{\emph{Lect. Notes Phys.}
  {\bf 906} (2015) pp.1--237}, [\href{https://arxiv.org/abs/1501.06570}{{\tt
  1501.06570}}].

\bibitem{Bardeen:1999px}
J.~M. Bardeen and G.~T. Horowitz, \emph{{The Extreme Kerr throat geometry: A
  Vacuum analog of AdS(2) x S**2}},
  \href{http://dx.doi.org/10.1103/PhysRevD.60.104030}{\emph{Phys. Rev.} {\bf
  D60} (1999) 104030}, [\href{https://arxiv.org/abs/hep-th/9905099}{{\tt
  hep-th/9905099}}].

\bibitem{Breitenlohner:1982jf}
P.~Breitenlohner and D.~Z. Freedman, \emph{{Stability in Gauged Extended
  Supergravity}},
  \href{http://dx.doi.org/10.1016/0003-4916(82)90116-6}{\emph{Annals Phys.}
  {\bf 144} (1982) 249}.

\bibitem{Breitenlohner:1982bm}
P.~Breitenlohner and D.~Z. Freedman, \emph{{Positive Energy in anti-De Sitter
  Backgrounds and Gauged Extended Supergravity}},
  \href{http://dx.doi.org/10.1016/0370-2693(82)90643-8}{\emph{Phys. Lett.} {\bf
  115B} (1982) 197--201}.

\bibitem{Klebanov:1999tb}
I.~R. Klebanov and E.~Witten, \emph{{AdS / CFT correspondence and symmetry
  breaking}},
  \href{http://dx.doi.org/10.1016/S0550-3213(99)00387-9}{\emph{Nucl. Phys.}
  {\bf B556} (1999) 89--114}, [\href{https://arxiv.org/abs/hep-th/9905104}{{\tt
  hep-th/9905104}}].

\bibitem{Ishibashi:2004wx}
A.~Ishibashi and R.~M. Wald, \emph{{Dynamics in nonglobally hyperbolic static
  space-times. 3. Anti-de Sitter space-time}},
  \href{http://dx.doi.org/10.1088/0264-9381/21/12/012}{\emph{Class. Quant.
  Grav.} {\bf 21} (2004) 2981--3014},
  [\href{https://arxiv.org/abs/hep-th/0402184}{{\tt hep-th/0402184}}].

\bibitem{Gubser:2008px}
S.~S. Gubser, \emph{{Breaking an Abelian gauge symmetry near a black hole
  horizon}}, \href{http://dx.doi.org/10.1103/PhysRevD.78.065034}{\emph{Phys.
  Rev.} {\bf D78} (2008) 065034}, [\href{https://arxiv.org/abs/0801.2977}{{\tt
  0801.2977}}].

\bibitem{hartnoll2008building}
S.~A. Hartnoll, C.~P. Herzog and G.~T. Horowitz, \emph{Building a holographic
  superconductor}, {\emph{Physical Review Letters} {\bf 101} (2008) 031601}.

\bibitem{Faulkner:2009wj}
T.~Faulkner, H.~Liu, J.~McGreevy and D.~Vegh, \emph{{Emergent quantum
  criticality, Fermi surfaces, and AdS(2)}},
  \href{http://dx.doi.org/10.1103/PhysRevD.83.125002}{\emph{Phys. Rev.} {\bf
  D83} (2011) 125002}, [\href{https://arxiv.org/abs/0907.2694}{{\tt
  0907.2694}}].

\bibitem{Dias:2010ma}
O.~J.~C. Dias, R.~Monteiro, H.~S. Reall and J.~E. Santos, \emph{{A Scalar field
  condensation instability of rotating anti-de Sitter black holes}},
  \href{http://dx.doi.org/10.1007/JHEP11(2010)036}{\emph{JHEP} {\bf 11} (2010)
  036}, [\href{https://arxiv.org/abs/1007.3745}{{\tt 1007.3745}}].

\bibitem{dias2012hairyBHs}
O.~J. Dias, P.~Figueras, S.~Minwalla, P.~Mitra, R.~Monteiro and J.~E. Santos,
  \emph{Hairy black holes and solitons in global ads 5}, {\emph{Journal of High
  Energy Physics} {\bf 2012} (2012) 117}.

\bibitem{Dias:2016pma}
O.~J.~C. Dias and R.~Masachs, \emph{{Hairy black holes and the endpoint of
  AdS$_4$ charged superradiance}},
  \href{http://dx.doi.org/10.1007/JHEP02(2017)128}{\emph{JHEP} {\bf 02} (2017)
  128}, [\href{https://arxiv.org/abs/1610.03496}{{\tt 1610.03496}}].

\bibitem{Dias:2018zjg}
O.~J.~C. Dias and R.~Masachs, \emph{{Charged black hole bombs in a Minkowski
  cavity}}, \href{http://dx.doi.org/10.1088/1361-6382/aad70b}{\emph{Class.
  Quant. Grav.} {\bf 35} (2018) 184001},
  [\href{https://arxiv.org/abs/1801.10176}{{\tt 1801.10176}}].

\bibitem{Klein:1929zz}
O.~Klein, \emph{{Die Reflexion von Elektronen an einem Potentialsprung nach der
  relativistischen Dynamik von Dirac}},
  \href{http://dx.doi.org/10.1007/BF01339716}{\emph{Z. Phys.} {\bf 53} (1929)
  157}.

\bibitem{Sauter:1931zz}
F.~Sauter, \emph{{Uber das Verhalten eines Elektrons im homogenen elektrischen
  Feld nach der relativistischen Theorie Diracs}},
  \href{http://dx.doi.org/10.1007/BF01339461}{\emph{Z. Phys.} {\bf 69} (1931)
  742--764}.

\bibitem{Winter:1959}
R.~G. Winter, \emph{{Klein Paradox for the Klein-Gordon Equation}},
  \href{http://dx.doi.org/10.1119/1.1934851}{\emph{American Journal of Physics}
  {\bf 27} (1959) 355}.

\bibitem{Unruh:1973bda}
W.~Unruh, \emph{{Separability of the Neutrino Equations in a Kerr Background}},
  \href{http://dx.doi.org/10.1103/PhysRevLett.31.1265}{\emph{Phys. Rev. Lett.}
  {\bf 31} (1973) 1265--1267}.

\bibitem{Holstein:1998}
B.~R. Holstein, \emph{{Klein's Paradox}},
  \href{http://dx.doi.org/10.1119/1.18891}{\emph{American Journal of Physics}
  {\bf 66} (1998) 507}.

\bibitem{Iqbal:2009fd}
N.~Iqbal and H.~Liu, \emph{{Real-time response in AdS/CFT with application to
  spinors}}, \href{http://dx.doi.org/10.1002/prop.200900057}{\emph{Fortsch.
  Phys.} {\bf 57} (2009) 367--384},
  [\href{https://arxiv.org/abs/0903.2596}{{\tt 0903.2596}}].

\bibitem{Lee:2008xf}
S.-S. Lee, \emph{{A Non-Fermi Liquid from a Charged Black Hole: A Critical
  Fermi Ball}}, \href{http://dx.doi.org/10.1103/PhysRevD.79.086006}{\emph{Phys.
  Rev.} {\bf D79} (2009) 086006}, [\href{https://arxiv.org/abs/0809.3402}{{\tt
  0809.3402}}].

\bibitem{Liu:2009dm}
H.~Liu, J.~McGreevy and D.~Vegh, \emph{{Non-Fermi liquids from holography}},
  \href{http://dx.doi.org/10.1103/PhysRevD.83.065029}{\emph{Phys. Rev.} {\bf
  D83} (2011) 065029}, [\href{https://arxiv.org/abs/0903.2477}{{\tt
  0903.2477}}].

\bibitem{Guarrera:2011my}
D.~Guarrera and J.~McGreevy, \emph{{Holographic Fermi surfaces and bulk dipole
  couplings}},  \href{https://arxiv.org/abs/1102.3908}{{\tt 1102.3908}}.

\bibitem{Iqbal:2011ae}
N.~Iqbal, H.~Liu and M.~Mezei, \emph{{Lectures on holographic non-Fermi liquids
  and quantum phase transitions}},  in \emph{{Proceedings, Theoretical Advanced
  Study Institute in Elementary Particle Physics (TASI 2010). String Theory and
  Its Applications: From meV to the Planck Scale: Boulder, Colorado, USA, June
  1-25, 2010}}, pp.~707--816, 2011.
\newblock \href{https://arxiv.org/abs/1110.3814}{{\tt 1110.3814}}.
\newblock \href{http://dx.doi.org/10.1142/9789814350525_0013}{DOI}.

\bibitem{Hartnoll:2016apf}
S.~A. Hartnoll, A.~Lucas and S.~Sachdev, \emph{{Holographic quantum matter}},
  \href{https://arxiv.org/abs/1612.07324}{{\tt 1612.07324}}.

\bibitem{Cubrovic:2009ye}
M.~Cubrovic, J.~Zaanen and K.~Schalm, \emph{{String Theory, Quantum Phase
  Transitions and the Emergent Fermi-Liquid}},
  \href{http://dx.doi.org/10.1126/science.1174962}{\emph{Science} {\bf 325}
  (2009) 439--444}, [\href{https://arxiv.org/abs/0904.1993}{{\tt 0904.1993}}].

\bibitem{Cardoso:2017soq}
V.~Cardoso, J.~a.~L. Costa, K.~Destounis, P.~Hintz and A.~Jansen,
  \emph{{Quasinormal modes and Strong Cosmic Censorship}},
  \href{http://dx.doi.org/10.1103/PhysRevLett.120.031103}{\emph{Phys. Rev.
  Lett.} {\bf 120} (2018) 031103},
  [\href{https://arxiv.org/abs/1711.10502}{{\tt 1711.10502}}].

\bibitem{Dias:2018ynt}
O.~J.~C. Dias, F.~C. Eperon, H.~S. Reall and J.~E. Santos, \emph{{Strong cosmic
  censorship in de Sitter space}},
  \href{http://dx.doi.org/10.1103/PhysRevD.97.104060}{\emph{Phys. Rev.} {\bf
  D97} (2018) 104060}, [\href{https://arxiv.org/abs/1801.09694}{{\tt
  1801.09694}}].

\bibitem{Dias:2018etb}
O.~J.~C. Dias, H.~S. Reall and J.~E. Santos, \emph{{Strong cosmic censorship:
  taking the rough with the smooth}},
  \href{http://dx.doi.org/10.1007/JHEP10(2018)001}{\emph{JHEP} {\bf 10} (2018)
  001}, [\href{https://arxiv.org/abs/1808.02895}{{\tt 1808.02895}}].

\bibitem{Wang:2017fie}
M.~Wang, C.~Herdeiro and J.~Jing, \emph{{Dirac perturbations on
  Schwarzschild--anti--de Sitter spacetimes: Generic boundary conditions and
  new quasinormal modes}},
  \href{http://dx.doi.org/10.1103/PhysRevD.96.104035}{\emph{Phys. Rev.} {\bf
  D96} (2017) 104035}, [\href{https://arxiv.org/abs/1710.10461}{{\tt
  1710.10461}}].

\bibitem{Wang:2019qja}
M.~Wang, C.~Herdeiro and J.~Jing, \emph{{Charged Dirac perturbations on
  Reissner-Nordstr\"om-Anti-de Sitter spacetimes: quasinormal modes with Robin
  boundary conditions}},  \href{https://arxiv.org/abs/1910.14305}{{\tt
  1910.14305}}.

\bibitem{Mueck:1998iz}
W.~Mueck and K.~S. Viswanathan, \emph{{Conformal field theory correlators from
  classical field theory on anti-de Sitter space. 2. Vector and spinor
  fields}}, \href{http://dx.doi.org/10.1103/PhysRevD.58.106006}{\emph{Phys.
  Rev.} {\bf D58} (1998) 106006},
  [\href{https://arxiv.org/abs/hep-th/9805145}{{\tt hep-th/9805145}}].

\bibitem{Henningson:1998cd}
M.~Henningson and K.~Sfetsos, \emph{{Spinors and the AdS / CFT
  correspondence}},
  \href{http://dx.doi.org/10.1016/S0370-2693(98)00559-0}{\emph{Phys. Lett.}
  {\bf B431} (1998) 63--68}, [\href{https://arxiv.org/abs/hep-th/9803251}{{\tt
  hep-th/9803251}}].

\bibitem{Henneaux:1998ch}
M.~Henneaux, \emph{{Boundary terms in the AdS / CFT correspondence for spinor
  fields}},  in \emph{{Mathematical methods in modern theoretical physics.
  Proceedings, International Meeting, School and Workshop, ISPM'98, Tbilisi,
  Georgia, September 5-18, 1998}}, pp.~161--170, 1998.
\newblock \href{https://arxiv.org/abs/hep-th/9902137}{{\tt hep-th/9902137}}.

\bibitem{Contino:2004vy}
R.~Contino and A.~Pomarol, \emph{{Holography for fermions}},
  \href{http://dx.doi.org/10.1088/1126-6708/2004/11/058}{\emph{JHEP} {\bf 11}
  (2004) 058}, [\href{https://arxiv.org/abs/hep-th/0406257}{{\tt
  hep-th/0406257}}].

\bibitem{Amsel:2008iz}
A.~J. Amsel and D.~Marolf, \emph{{Supersymmetric Multi-trace Boundary
  Conditions in AdS}},
  \href{http://dx.doi.org/10.1088/0264-9381/26/2/025010}{\emph{Class. Quant.
  Grav.} {\bf 26} (2009) 025010}, [\href{https://arxiv.org/abs/0808.2184}{{\tt
  0808.2184}}].

\bibitem{Amsel:2009rr}
A.~J. Amsel and G.~Compere, \emph{{Supergravity at the boundary of AdS
  supergravity}},
  \href{http://dx.doi.org/10.1103/PhysRevD.79.085006}{\emph{Phys. Rev.} {\bf
  D79} (2009) 085006}, [\href{https://arxiv.org/abs/0901.3609}{{\tt
  0901.3609}}].

\bibitem{Andrade:2011dg}
T.~Andrade and D.~Marolf, \emph{{AdS/CFT beyond the unitarity bound}},
  \href{http://dx.doi.org/10.1007/JHEP01(2012)049}{\emph{JHEP} {\bf 01} (2012)
  049}, [\href{https://arxiv.org/abs/1105.6337}{{\tt 1105.6337}}].

\bibitem{Michalogiorgakis:2006jc}
G.~Michalogiorgakis and S.~S. Pufu, \emph{{Low-lying gravitational modes in the
  scalar sector of the global AdS(4) black hole}},
  \href{http://dx.doi.org/10.1088/1126-6708/2007/02/023}{\emph{JHEP} {\bf 02}
  (2007) 023}, [\href{https://arxiv.org/abs/hep-th/0612065}{{\tt
  hep-th/0612065}}].

\bibitem{Dias:2013sdc}
O.~J.~C. Dias and J.~E. Santos, \emph{{Boundary Conditions for Kerr-AdS
  Perturbations}}, \href{http://dx.doi.org/10.1007/JHEP10(2013)156}{\emph{JHEP}
  {\bf 10} (2013) 156}, [\href{https://arxiv.org/abs/1302.1580}{{\tt
  1302.1580}}].

\bibitem{Cardoso:2013pza}
V.~Cardoso, O.~J.~C. Dias, G.~S. Hartnett, L.~Lehner and J.~E. Santos,
  \emph{{Holographic thermalization, quasinormal modes and superradiance in
  Kerr-AdS}}, \href{http://dx.doi.org/10.1007/JHEP04(2014)183}{\emph{JHEP} {\bf
  04} (2014) 183}, [\href{https://arxiv.org/abs/1312.5323}{{\tt 1312.5323}}].

\bibitem{Giammatteo:2004wp}
M.~Giammatteo and J.-l. Jing, \emph{{Dirac quasinormal frequencies in
  Schwarzschild-AdS space-time}},
  \href{http://dx.doi.org/10.1103/PhysRevD.71.024007}{\emph{Phys. Rev.} {\bf
  D71} (2005) 024007}, [\href{https://arxiv.org/abs/gr-qc/0403030}{{\tt
  gr-qc/0403030}}].

\bibitem{Jing:2005ux}
J.-l. Jing, \emph{{Quasinormal modes of Dirac field perturbation in
  Schwarzschild-anti-de Sitter black hole}},
  \href{https://arxiv.org/abs/gr-qc/0502010}{{\tt gr-qc/0502010}}.

\bibitem{Birrell:1982ix}
N.~D. Birrell and P.~C.~W. Davies, \emph{{Quantum Fields in Curved Space}}.
\newblock Cambridge Monographs on Mathematical Physics. Cambridge Univ. Press,
  Cambridge, UK, 1984,
  \href{http://dx.doi.org/10.1017/CBO9780511622632}{10.1017/CBO9780511622632}.

\bibitem{Tong:2007}
D.~Tong, \emph{{Lectures on Quantum Field Theory (Part III)}}.
\newblock Cambridge Univ., Cambridge, UK, 2007.

\bibitem{Pollock:2010zz}
M.~D. Pollock, \emph{{On the Dirac equation in curved space-time}}, {\emph{Acta
  Phys. Polon.} {\bf B41} (2010) 1827--1846}.

\bibitem{Yepez:2011bw}
J.~Yepez, \emph{{Einstein's vierbein field theory of curved space}},
  \href{https://arxiv.org/abs/1106.2037}{{\tt 1106.2037}}.

\bibitem{Cotaescu:2003be}
I.~I. Cotaescu, \emph{{Discrete quantum modes of the Dirac field in AdS (d+1)
  backgrounds}}, \href{http://dx.doi.org/10.1142/S0217751X04017690}{\emph{Int.
  J. Mod. Phys.} {\bf A19} (2004) 2217--2232},
  [\href{https://arxiv.org/abs/gr-qc/0306127}{{\tt gr-qc/0306127}}].

\bibitem{McKellar:1993ej}
B.~H.~J. McKellar, M.~J. Thomson and G.~J. Stephenson, \emph{{The Dirac
  equation in Kerr space-time, spheroidal coordinates and the MIT bag model of
  hadrons}}, \href{http://dx.doi.org/10.1088/0305-4470/26/14/025}{\emph{J.
  Phys.} {\bf A26} (1993) 3649--3657}.

\bibitem{Dolan:2015eua}
S.~R. Dolan and D.~Dempsey, \emph{{Bound states of the Dirac equation on Kerr
  spacetime}},
  \href{http://dx.doi.org/10.1088/0264-9381/32/18/184001}{\emph{Class. Quant.
  Grav.} {\bf 32} (2015) 184001}, [\href{https://arxiv.org/abs/1504.03190}{{\tt
  1504.03190}}].

\bibitem{Dias:2011tj}
O.~J.~C. Dias, P.~Figueras, S.~Minwalla, P.~Mitra, R.~Monteiro and J.~E.
  Santos, \emph{{Hairy black holes and solitons in global $AdS_5$}},
  \href{http://dx.doi.org/10.1007/JHEP08(2012)117}{\emph{JHEP} {\bf 08} (2012)
  117}, [\href{https://arxiv.org/abs/1112.4447}{{\tt 1112.4447}}].

\bibitem{Wang:2000gsa}
B.~Wang, C.-Y. Lin and E.~Abdalla, \emph{{Quasinormal modes of
  Reissner-Nordstrom anti-de Sitter black holes}},
  \href{http://dx.doi.org/10.1016/S0370-2693(00)00409-3}{\emph{Phys. Lett.}
  {\bf B481} (2000) 79--88}, [\href{https://arxiv.org/abs/hep-th/0003295}{{\tt
  hep-th/0003295}}].

\bibitem{Berti:2003ud}
E.~Berti and K.~D. Kokkotas, \emph{{Quasinormal modes of
  Reissner-Nordstrom-anti-de Sitter black holes: Scalar, electromagnetic and
  gravitational perturbations}},
  \href{http://dx.doi.org/10.1103/PhysRevD.67.064020}{\emph{Phys. Rev.} {\bf
  D67} (2003) 064020}, [\href{https://arxiv.org/abs/gr-qc/0301052}{{\tt
  gr-qc/0301052}}].

\bibitem{Uchikata:2011zz}
N.~Uchikata and S.~Yoshida, \emph{{Quasinormal modes of a massless charged
  scalar field on a small Reissner-Nordstrom-anti-de Sitter black hole}},
  \href{http://dx.doi.org/10.1103/PhysRevD.83.064020}{\emph{Phys. Rev.} {\bf
  D83} (2011) 064020}, [\href{https://arxiv.org/abs/1109.6737}{{\tt
  1109.6737}}].

\bibitem{Dias:2018ufh}
O.~J.~C. Dias, H.~S. Reall and J.~E. Santos, \emph{{Strong cosmic censorship
  for charged de Sitter black holes with a charged scalar field}},
  \href{http://dx.doi.org/10.1088/1361-6382/aafcf2}{\emph{Class. Quant. Grav.}
  {\bf 36} (2019) 045005}, [\href{https://arxiv.org/abs/1808.04832}{{\tt
  1808.04832}}].

\bibitem{Basu:2010uz}
P.~Basu, J.~Bhattacharya, S.~Bhattacharyya, R.~Loganayagam, S.~Minwalla and
  V.~Umesh, \emph{{Small Hairy Black Holes in Global AdS Spacetime}},
  \href{http://dx.doi.org/10.1007/JHEP10(2010)045}{\emph{JHEP} {\bf 10} (2010)
  045}, [\href{https://arxiv.org/abs/1003.3232}{{\tt 1003.3232}}].

\bibitem{Arias:2016aig}
R.~Arias, J.~Mas and A.~Serantes, \emph{{Stability of charged global AdS$_{4}$
  spacetimes}}, \href{http://dx.doi.org/10.1007/JHEP09(2016)024}{\emph{JHEP}
  {\bf 09} (2016) 024}, [\href{https://arxiv.org/abs/1606.00830}{{\tt
  1606.00830}}].

\bibitem{lopez2006Dirac_dS}
A.~Lopez-Ortega, \emph{Absorption and quasinormal modes of classical fields
  propagating on 3d and 4d de sitter spacetime}, {\emph{General Relativity and
  Gravitation} {\bf 38} (2006) 743}.

\bibitem{1965handbook}
M.~Abramowitz and I.~A. Stegun, \emph{Handbook of mathematical functions: with
  formulas, graphs, and mathematical tables}, vol.~55.
\newblock Courier Corporation, 1965.

\bibitem{Cotaescu:1998ts}
I.~I. Cotaescu, \emph{{The Dirac particle on central backgrounds and the
  anti-de Sitter oscillator}},
  \href{http://dx.doi.org/10.1142/S0217732398003107}{\emph{Mod. Phys. Lett.}
  {\bf A13} (1998) 2923--2936},
  [\href{https://arxiv.org/abs/gr-qc/9803042}{{\tt gr-qc/9803042}}].

\bibitem{BVR2009real}
B.~C. van Rees, \emph{Real-time gauge/gravity duality and ingoing boundary
  conditions}, {\emph{Nuclear Physics B-Proceedings Supplements} {\bf 192}
  (2009) 193--196}.

\bibitem{skenderisBVR2009realholog}
K.~Skenderis and B.~C. van Rees, \emph{Real-time gauge/gravity duality:
  Prescription, renormalization and examples}, {\emph{Journal of High Energy
  Physics} {\bf 2009} (2009) 085}.

\bibitem{Dias:2009iu}
O.~J.~C. Dias, P.~Figueras, R.~Monteiro, J.~E. Santos and R.~Emparan,
  \emph{{Instability and new phases of higher-dimensional rotating black
  holes}}, \href{http://dx.doi.org/10.1103/PhysRevD.80.111701}{\emph{Phys.
  Rev.} {\bf D80} (2009) 111701}, [\href{https://arxiv.org/abs/0907.2248}{{\tt
  0907.2248}}].

\bibitem{Dias:2010maa}
O.~J.~C. Dias, P.~Figueras, R.~Monteiro and J.~E. Santos, \emph{{Ultraspinning
  instability of rotating black holes}},
  \href{http://dx.doi.org/10.1103/PhysRevD.82.104025}{\emph{Phys. Rev.} {\bf
  D82} (2010) 104025}, [\href{https://arxiv.org/abs/1006.1904}{{\tt
  1006.1904}}].

\bibitem{Dias:2010eu}
O.~J.~C. Dias, P.~Figueras, R.~Monteiro, H.~S. Reall and J.~E. Santos,
  \emph{{An instability of higher-dimensional rotating black holes}},
  \href{http://dx.doi.org/10.1007/JHEP05(2010)076}{\emph{JHEP} {\bf 05} (2010)
  076}, [\href{https://arxiv.org/abs/1001.4527}{{\tt 1001.4527}}].

\bibitem{Dias:2010gk}
O.~J.~C. Dias, P.~Figueras, R.~Monteiro and J.~E. Santos, \emph{{Ultraspinning
  instability of anti-de Sitter black holes}},
  \href{http://dx.doi.org/10.1007/JHEP12(2010)067}{\emph{JHEP} {\bf 12} (2010)
  067}, [\href{https://arxiv.org/abs/1011.0996}{{\tt 1011.0996}}].

\bibitem{Dias:2011jg}
O.~J.~C. Dias, R.~Monteiro and J.~E. Santos, \emph{{Ultraspinning instability:
  the missing link}},
  \href{http://dx.doi.org/10.1007/JHEP08(2011)139}{\emph{JHEP} {\bf 08} (2011)
  139}, [\href{https://arxiv.org/abs/1106.4554}{{\tt 1106.4554}}].

\bibitem{Dias:2014eua}
O.~J.~C. Dias, G.~S. Hartnett and J.~E. Santos, \emph{{Quasinormal modes of
  asymptotically flat rotating black holes}},
  \href{http://dx.doi.org/10.1088/0264-9381/31/24/245011}{\emph{Class. Quant.
  Grav.} {\bf 31} (2014) 245011}, [\href{https://arxiv.org/abs/1402.7047}{{\tt
  1402.7047}}].

\bibitem{Dias:2015wqa}
O.~J.~C. Dias, M.~Godazgar and J.~E. Santos, \emph{{Linear Mode Stability of
  the Kerr-Newman Black Hole and Its Quasinormal Modes}},
  \href{http://dx.doi.org/10.1103/PhysRevLett.114.151101}{\emph{Phys. Rev.
  Lett.} {\bf 114} (2015) 151101},
  [\href{https://arxiv.org/abs/1501.04625}{{\tt 1501.04625}}].

\bibitem{Dias:2015nua}
O.~J.~C. Dias, J.~E. Santos and B.~Way, \emph{{Numerical Methods for Finding
  Stationary Gravitational Solutions}},
  \href{http://dx.doi.org/10.1088/0264-9381/33/13/133001}{\emph{Class. Quant.
  Grav.} {\bf 33} (2016) 133001}, [\href{https://arxiv.org/abs/1510.02804}{{\tt
  1510.02804}}].

\bibitem{Murata:2010dx}
K.~Murata, S.~Kinoshita and N.~Tanahashi, \emph{{Non-equilibrium Condensation
  Process in a Holographic Superconductor}},
  \href{http://dx.doi.org/10.1007/JHEP07(2010)050}{\emph{JHEP} {\bf 07} (2010)
  050}, [\href{https://arxiv.org/abs/1005.0633}{{\tt 1005.0633}}].

\bibitem{Bosch:2016vcp}
P.~Bosch, S.~R. Green and L.~Lehner, \emph{{Nonlinear Evolution and Final Fate
  of Charged Anti-de Sitter Black Hole Superradiant Instability}},
  \href{http://dx.doi.org/10.1103/PhysRevLett.116.141102}{\emph{Phys. Rev.
  Lett.} {\bf 116} (2016) 141102},
  [\href{https://arxiv.org/abs/1601.01384}{{\tt 1601.01384}}].

\bibitem{Wald:106274}
R.~M. Wald, \emph{{General relativity}}.
\newblock Chicago Univ. Press, Chicago, IL, 1984.

\bibitem{Hartnoll:2010gu}
S.~A. Hartnoll and A.~Tavanfar, \emph{{Electron stars for holographic metallic
  criticality}},
  \href{http://dx.doi.org/10.1103/PhysRevD.83.046003}{\emph{Phys. Rev.} {\bf
  D83} (2011) 046003}, [\href{https://arxiv.org/abs/1008.2828}{{\tt
  1008.2828}}].

\bibitem{Allais:2013lha}
A.~Allais and J.~McGreevy, \emph{{How to construct a gravitating quantum
  electron star}},
  \href{http://dx.doi.org/10.1103/PhysRevD.88.066006}{\emph{Phys. Rev.} {\bf
  D88} (2013) 066006}, [\href{https://arxiv.org/abs/1306.6075}{{\tt
  1306.6075}}].

\bibitem{Starobinsky:1973aij}
A.~A. Starobinsky, \emph{{Amplification of waves reflected from a rotating
  "black hole".}}, {\emph{Sov. Phys. JETP} {\bf 37} (1973) 28--32}.

\bibitem{PhysRevUnruh}
W.~G. Unruh, \emph{Absorption cross section of small black holes},
  \href{http://dx.doi.org/10.1103/PhysRevD.14.3251}{\emph{Phys. Rev. D} {\bf
  14} (Dec, 1976) 3251--3259}.

\bibitem{maldacena1997matching}
J.~Maldacena and A.~Strominger, \emph{Universal low-energy dynamics for
  rotating black holes}, {\emph{Physical Review D} {\bf 56} (1997) 4975}.

\bibitem{Cardoso:2004nk}
V.~Cardoso, O.~J.~C. Dias, J.~P.~S. Lemos and S.~Yoshida, \emph{{The Black hole
  bomb and superradiant instabilities}},
  \href{http://dx.doi.org/10.1103/PhysRevD.70.049903,
  10.1103/PhysRevD.70.044039}{\emph{Phys. Rev.} {\bf D70} (2004) 044039},
  [\href{https://arxiv.org/abs/hep-th/0404096}{{\tt hep-th/0404096}}].

\bibitem{Cardoso:2004hs}
V.~Cardoso and O.~J.~C. Dias, \emph{{Small Kerr-anti-de Sitter black holes are
  unstable}}, \href{http://dx.doi.org/10.1103/PhysRevD.70.084011}{\emph{Phys.
  Rev.} {\bf D70} (2004) 084011},
  [\href{https://arxiv.org/abs/hep-th/0405006}{{\tt hep-th/0405006}}].

\bibitem{Bhattacharyya:2010yg}
S.~Bhattacharyya, S.~Minwalla and K.~Papadodimas, \emph{{Small Hairy Black
  Holes in $AdS_5 x S^5$}},
  \href{http://dx.doi.org/10.1007/JHEP11(2011)035}{\emph{JHEP} {\bf 11} (2011)
  035}, [\href{https://arxiv.org/abs/1005.1287}{{\tt 1005.1287}}].

\bibitem{Dias:2011at}
O.~J.~C. Dias, G.~T. Horowitz and J.~E. Santos, \emph{{Black holes with only
  one Killing field}},
  \href{http://dx.doi.org/10.1007/JHEP07(2011)115}{\emph{JHEP} {\bf 07} (2011)
  115}, [\href{https://arxiv.org/abs/1105.4167}{{\tt 1105.4167}}].

\end{thebibliography}\endgroup
\bibliographystyle{JHEP}
\end{document}